\documentclass[preprint,12pt]{elsarticle}

\usepackage{amsfonts,amsmath,amsfonts}
\usepackage{mathtools}
\usepackage{bm}
\usepackage{subcaption}
\usepackage{hyperref}
\usepackage{tikz}
\usetikzlibrary{quantikz}

\newcommand{\di}{i} 

\graphicspath{{Figures/}} 

\journal{}

\begin{document}

\begin{frontmatter}



\title{Reduced-order Modeling on a Near-term Quantum Computer} 


\author[label1]{Katherine Asztalos}\ead{kasztalos@anl.gov}
\author[label2]{Ren\'e Steijl}\ead{Rene.Steijl@glasgow.ac.uk}
\author[label1]{Romit Maulik}\ead{rmaulik@anl.gov}

\affiliation[label1]{organization={Argonne National Laboratory},
            addressline={9700 S Cass Avenue}, 
            city={Lemont},
            postcode={60439}, 
            state={IL},
            country={USA}}


\affiliation[label2]{organization={University of Glasgow},
            addressline={University Avenue}, 
            city={Glasgow},
            postcode={G12 8QQ}, 
            country={Scotland}}

\begin{abstract}

Quantum computing is an advancing area of research in which computer hardware and algorithms are developed to take advantage of quantum mechanical phenomena. In recent studies, quantum algorithms have shown promise in solving linear systems of equations as well as systems of linear ordinary differential equations (ODEs) and partial differential equations (PDEs). Reduced-order modeling (ROM) algorithms for studying fluid dynamics have shown success in identifying linear operators that can describe flowfields, where dynamic mode decomposition (DMD) is a particularly useful method in which a linear operator is identified from data. In this work, DMD is reformulated as an optimization problem to propagate the state of the linearized dynamical system on a quantum computer. Quadratic unconstrained binary optimization (QUBO), a technique for optimizing quadratic polynomials in binary variables, allows for quantum annealing algorithms to be applied. A quantum circuit model (quantum approximation optimization algorithm, QAOA) is utilized to obtain predictions of the state trajectories. Results are shown for the quantum-ROM predictions for flow over a 2D cylinder at Re = 220 and flow over a NACA0009 airfoil at Re = 500 and $\alpha = 15^\circ{}$. The quantum-ROM predictions are found to depend on the number of bits utilized for a fixed point representation and the truncation level of the DMD model. Comparisons with DMD predictions from a classical computer algorithm are made, as well as an analysis of the computational complexity and prospects for future, more fault-tolerant quantum computers. 
\end{abstract}




\end{frontmatter}


\section{Introduction}\label{intro}

Quantum computing (QC)~\cite{Chuang2010} is a fast-developing area of research investigating computing hardware and algorithms that take advantage of quantum mechanical phenomena. Broadly speaking, the longer term aims are to perform computations which may typically be intractable on so-called ``classical" computers. In recent years, significant progress has been made in building quantum computers. Despite this progress, a significant amount of research and development work remains before quantum computers can become practical for everyday use by computational scientist. The current and near-future quantum computers are often referred to as Noisy Intermediate-Scale Quantum (NISQ)-era hardware\cite{Preskill2018}. In the context of the Quantum Circuit Model this means that the relatively small number of qubits restricts the circuit width, while the sensitivity to quantum decoherence and noise means that the quantum circuit depth is severely restricted due to the limited time the qubit register will remain in a coherent state. Limited qubit connectivity imposes further constraints on the type of multi-qubit gate operations used.

Quantum gate operations used in quantum computers are significantly slower than operations in silicon-based processors. Therefore, achieving quantum algorithms with an exponential computational complexity improvement over the best classical approach is a key aim of quantum computing research. If that proves impossible, then a significant polynomial speed-up is the target. Following more than two decades of research, the number of (practically relevant) applications for which algorithms with exponential improvement exist is still very limited, e.g. simulation of quantum many-body systems, number theoretic problems such as integer factoring, and solving certain types of linear systems. The well-known Grover's search algorithm provides a quadratic improvement in complexity.

Computational quantum chemistry and quantum machine learning are among the areas of application receiving much recent research activity. Important developments for a wider range of applications include quantum algorithms for linear systems, e.g. the HHL algorithm~\cite{HHL2009} and its further developments, and quantum algorithms for the solution of (systems of) linear ordinary differential equations (ODEs) and linear partial differential equations (PDEs), e.g. \cite{Berry2014, Berry2017, Zanger2021}. A much smaller number of published works deal with nonlinear ODEs and PDEs, e.g.\cite{Leyton2008, Childs2021}.

In recent years, applications to computational science and engineering problems beyond quantum chemistry have appeared~\cite{OMalley2018, ray2019towards, Borle2018, Daley2018, Steijl2018, Steijl2020a, Steijl2020b, Gaitan2020, Budinski2021, Budinski2022, Moawad2022}. Despite this research effort, quantum computing applications in computational engineering have so far been limited.

A key aspect of these research works is the type of target quantum hardware.
In broad terms, quantum computing hardware can be categorized as an {\it adiabatic quantum computer} or as a {\it universal/digital quantum computer}. 

The first category involves works based on the concept of adiabatic quantum computing, or its more restricted form termed \textit{quantum annealing}. This approach is based on re-casting the computational problem in terms of a Quadratic Unconstrained Binary Optimization (QUBO) problem, for which then the quantum computer performs quantum annealing. A significant driving force behind this approach has been the availability of the DWave quantum annealers. The annealing approach fits well with optimization problems, and so far, only limited attempts were reported where the focus was directly on performing fluid simulations. The work of Ray and co-workers\cite{Ray2022} casts the (linear) equations for a fully-developed channel flow in terms of a QUBO and considers in detail the use of a fixed-point representation of the data in the QUBO formulation.

The second category involves efforts targeting \textit{universal quantum computers}, typically using the quantum circuit model. As detailed later, representing nonlinear operations in terms of the quantum circuit model poses fundamental challenges due to the linearity and reversibility of quantum mechanics. This challenge, along with the limitations of NISQ-era computers in terms of qubit count and circuit depth means that typically hybrid quantum/classical approaches were used with a tight coupling and low-depth quantum circuits. In the work of Gaitan\cite{Gaitan2020} the quasi one-dimensional inviscid flow in a converging-diverging duct with a normal shock wave was considered. The algorithm presented by Gaitan uses Kacewiz's quantum amplitude estimation ODE algorithm \cite{Kacewicz2006} as applied to the set of nonlinear ODE's resulting from standard discretization of the Navier-Stokes equations.
For the incompressible-flow Navier-Stokes equations, Steijl and Barakos\cite{Steijl2018} presented a vortex-in-cell based algorithm with a Poisson solver based on Quantum Fourier Transform. Based on the Lattice-Boltzann method, Budinksi\cite{Budinski2022} detailed a hybrid classical-quantum algorithm for the incompressible-flow Navier-Stokes equations in streamfunction-vorticity formulation. In a previous work, Budinski\cite{Budinski2021} presented a quantum algorithm based on the Lattice Boltzman equation for the linear convection/diffusion equations, where because of the linearity the quantum measurement and re-initialization at the end of each time step could be avoided when a suitable re-normalization is implemented to account for non-unitarity of operations.

Currently, the development of quantum-computer implementation of the Lattice Boltzmann method (LBM) forms an active area of research~\cite{Budinski2021,Succi2022,Moawad2022,Steijl2023a}, where the non-linearity of the governing equations in fluid mechanics forms the main challenge.

In applications targeting universal quantum computers using the quantum circuit model, the considered hardware needs to be examined in more detail.
Specifically, in general terms, two different lines of research can be discerned:
\begin{itemize}
    \item Work current and near-future NISQ-era quantum hardware. In this context, modelling fluid dynamics using quantum variational approaches has also been considered\cite{Lubasch2020}. In such approaches, the low-depth quantum circuits are used involving parameters that need to be obtained by optimization on a classical computer\cite{Peruzzo2014, McClean2016};
    \item Algorithm development work targeting more fault-tolerant universal quantum computers with a much higher level quantum error correction than available on NISQ-era hardware. The works by Gaitan, Budinski as well as Steijl and co-workers belong to this category.
\end{itemize}

\textbf{A key motivation behind the current work is to assess the potential for reduced-order models for flows of varying levels of complexity on NISQ-era hardware, taking into account the limited qubit count and restricted quantum circuit depth in the quantum circuit model.}

For the referenced works involving hybrid classical-quantum algorithms it remains largely unclear for what flow problems and on what future quantum computers a quantum speed-up can be achieved. With more \textit{fault-tolerant quantum computers} to appear further into the future, the overhead introduced by tightly coupling quantum and classical domains can be reduced, i.e. with more of the work performed by the quantum processor and with fewer quantum/classical data exchanges.
 

Further research work in quantum computing related to CFD, beyond ``traditional'' approaches, have also appeared in the literature and shows promise on NISQ-era hardware as well as future, more fault-tolerant hardware.
For applications in rarefied gas dynamics, Todorova and Steijl\cite{Steijl2020a} introduced an efficient quantum algorithm for the collisionless Boltzmann equation based on a discrete-velocity discretization and quantum walks. Because of the linearity of the governing equations as well as formulation in terms of strictly unitary operations on the quantum state encoding the flow solution, the presented algorithm facilitated multiple time-steps to be performed between quantum state initialization and quantum-measurement steps, i.e. no tight coupling of the quantum and classical domains was used.

A further alternative approach involves simulation methods based on machine learning\cite{Kyriienko2021} and reduced-order modeling (ROM). In recent years, investigations into machine learning (ML) and ROM for fluid dynamics has been a highly dynamic area of research. In particular, discussions on incorporating physical knowledge into the process of utilizing ML to build data-driven models for fluid mechanics highlights the opportunities of solving a variety of issues, especially when reformulated as optimization problems~\cite{brunton2021applying,vinuesa2021potential,vinuesa2022enhancing}. ROMs have been utilized in a variety of applications, such as flow modeling~\cite{aubry1988dynamics,noack2011reduced}, flow control~\cite{rabault2019artificial,tang2020robust,zheng2021active}, and optimization~\cite{legresley2000airfoil,loiseau2018constrained}. 


The present works introduces a novel approach based on a reformulation of DMD as an optimization problem on a quantum computer. As mentioned before, this approach was motivated by the limitations of NISQ-era quantum computing hardware. In taking this first step towards quantum-ROM modelling of flows, the current work makes the following contributions:
\begin{itemize}
    \item The formulation of DMD as an optimization problem in the form of a QUBO problem;
    \item The re-construction of 2D flow fields using this QUBO-based DMD with quantum annealing as well as application of the QAOA algorithms;
    \item Analysis of the sensitivity of the approach to precision of variables used in the transformation of the ROM to QUBO;
\end{itemize}
In future work, the computational complexity of the presented approach will be compared with alternative quantum algorithm approaches relying on a direct implementation of the matrix-vector multiplication.




The rest of this manuscript is structured as follows. Section \ref{sect_review} provides a more detailed review of existing works on quantum computing applications to fluid dynamics. A concise review of key principles of quantum computing as relevant to the present work is presented in Section \ref{background_qc}. The reformulation of DMD as an optimization problem to facilitate quantum computer implementation is discussed in Section \ref{qc-dmd_methods}. Section \ref{results_cylflow} then presents results obtained from the flow over a 2D cylinder. The application to a more demanding flow problem is presented in Section \ref{results_airfoil} where the flow over a NACA0009 airfoil is considered. The computational complexity is analysed in Section \ref{sect_complexity}. Finally, conclusions are drawn in Section \ref{sect_conclusions} along with an outline for future work.

\section{Review of related work in quantum computing}
\label{sect_review}
%
For many CFD applications, solving systems of linear equations or inverting matrices forms an important part of the solution process. A number of previous works have also considered the solution of a linear system of equations through applications of quantum annealing.

Rogers and Singleton~\cite{Rogers2020} investigated the application of quantum annealing to inverting (small) matrices ($2\times2$ and $3\times3$). As part of their work the division of two real numbers was also analyzed in detail. This division was reformulated as a QUBO problem with the output expressed as a $4$-bit fixed-point number. Similarly, the matrix inversion is re-cast in the form of QUBO with the output a vector of fixed-point numbers. 
They emphasize that their algorithm provides the full solution to the matrix problem, while HHL (or quantum algorithm for linear systems of equations~\cite{HHL2009}) provides only an expectation value. Furthermore, the algorithm as introduced in this work places no constraints on the matrix that is inverted, such as a sparsity condition. Yun and Yu~\cite{Jun2021qubo} proposed two representative QUBO models for $n\times n$ linear systems. They also provided Python code to create QUBO models that can be used in D-Wave quantum annealers.

For the solution of least-squares problems, there have been a few works related to quantum annealing. O’Malley and Vesselinov~\cite{OMalley2016} briefly explored using a quantum annealing machine for solving linear least squares problems for real numbers. They suggested that it is best suited for binary and sparse versions of the problem. Following from that work, Borle and Lomonaco~\cite{Borle2018} proposed a more compact way to represent variables using two’s and one’s complement on a quantum annealer. An in-depth theoretical analysis of this approach was presented, showing the conditions for which this method may be able to outperform the traditional classical methods for solving general linear least squares problems. Finally, based on their analysis and observations, the authors discussed potentially promising areas of further research where quantum annealing can be especially beneficial.

Zanger et al.~\cite{Zanger2021} explored the utilization of quantum computers for the purpose of solving differential equations. They consider two approaches: (i) basis encoding and fixed-point arithmetic on a digital quantum computer, and (ii) representing and solving high-order Runge-Kutta methods as optimization problems on quantum annealers. As realizations applied to two-dimensional linear ordinary differential equations, the authors devised and simulated corresponding digital quantum circuits, and implemented and ran a $6$th-order Gauss-Legendre collocation method on a D-Wave 2000Q system, showing good agreement with the reference solution. They found that the quantum annealing approach exhibits the largest potential for high-order implicit integration methods.

Fluid dynamics applications of quantum annealing have been relatively limited so far.
O'Malley applied the D-Wave 2X quantum annealer to solve 1D and 2D hydrologic inverse problems~\cite{OMalley2018}. This work relates to the flow and transport in an aquifer and requires knowledge of the heterogeneous properties of the aquifer, such as permeability. Computational methods for inverse analysis are commonly used to infer these properties from quantities that are more readily observable such as hydraulic head. While quantum computing is in an early stage compared to classical computing, this work demonstrated that it is sufficiently developed and can be used to solve certain subsurface flow problems. By modern standards, the considered problems were relatively small. As noted by the author, the presented results and the rapid progress being made with quantum computing hardware indicate that the era of quantum-computational hydrology may not be too far in the future.

Ray et al.~\cite{Ray2022} explored the suitability of adiabatic annealing based quantum computers, to solve fluid dynamics problems that form a critical component of several scientific and engineering applications. In their experiments, the authors start with a well-studied one-dimensional simple flow problem, and provide a framework to convert such problems in a continuum to a form amenable for deployment on such quantum annealers. The DWave annealer used in this work returns multiple states sampling the energy landscape of the problem. To address this, the authors explored multiple solution selection strategies to approximate the solution of the problem. In the proof-of-concept experiments, they analyze the continuum solutions obtained both qualitatively and quantitatively as well as their sensitivities to the particular solution selection scheme.

Quantum algorithms based on the finite-difference paradigm for linear high-dimensional and multiscale PDEs were investigated by Jin et al.\cite{Jin2022}. In their work, explicit and implicit time-discretization approaches were analyzed in terms of the potential for quantum advantages over classical difference methods. It was found that for multi-scale problems defined by a scaling factor $\epsilon$ the time complexity for the classical as well as quantum difference method scales as $O(1/\epsilon)$, i.e. in this context there was no quantum advantage relative to the equivalent classical finite difference approach.

Recently, Jin et al.\cite{Jin2023} constructed quantum algorithms to compute the solution and/or physical observables of nonlinear ODEs and nonlinear Hamilton-Jacobi equations (HJE) via linear representation or exact mapping between nonlinear ODEs/HJEs and linear partial differential equations. Specifically, the linear partial differential equations considered were the Liouville equation and the Koopman-von Neumann equation.

\section{Background on Quantum Computing Principles}\label{background_qc}

A thorough introduction and review of quantum computing is beyond the scope of this work. A brief review of some key concepts relevant to this work is presented here, with quantum computing approaches discussed in section~\ref{qc_approaches}, quantum annealing and adiabatic computing discussed in section~\ref{qc_anneal}, and quantum approximate optimization algorithms discussed in section~\ref{intro_qaoa}.

\subsection{Quantum Computing approaches}\label{qc_approaches}

Within Quantum Computing, there are a number of different approaches to be discerned:

\begin{itemize}
    \item Adiabatic quantum computing and its more restricted form ``quantum annealing". This approach has been thoroughly investigated in recent years and for a number of computational problems, generally in the form of ``optimization problems", in which significant progress has been made. A key challenges is the need to re-cast the numerical problem at hand into a Quadratic Unconstrained Binary Optimization (QUBO) problem in case annealing is used;
    \item Universal/digital quantum computing based on the Quantum Circuit model. This approach is more general than Quantum Annealing; however, there are many key challenges. In the Quantum Circuit model, the quantum gates acting on the quantum state vector perform unitary transformations. The requirement for unitary transformation (linear and reversible), as well as the no-cloning theorem in quantum mechanics, greatly complicate the implementation of computational algorithms in terms of quantum circuits. In particular, how to include non-linear terms (typically found in the governing equations of fluid mechanics) in the quantum-circuit model is a largely unresolved challenge; 
    \item Continuous-Variable Quantum computing. This approach originally introduced by Seth Lloyd in 1999~\cite{lloyd1999quantum} has not been widely used. However, more recently it has been gaining increased interest.
\end{itemize}

In this work, quantum annealing, as well as the quantum-circuit model, is considered.

\subsection{Quantum Annealing and Adiabatic Quantum Computing}\label{qc_anneal}

Quantum annealing was proposed as a quantum version of simulated annealing~\cite{Finnila1994,Kadowaki1998} and shortly thereafter, the related notion of adiabatic quantum computation was introduced~\cite{Farhi2000, Childs2001}.

D-Wave quantum annealers represent a novel computational architecture and have attracted significant interest, but have been used for few real-world computations. Machine learning has been identified as an area where quantum annealing may be useful. O'Malley et al.~\cite{O_Malley_2018} show that the D-Wave 2X can be effectively used as part of an unsupervised machine learning method. As suggested by the authors, their method can also be used to analyze large datasets. It was found that the D-Wave only limits the number of features that can be extracted from the dataset.

\subsection{Quantum Approximate Optimization Algorithm (QAOA)}\label{intro_qaoa}

The quantum approximate optimization algorithm (QAOA) is a variational method for solving combinatorial optimization problems on a gate-based quantum computer~\cite{FarhiQAOA}. In general terms, combinatorial optimization is the task of finding, from a finite number of objects, that object which minimizes a cost function.

The QAOA is based on a reformulation of the combinatorial optimization in terms of finding an approximation to the ground state of a Hamiltonian by adopting a specific variational ansatz for the trial wave function. This ansatz is specified in terms of a gate circuit and involves $2 p$ parameters which have to be optimized by running a minimization algorithm on a classical computer~\cite{Willsch2020}. In general, for finite $p$, there is no guarantee that the QAOA solution corresponds to the solution of the original combinatorial optimization problem. By viewing the QAOA as a form of quantum annealing using discrete time steps, it can be shown that for $p\rightarrow \infty$ (or vanishingly small time steps), the adaiabatic theorem of quantum mechanics guarantees that QAOA yields the correct answer if adiabatic conditions are satisfied. In addition, there exists a special class of models for which QAOA with $p=1$ solves the optimization problem exactly~\cite{Streif2019}.

Interest in the QAOA has increased dramatically in the past few years as it may, in contrast to Shor's factoring algorithm, lead to useful results even when used on NISQ devices. Moreover, the field of application, which is optimization, is much larger than, for example, factoring, rendering the QAOA a possible valuable application for gate-based quantum computers in general~\cite{Willsch2020}. A key feature making QAOA useful for current and near-future noisy quantum computers is the hybrid quantum/classical approach leading to relatively short time intervals during which a quantum state needs to stay coherent. Furthermore, variational approaches create a useful level of tolerance to noise.

Willsch et al.~\cite{Willsch2020} presented a critical assessment of the QAOA, based on results obtained by simulation, running the QAOA on the IBM Q Experience, and a comparison with data produced by the D-Wave 2000Q quantum annealer. In this comparison, 2-SAT and MaxCut problems were used as test cases. For benchmarking purposes, the authors only consider problems for which the solution, i.e. the true ground state of the problem Hamiltonian, is known. In this case, the success probability, i.e. the probability to sample the true ground state, can be used as the function to be minimized. The presented simulation data shows that the success of the QAOA based on minimizing the expectation value of the problem Hamiltonian strongly depends on the problem instance. In contrast, the authors concluded that for quantum annealing the dependency on the problem instance was significantly smaller.

Applying QAOA starts with identifying a ``cost" Hamiltonian typically defined as:

\begin{equation}
C = \sum_i h_i Z_i + \sum_{i,j} J_{i,j} |_i Z_j
\label{eq_QAOA_1}
\end{equation}
where $h_i$ and $J_{i,j}$ are real-valued coefficients that encode a QUBO problem in the eigenspectrum of $C$. The scaling of the QAOA was recently investigated by Lotshaw et al.\cite{Lotshaw2022}.

The conjugate transpose (or Hermitian adjoint matrix) of a complex matrix is the result of transposing the matrix and replacing its elements by their conjugates. A Hermitian matrix is a square matrix with complex entries that is equal to its own conjugate transpose. A real matrix is Hermitian if it is symmetric. A unitary matrix is a matrix whose inverse equals it conjugate transpose. Unitary matrices are particularly advantageous for quantum computing because they preserve norms, which in turn preserves probability amplitudes. Additionally, the eigenvalues of a unitary matrix all lie on the unit circle in the complex plane.

\section{Reduced-order Modeling on a Quantum Computer}\label{qc-dmd_methods}

There is an ever increasing ability to produce more high-fidelity data for researchers, which has led to an increase in dataset size and quantity. The ability to gain insight into complex flowfields through a simplified model gives rise to the increased interest in reduced-complexity modeling. It is becoming more common to utilize appropriate data analysis techniques that can extract physically relevant features from data, as well as develop a low-dimensional approximation of the dynamics of the full system in which the dimensionality is reduced but the dynamical evolution of the state is preserved. Not only do lower-dimensionality models aid in efficient and compact representations of the dynamics, but they are often required to develop efficient and accurate models for prediction, estimation, and control of the systems of interest \cite{rowley2017model,taira2017modal,taira2020modal}. As quantum computing has relatively reduced resources available, ROMs are advantageous to utilize due to their reduced dimensionality. 

Dynamic mode decomposition (DMD) \cite{schmid2010dynamic,schmid2011applications,schmid2011application} is a purely data-driven technique which has been used extensively in fluids applications \cite{ahuja2010feedback,hemati2017biasing,dawson2016characterizing}. DMD is commonly utilized to identify and analyze the dynamical features of fluid flows, in which spatial modes that evolve with their own characteristic frequency and growth/decay rates are identified. DMD also identifies a linear model that propagates the system forward in time, and has been shown to have a strong connection to the Koopman operator \cite{koopman1931hamiltonian}, an infinite-dimensional linear operator that fully captures the systems' nonlinear dynamics through a linear evolution of the functions of the state space. The link between DMD and the Koopman operator is valid for sufficiently rich datasets with a large set of observables in the system, where DMD can be interpreted as an approximation to the Koopman spectrum \cite{tu2014dynamic}. 

From the brief review in the previous sections, it is clear that effective techniques exist to create quantum circuit implementations of ODE and PDE solution methods, provided that the matrices resulting from time-integration (for ODEs) or space- and time-integration (PDEs) have a particular structure (e.g. Hankel, Toeplitz or circulant matrices). In general the operator obtained from DMD does not have a structure as defined for the Hankel or Toeplitz matrix. For the DMD model, the approach to integrate the system from time level $n$ to $n+1$ is to reformulate the problem as an optimization problem:

\begin{itemize}
    \item Using Quantum Annealing, the resulting optimization problem needs to be formulated in terms of a QUBO;
    \item Using Universal Quantum Computing (e.g. quantum circuit model), QAOA introduced by Farhi et al.\cite{FarhiQAOA} is used here. This quantum algorithm was designed to solve combinatorial problems. In general, combinatorial optimization problems involve finding an optimal object out of a finite set of objects. Typically, the problem is defined such that the solution follows from finding {\it optimal} bitstrings composed of 0's and 1's among a finite set of bitstrings.
\end{itemize}
In this work, both approaches are considered. 

\subsection{Leveraging Dynamic Mode Decomposition}

Dynamic mode decomposition (DMD) is a useful technique to study the evolution of dynamically important flowfield features. DMD obtains a wavespeed and frequency for each DMD mode $\psi_{j}$, 
\begin{equation}
    \textbf{u}(x,t) = \sum_{j=1}^{r} c_{j} e^{\lambda_{j}t}\psi_{j}, \label{DMD}
\end{equation}
where $\lambda_{j} = s_{j} + \di \omega_{j}$ are the DMD eigenvalues. This enables one to study the time evolution of relevant coherent structures in the system of interest, in addition to identifying a reduced-order model of the system which enables predictive abilities about the dynamics of the system.

To obtain a reduced-order model of the system, DMD identifies a linear operator, $\mathbf{A}$, that propagates the state of the system forward in time. Typically, given a state vector $\mathbf{x} \in \mathbb{R}^N$, DMD finds an operator $\mathbf{A} \in \mathbb{R}^{N \times N}$ i.e., 
\begin{align}
    \mathbf{x}^{n+1} = \mathbf{A} \mathbf{x}^{n}.
\end{align}
One may approximate $\mathbf{A}$ with
\begin{align}
    \mathbf{A} = \mathbf{Y} \mathbf{X}^\dagger
\end{align}
where $\mathbf{X} \in \mathbb{R}^{N \times M}$ and $\mathbf{Y} \in \mathbb{R}^{N \times M}$ are matrices with columns corresponding to snapshots of data and $^\dagger$ denotes the Moorse-Penrose pseudoinverse. To perform DMD, the snapshot matrices are formed such that $\mathbf{X}$ contains snapshots from $n=0$ to $n=M-1$ and $\mathbf{Y}$ contains snapshots from $n=1$ to $n=M$. Therefore, $\mathbf{A}$ can be precomputed from training data. Once available, one may construct an implicit computation of the state evolution through
\begin{align}
    \mathbf{A}^T \mathbf{A} \mathbf{x}^n = \mathbf{A}^T \mathbf{x}^{n+1} \\
    (\mathbf{A}^T \mathbf{A})^{-1} \mathbf{A}^T \mathbf{x}^{n+1} = \mathbf{x}^n \label{single_QUBO}
\end{align}
which leads to the following optimization problem
\begin{align}
    \label{OF_Equation}
    J = \min_{\mathbf{x}^{n+1}} || (\mathbf{A}^T \mathbf{A})^{-1} A^T \mathbf{x}^{n+1} - \mathbf{x}^n||
\end{align}

An effective technique in computing DMD is to use proper orthogonal decomposition (POD) as a truncated subspace. Data can be collected from any system of interest and arranged to perform POD; for each snapshot in time, relevant state information is collected and POD is performed on the state matrix $\mathbf{X}$. The singular value decomposition (SVD) of the state matrix
\begin{equation}
    \textbf{X} = \textbf{U} {\bf\Sigma} \textbf{V}^{T} = \sum_{j=1}^{r}\sigma_{j}u_{j}v_{j}^{T} \label{svd}
\end{equation} 
yields the orthonormal POD spatial modes $u_{j}$ contained as columns in $\textbf{U}$ for an $r$-dimensional subspace, the singular values $\sigma_{j}$ ordered in monotonically decreasing order in the pseudo-diagonal matrix $\bf\Sigma$, and the right singular vectors $v_{j}$ as columns of $\mathbf{V}$. The right singular vectors contain information about the time-dependent evolution of the POD mode coefficients. POD produces an optimal fit of the data in the minimization of the $l_2$-norm of the error between the original data and a low-rank approximation using a subspace of POD modes. To perform DMD in the space of POD mode coefficients, the snapshot matrices $\mathbf{X}$ and $\mathbf{Y}$ are defined as follows: 
\begin{equation}
\tilde{\textbf{X}} = \textbf{U}_r^* \textbf{X}, \ \ \ \ \tilde{\textbf{Y}} = \textbf{U}_r^* \textbf{Y},
\end{equation}
from which the DMD operator can be computed as
\begin{equation}
    \tilde{\textbf{A}} = \tilde{\textbf{Y}} \tilde{\textbf{X}}^\dagger \label{eq:DMD}.
\end{equation}
For most high-dimensional fluid dynamical simulations, equation~(\ref{eq:DMD}) is not typically explicitly computed from full state data, as $n \gg m$ usually. Instead, the equivalent operator $\tilde{\textbf{A}}$ is computed in the space of POD mode coefficients, with the resultant operators having rank at most of min($m,n$), a much more efficient representation in a lower dimension. The operator $\tilde{\textbf{A}}$ is related to the full state propagation matrix $\textbf{A}$ via a projection onto the POD modes. The DMD modes and eigenvalues can then be computed through an eigendecomposition of the matrix $\textbf{A}$. For further details on DMD and general algorithm formulations, the interested reader is referred to \cite{tu2014dynamic,dawson2017reduced}.

\subsection{Quadratic Unconstrained Binary Optimization}

Equation~(\ref{OF_Equation}) represents an unconstrained optimization problem, the result of which updates the state of the linearized dynamical system. This presents an opportunity for using quadratic unconstrained binary optimization (QUBO), a technique for optimizing quadratic polynomials in binary variables, for solving this system. Following~\cite{Ray2022}, we can formulate the optimization problem in QUBO form as follows. First, we seek a binary representation of the state vector $\mathbf{x}$ given by $\tilde{\mathbf{x}}$, where one has
\begin{align}
    \mathbf{x} \approx B \tilde{\mathbf{x}}
\end{align}
as a relation to relate the state in real space to binary space. Here $B$ represents a linear expansion determined by the precision retained for the real-state representation. Subsequently our modified optimization problem becomes
\begin{align}
    \label{OF_Equation_QUBO}
    J_{\tilde{\mathbf{x}}} = \min_{\tilde{\mathbf{x}}^{n+1}} || \mathbf{A}^d \tilde{\mathbf{x}}^{n+1} - \mathbf{b}||.
\end{align}
where 
\begin{align}
    \mathbf{A}^d = (\mathbf{A}^T \mathbf{A})^{-1} \mathbf{A}^T B \\
    \mathbf{b} = B \tilde{\mathbf{x}}^n.
\end{align}
Note that $\tilde{\mathbf{x}}$ is now a binary-valued vector. A quantum annealer can be provided with the following form of the objective function given by
\begin{align}
    J_{\tilde{\mathbf{x}}} = \sum_i v_i \tilde{\mathbf{x}}_i + \sum_{i < j} w_{ij} \tilde{\mathbf{x}}_i \tilde{\mathbf{x}}_j
\end{align}
with 
\begin{align}
    v_j = \sum_i \mathbf{A}^d_{ij} (\mathbf{A}^d_{ij} -2 b_i) \\
    w_{jk} = 2 \sum_i \mathbf{A}^d_{ij} \mathbf{A}^d_{ik}
\end{align}

Once a QUBO formulation is constructed as shown above, an annealing based algorithm may be used to solve for the binary valued state vector at each time step. In practice, several iterations are used to resolve the QUBO problem on an annealer and the solution corresponding to the lowest objective value is selected for the state update. 

\subsection{Performing Multiple Steps in QUBO}

To obtain predictions for $n$ number of timesteps, one can extend equation~(\ref{single_QUBO}) $n$ times as:

\begin{eqnarray}
&& (\mathbf{A}^T \mathbf{A})^{-1} \mathbf{A}^T \mathbf{x}^{n+1} - \mathbf{x}^n = 0 \nonumber\\
&& (\mathbf{A}^T \mathbf{A})^{-1} \mathbf{A}^T \mathbf{x}^{n+2} - \mathbf{x}^{n+1} = 0\nonumber
\end{eqnarray}

By creating a $2N\times 2N$ matrix, this can be re-written as,

\begin{equation}
\left(\begin{array}{cc}
(\mathbf{A}^T \mathbf{A})^{-1} \mathbf{A}^T & -I\\
0 &(\mathbf{A}^T \mathbf{A})^{-1} A^T
\end{array}\right)
\left(\begin{array}{c}
\mathbf{x}^{n+2}\\
\mathbf{x}^{n+1}
\end{array}\right)
=
\left(\begin{array}{c}
0\\
\mathbf{x}^n
\end{array}\right)
\end{equation}

The results were simulated using Qiskit available within Python. The repository for the code is available on Github. \footnote{Code can be found at \url{test}.} In the following sections, DMD will be applied to two different test cases; flow over a 2D cylinder and flow over a 2D NACA0009 airfoil. In both cases, QUBO and QUBO-QAOA are applied, with results shown for increasing complexity (i.e., increasing the number of bits required for a fixed-point representation). 

Related work to investigating a particular structure associated with DMD matrices is the physics-informed dynamic mode decomposition method (piDMD) proposed by Baddoo et al.~\cite{baddoo2023physics}. The authors investigate how physical principals, such as symmetries and conservation laws, can be integrated into DMD and by restricting the family of admissible models to the matrix manifold, a model which respects the physical structures of the system is obtained. As it was shown that piDMD identifies a conservative (i.e., unitary) operator for flow over a cylinder, it is of potential interest to investigate variants of DMD to incorporate with quantum computing to utilize advantages available given a specific matrix structure associated with physical problems in fluid mechanics. 

\section{Quantum-ROM Predictions for Flow over a 2D Cylinder}\label{results_cylflow}

Experiments were performed for varying truncation levels and precision levels corresponding to the number of bits required for a fixed-point representation. A ROM of the system was first identified using DMD, and the results from implicit DMD was compared with predictions generated using the quantum-ROM with both QUBO and QUBO-QAOA methodologies.

\begin{figure}[!htb]
\begin{subfigure}[!htb]{.5\textwidth}
 \centering
  \includegraphics[trim=0cm 0cm 0cm 0cm,clip,width=1\linewidth]{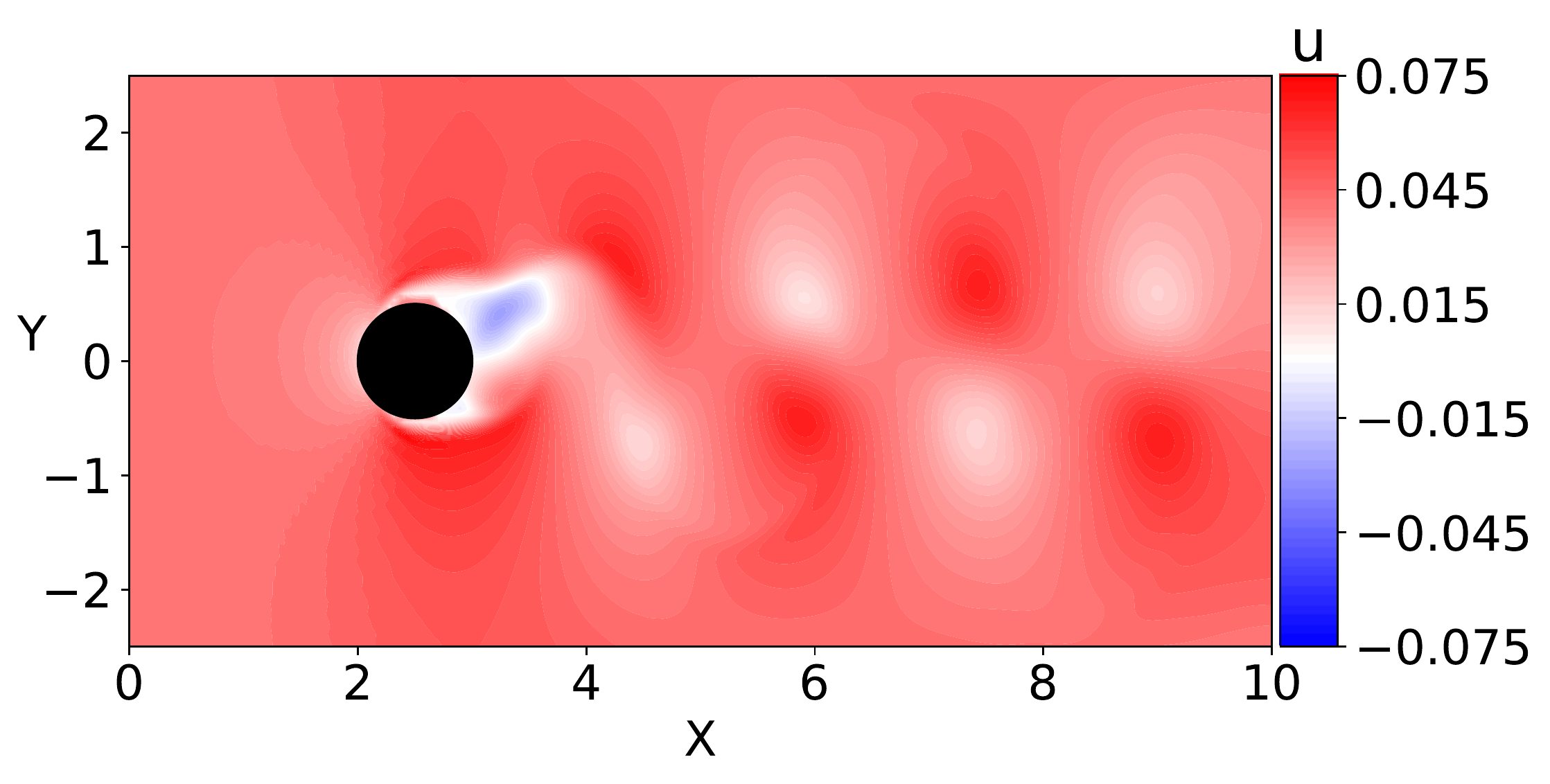}
  \caption{}
  \label{fig:true_u1}
\end{subfigure}
\begin{subfigure}[!htb]{.5\textwidth}
 \centering
  \includegraphics[trim=0cm 0cm 0cm 0cm,clip,width=1\linewidth]{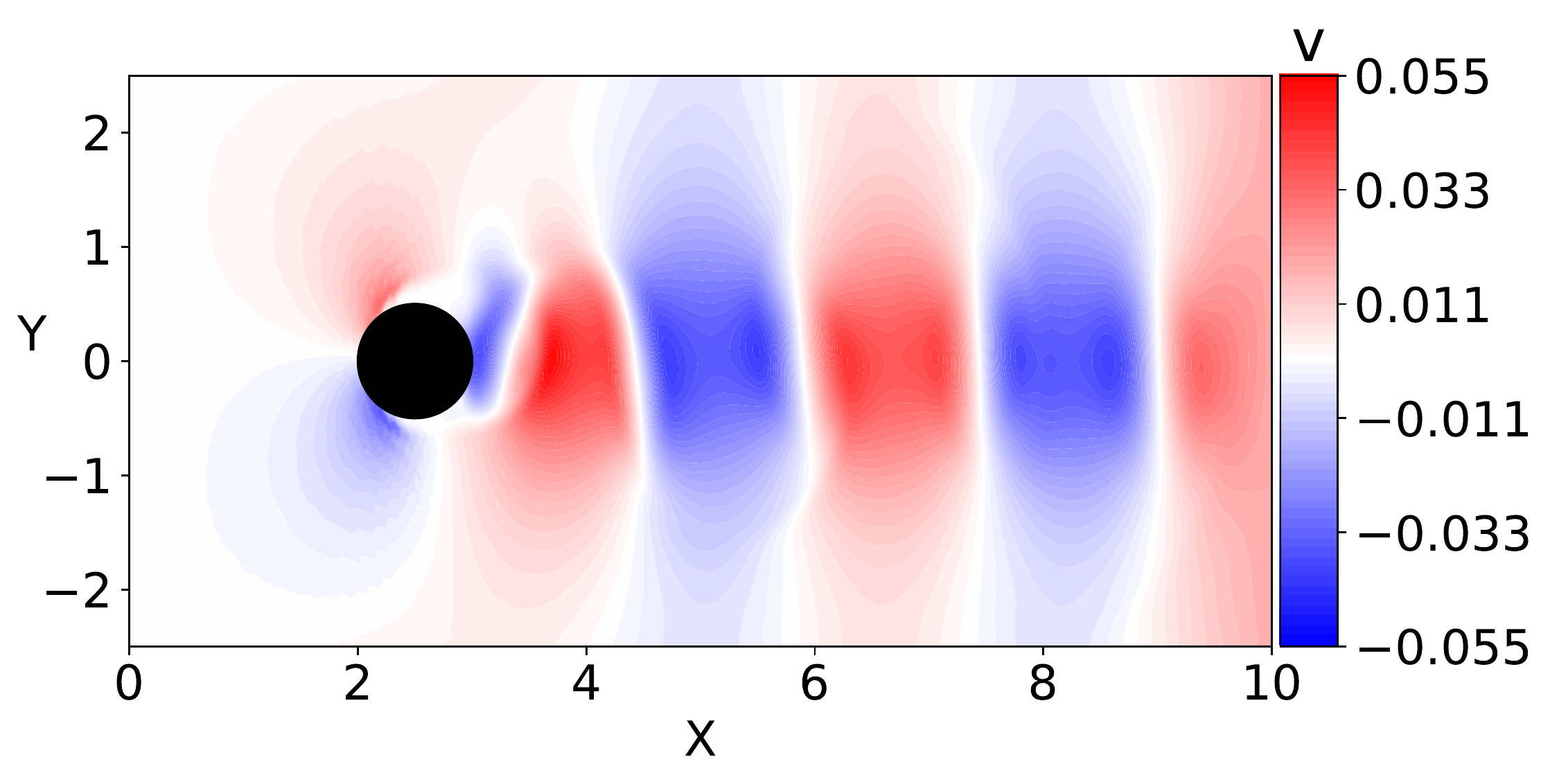}
  \caption{}
  \label{fig:true_v1}
\end{subfigure}
\caption{The true flowfield for steady flow over a 2D cylinder showing (a) the u-component and (b) the v-component of velocity.}
\label{fig:true_flowfield_cyl_flow}
\end{figure}

\begin{figure}[!htb]
\begin{subfigure}[!htb]{.5\textwidth}
 \centering
  \includegraphics[trim=0cm 0cm 0cm 0cm,clip,width=1\linewidth]{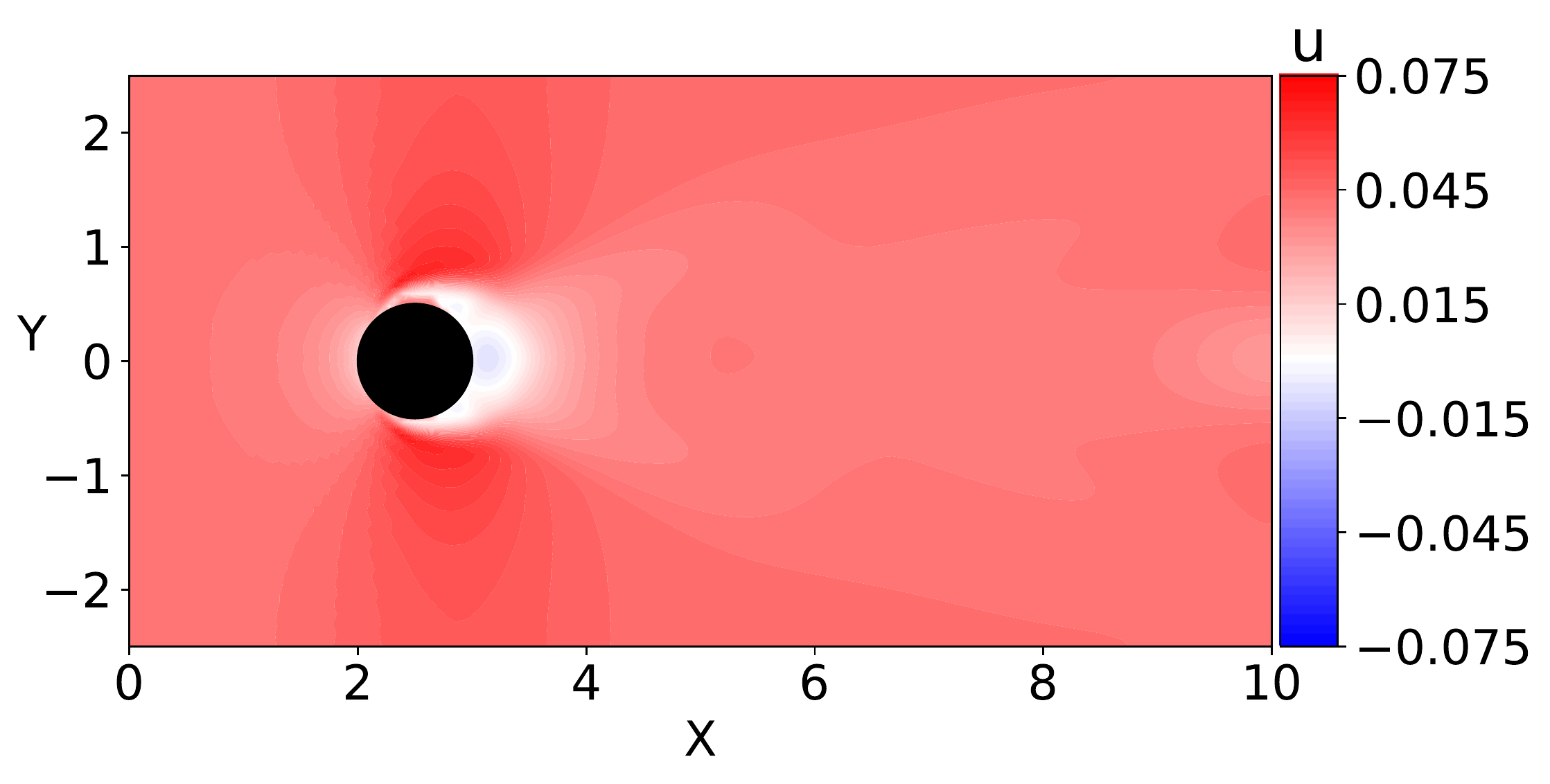}
  \caption{}
  \label{fig:true_umean}
\end{subfigure}
\begin{subfigure}[!htb]{.5\textwidth}
 \centering
  \includegraphics[trim=0cm 0cm 0cm 0cm,clip,width=1\linewidth]{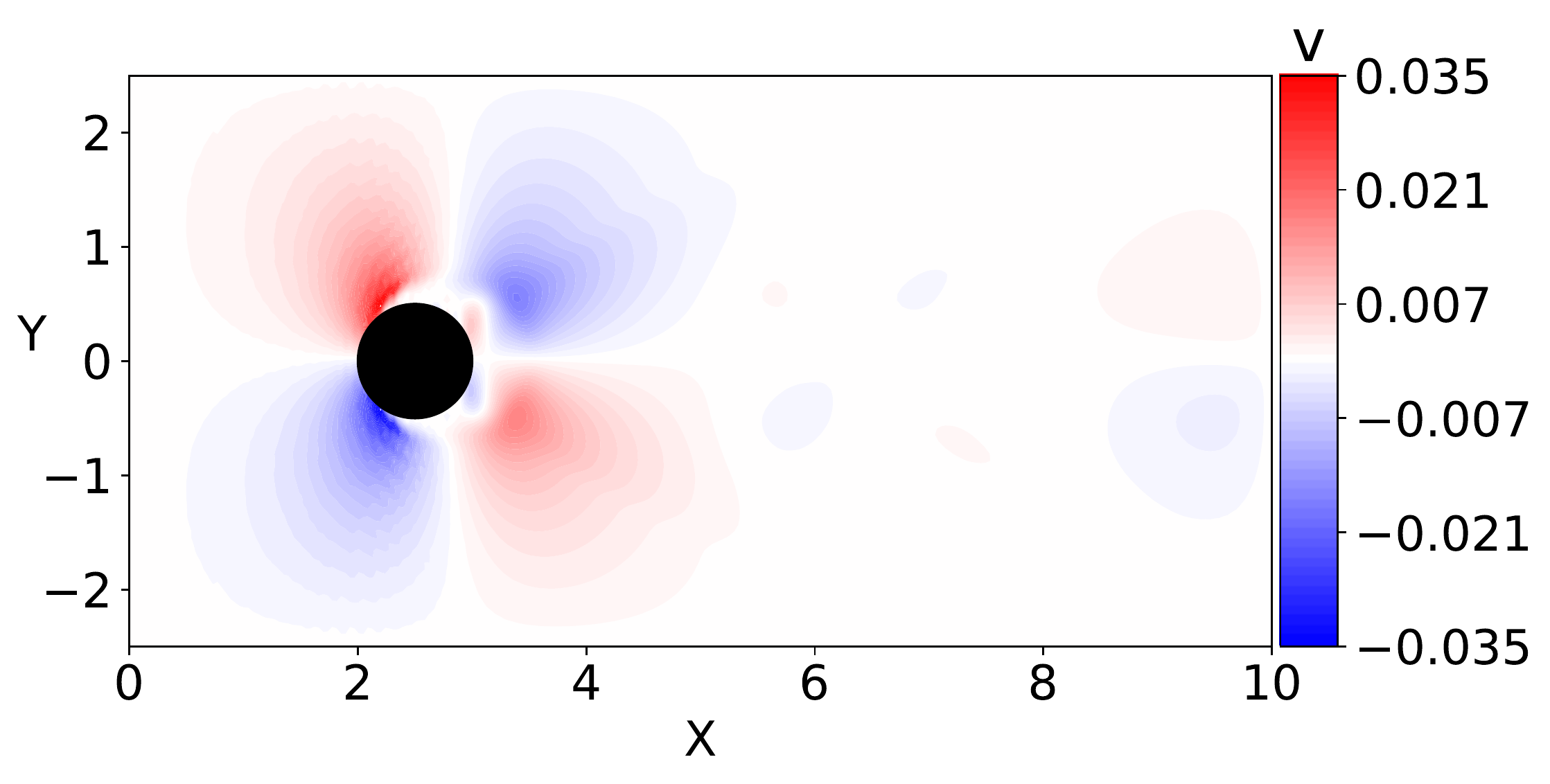}
  \caption{}
  \label{fig:true_vmean}
\end{subfigure}
\caption{The mean flowfield for flow over a 2D cylinder showing (a) the u-component and (b) the v-component of velocity.}
\label{fig:mean_flowfield_cyl_flow}
\end{figure}

Flow over a 2D cylinder at a Reynolds number of 220\footnote{Data can be visualized at \url{https://www.youtube.com/watch?v=M2PqI2JD2jo}.} was used as a test case to compute predictions using QUBO and annealing-based DMD. The true steady flowfield at a timestep corresponding to t = 10 is shown in figure~\ref{fig:true_flowfield_cyl_flow}, with the mean flow shown in figure~\ref{fig:mean_flowfield_cyl_flow}. It can clearly be seen that there exists a steady-state vortex street in the wake downstream of the cylinder. 

\begin{figure}[!htb]
\centering
\begin{subfigure}[!htb]{.45\textwidth}
 \centering
  \includegraphics[trim=0cm 0cm 0cm 0cm,clip,width=1\linewidth]{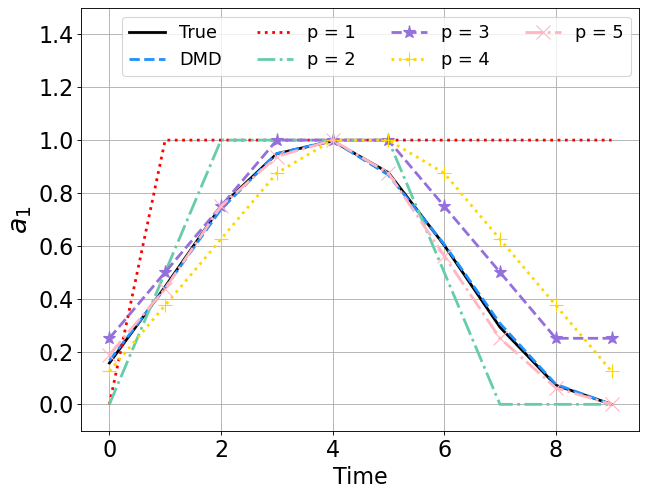}
  \caption{}
  \label{fig:QUBO_r5_a1}
\end{subfigure}
\begin{subfigure}[!htb]{.45\textwidth}
 \centering
  \includegraphics[trim=0cm 0cm 0cm 0cm,clip,width=1\linewidth]{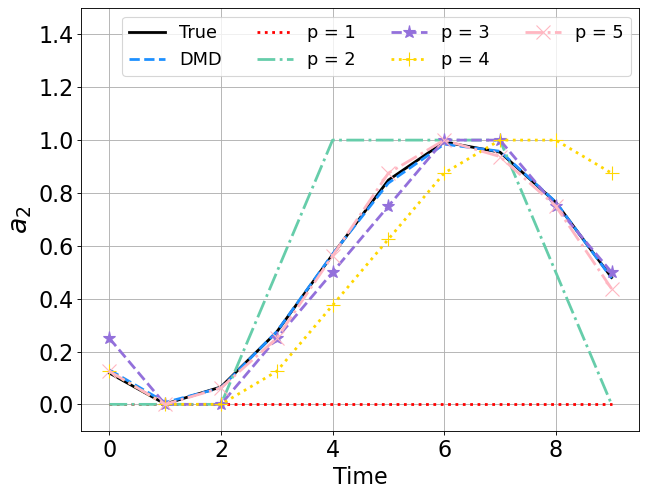}
  \caption{}
  \label{fig:QUBO_r5_a2}
\end{subfigure}
\begin{subfigure}[!htb]{.45\textwidth}
 \centering
  \includegraphics[trim=0cm 0cm 0cm 0cm,clip,width=1\linewidth]{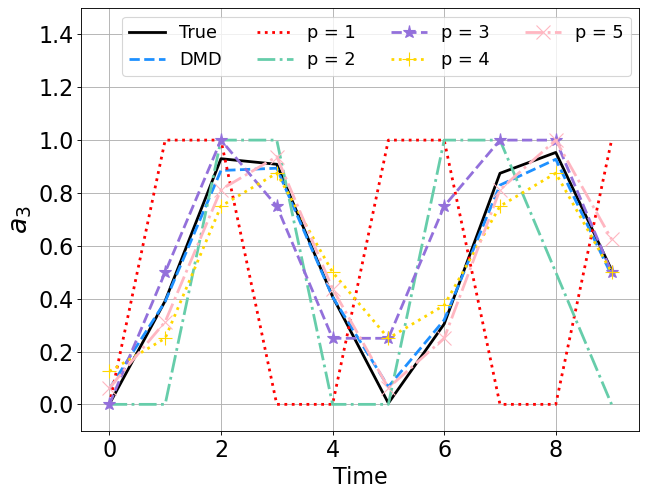}
  \caption{}
  \label{fig:QUBO_r5_a3}
\end{subfigure}
\begin{subfigure}[!htb]{.45\textwidth}
 \centering
  \includegraphics[trim=0cm 0cm 0cm 0cm,clip,width=1\linewidth]{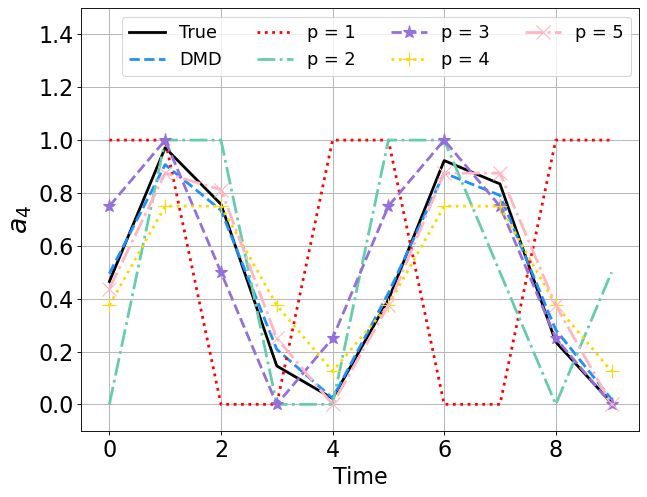}
  \caption{}
  \label{fig:QUBO_r5_a4}
\end{subfigure}
\caption{QUBO predictions compared to a DMD model with r = 5 for mode coefficients (a) $a_1$ - (d) $a_4$ for varying precision levels.}
\label{fig:QUBO_r5_results}
\end{figure}


\begin{figure}[!htb]
\centering
\begin{subfigure}[!htb]{.45\textwidth}
 \centering
  \includegraphics[trim=0cm 0cm 0cm 0cm,clip,width=1\linewidth]{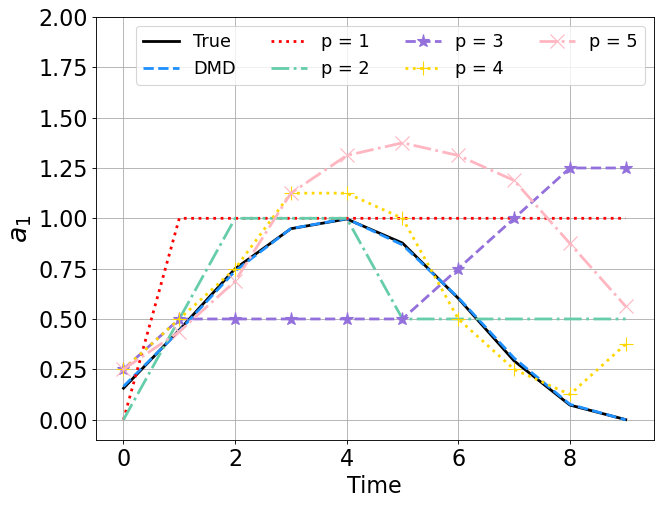}
  \caption{}
  \label{fig:QUBO-QAOA_r5_a1}
\end{subfigure}
\begin{subfigure}[!htb]{.45\textwidth}
 \centering
  \includegraphics[trim=0cm 0cm 0cm 0cm,clip,width=1\linewidth]{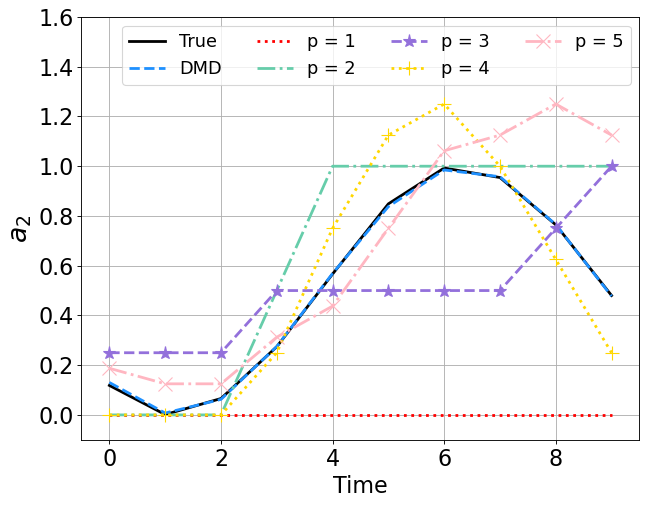}
  \caption{}
  \label{fig:QUBO-QAOA_r5_a2}
\end{subfigure}
\begin{subfigure}[!htb]{.45\textwidth}
 \centering
  \includegraphics[trim=0cm 0cm 0cm 0cm,clip,width=1\linewidth]{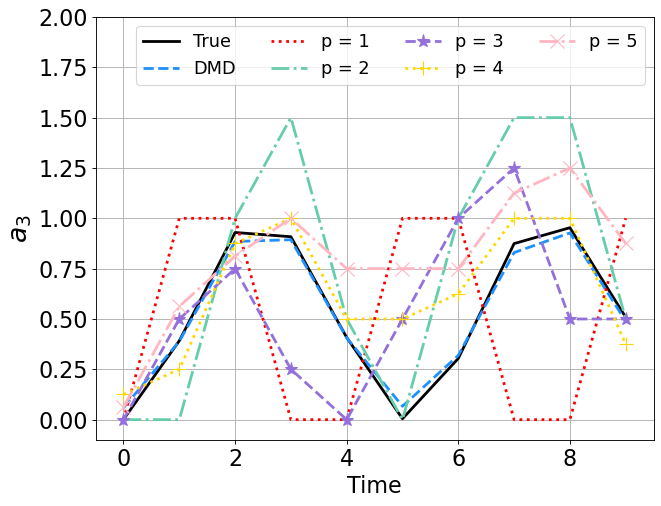}
  \caption{}
  \label{fig:QUBO-QAOA_r5_a3}
\end{subfigure}
\begin{subfigure}[!htb]{.45\textwidth}
 \centering
  \includegraphics[trim=0cm 0cm 0cm 0cm,clip,width=1\linewidth]{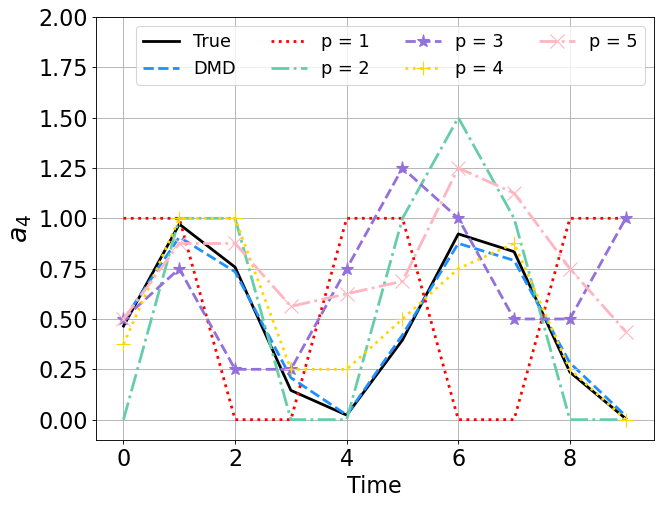}
  \caption{}
  \label{fig:QUBO-QAOA_r5_a4}
\end{subfigure}
\caption{QUBO-QAOA predictions compared to a DMD model with r = 5 for mode coefficients (a) $a_1$ - (d) $a_4$ for varying precision levels.}
\label{fig:QUBO-QAOA_r5_results}
\end{figure}

Results from predictions generated using a QUBO simulator for varying precision levels compared to predictions from an implicit DMD model with a truncation level of r = 5 for flow over a cylinder can be seen in figure~\ref{fig:QUBO_r5_results}. Increasing the number of bits (while maintaining a constant truncation level) has a similar effect as increasing the truncation level for implicit DMD does; namely, that the prediction becomes more accurate in capturing the mode coefficients for higher values of bits retained for a fixed-point representation. It is interesting to observe that for low values of bits utilized (p = 1 and 2), the QUBO prediction completely misses the dominant leading modes (see figure~\ref{fig:QUBO_r5_a1} -- \ref{fig:QUBO_r5_a2}), while it is able to pick up the higher-order modes with some level of accuracy (particularly in the amplitude, while there is a significant shift in the phase; see figures~\ref{fig:QUBO_r5_a3} -- \ref{fig:QUBO_r5_a4}). However, for higher number of bits tested (p = 5) the QUBO model is able to accurately capture all of the modes, including the leading modes. In this case, predictions from higher number of bits utilized (p $>$ 5) were not obtained due to their computational cost. Finally, it is possible that extending the number of snapshots in the model would yield a more accurate prediction for fewer bits.

Figure~\ref{fig:QUBO-QAOA_r5_results} shows the results for predictions generated from the quantum annealer QUBO-QAOA model for varying the number of bits utilized compared to predictions from the implicit DMD model with a truncation level of r = 5 for flow over a cylinder. It is interesting to note that although the QUBO model is able to obtain a very accurate prediction of the state for the highest number of bits tested (p = 5), the QAOA-QUBO model is unable to completely capture the evolution of the state, and for fewer number of bits tested (p = 1 and 2), the prediction is quite inaccurate. As observed with the QUBO model, the low number of bit models (p = 1 and 2) are not able to predict the leading modes $a_1$ and $a_2$ accurately (see figures~\ref{fig:QUBO-QAOA_r5_a1} -- \ref{fig:QUBO-QAOA_r5_a2}), while they are able to obtain a better prediction for the higher-order modes $a_3$ and $a_4$ (see figures~\ref{fig:QUBO-QAOA_r5_a3} -- \ref{fig:QUBO-QAOA_r5_a4}). Again, there is a significant phase shift between the predicted and true modes, but the amplitude is well-predicted.

\begin{figure}[!htb]
\begin{subfigure}[!htb]{.5\textwidth}
 \centering
  \includegraphics[trim=0cm 0cm 0cm 0cm,clip,width=1\linewidth]{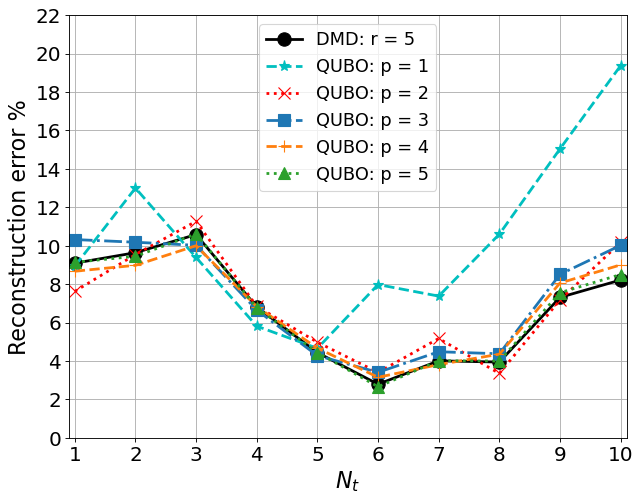}
  \caption{}
  \label{fig:Recon_error_QUBO_r5_pall}
\end{subfigure}
\begin{subfigure}[!htb]{.5\textwidth}
 \centering
  \includegraphics[trim=0cm 0cm 0cm 0cm,clip,width=1\linewidth]{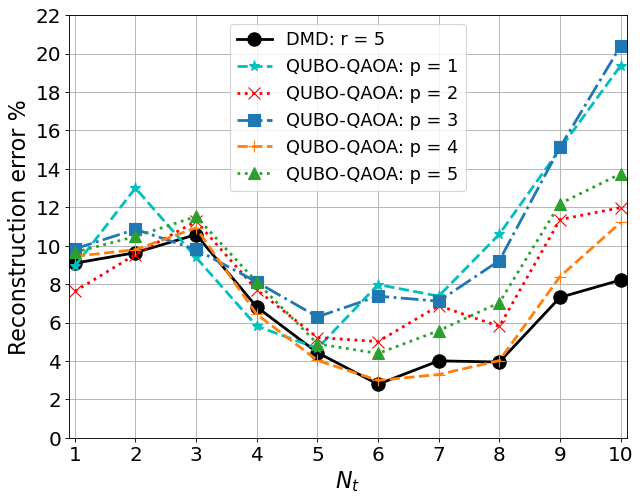}
  \caption{}
  \label{fig:Recon_error_QUBO-QAOA_r5_pall}
\end{subfigure}
\caption{Reconstruction error computed from (a) QUBO and (b) QUBO-QAOA predictions compared to a DMD model with r = 5 for varying number of bits utilized in the quantum-ROMs.}
\label{fig:QUBO_QUBO-QAOA_r5_recon_error_results}
\end{figure}

Figure~\ref{fig:QUBO_QUBO-QAOA_r5_recon_error_results} shows a comparison between the reconstruction error computed from QUBO (figure~\ref{fig:Recon_error_QUBO_r5_pall}) and QUBO-QAOA (figure~\ref{fig:Recon_error_QUBO-QAOA_r5_pall}) for a r = 5 model. It can be seen that as the number of bits included in the model increases, QUBO attains a lower reconstruction error, eventually recovering the same error as implicit DMD when using p = 5 bits. This is also true in general for QUBO-QAOA, with slight exceptions that might occur due to error accumulation when using more bits for a fixed-point representation.

\begin{figure}[!htb]
\centering
\begin{subfigure}[!htb]{.45\textwidth}
 \centering
  \includegraphics[trim=0cm 0cm 0cm 0cm,clip,width=1\linewidth]{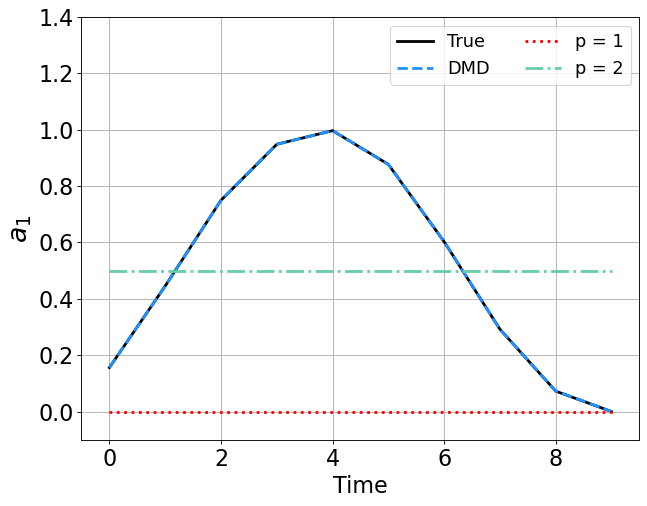}
  \caption{}
  \label{fig:QUBO_r8_a1}
\end{subfigure}
\begin{subfigure}[!htb]{.45\textwidth}
 \centering
  \includegraphics[trim=0cm 0cm 0cm 0cm,clip,width=1\linewidth]{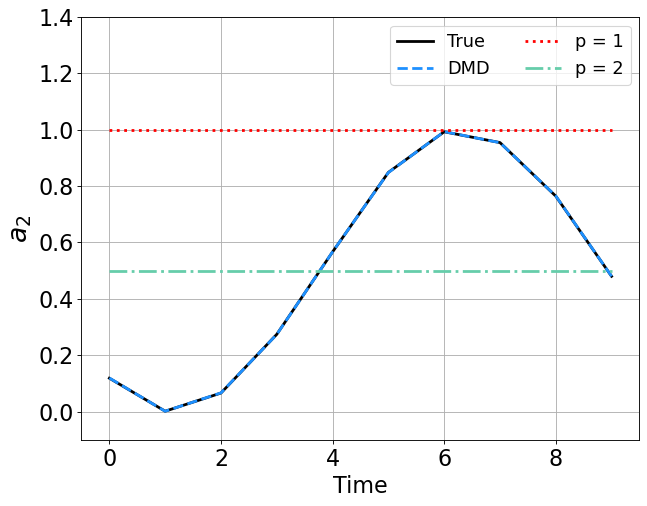}
  \caption{}
  \label{fig:QUBO_r8_a2}
\end{subfigure}
\begin{subfigure}[!htb]{.45\textwidth}
 \centering
  \includegraphics[trim=0cm 0cm 0cm 0cm,clip,width=1\linewidth]{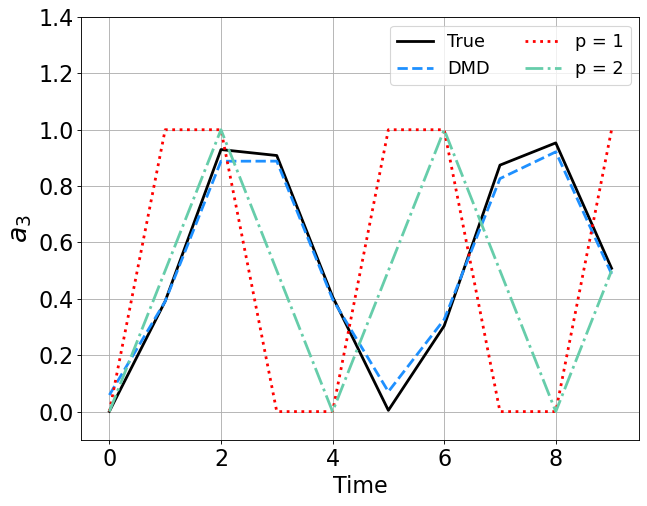}
  \caption{}
  \label{fig:QUBO_r8_a3}
\end{subfigure}
\begin{subfigure}[!htb]{.45\textwidth}
 \centering
  \includegraphics[trim=0cm 0cm 0cm 0cm,clip,width=1\linewidth]{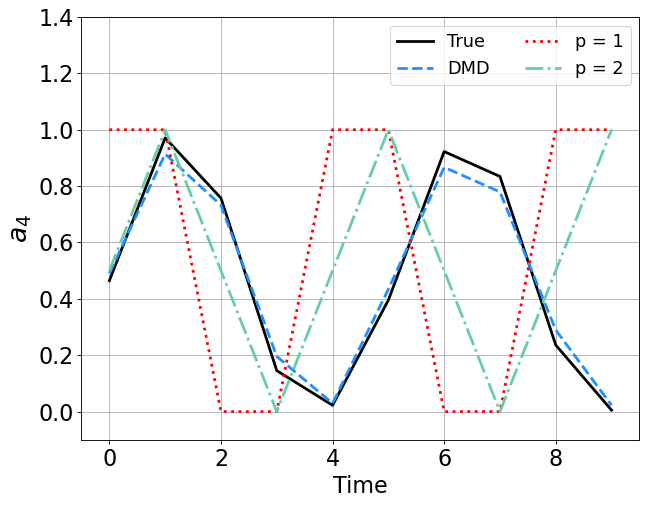}
  \caption{}
  \label{fig:QUBO_r8_a4}
\end{subfigure}
\caption{QUBO predictions compared to a DMD model with r = 8 for mode coefficients (a) $a_1$ - (d) $a_4$ for p = 1 and p = 2.}
\label{fig:QUBO_r8_results}
\end{figure}

In order to gain an understanding of the importance of the truncation level on model predictions from the QUBO and QUBO-QAOA models, the truncation level was increased to r = 8. In general, r = 5 modes captures $\approx 90\%$ of the kinetic energy in the flowfield for this particular configuration. Figure~\ref{fig:QUBO_r8_results} shows the results for predictions generated from a QUBO simulator for different numbers of bits compared to predictions from an implicit DMD model with a truncation level of r = 8 for flow over a cylinder. Results were obtained for low number of bits (p = 1 and 2) due to high computational costs with the larger truncation level.

\begin{figure}[htb!]
\centering
\begin{subfigure}[!htb]{.45\textwidth}
 \centering
  \includegraphics[trim=0cm 0cm 0cm 0cm,clip,width=1\linewidth]{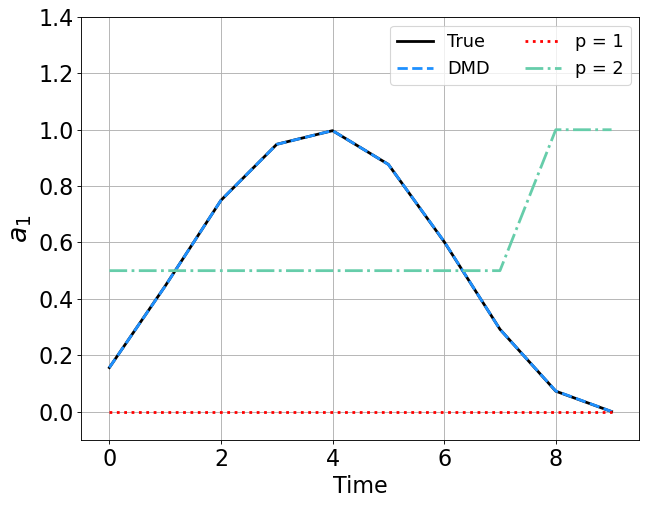}
  \caption{}
  \label{fig:QUBO-QAOA_r8_a1}
\end{subfigure}
\begin{subfigure}[!htb]{.45\textwidth}
 \centering
  \includegraphics[trim=0cm 0cm 0cm 0cm,clip,width=1\linewidth]{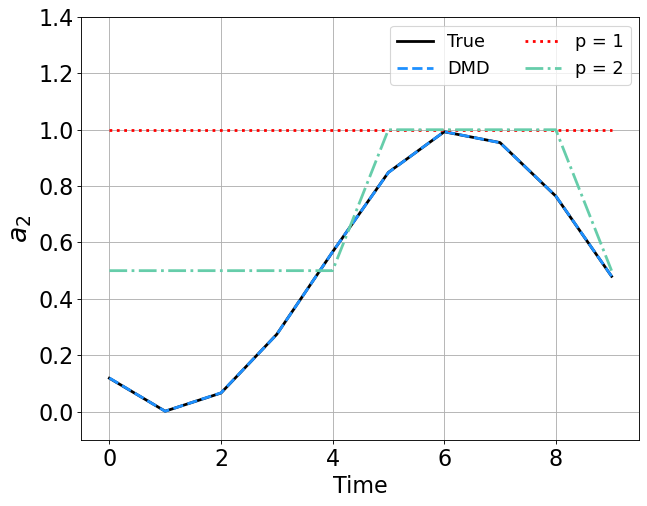}
  \caption{}
  \label{fig:QUBO-QAOA_r8_a2}
\end{subfigure}
\begin{subfigure}[!htb]{.45\textwidth}
 \centering
  \includegraphics[trim=0cm 0cm 0cm 0cm,clip,width=1\linewidth]{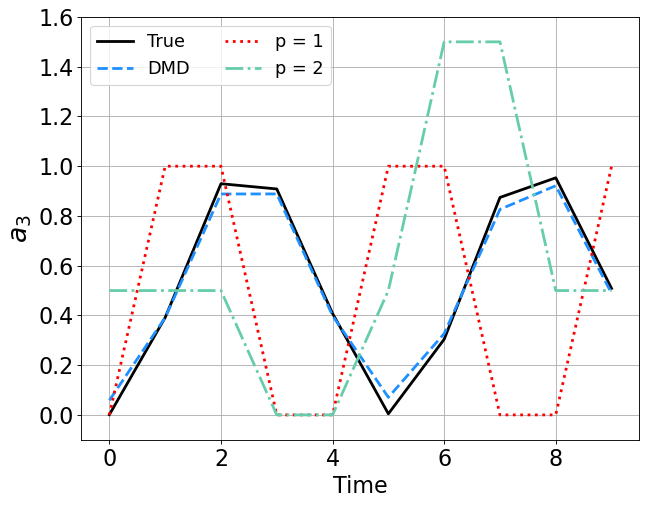}
  \caption{}
  \label{fig:QUBO-QAOA_r8_a3}
\end{subfigure}
\begin{subfigure}[!htb]{.45\textwidth}
 \centering
  \includegraphics[trim=0cm 0cm 0cm 0cm,clip,width=1\linewidth]{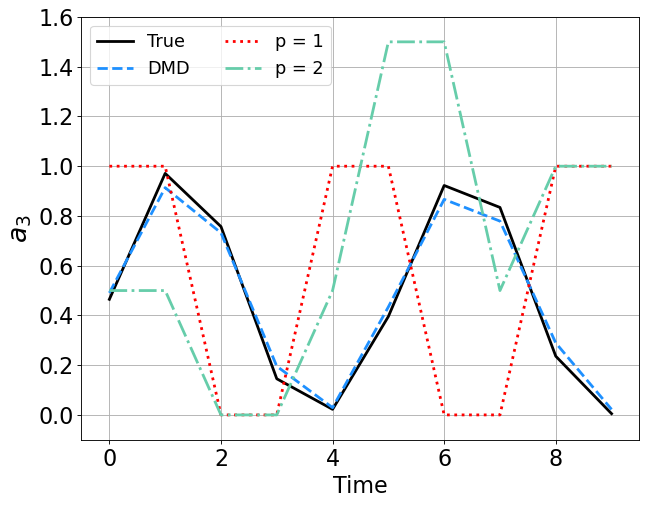}
  \caption{}
  \label{fig:QUBO-QAOA_r8_a4}
\end{subfigure}
\caption{QUBO-QAOA predictions compared to a DMD model with r = 8 for mode coefficients (a) $a_1$ - (d) $a_4$ for p = 1 and p = 2.}
\label{fig:QUBO-QAOA_r8_results}
\end{figure}

The model fails to predict the leading modes (figures~\ref{fig:QUBO_r8_a1} -- \ref{fig:QUBO_r8_a2}) for both the p = 1 and p = 2 case, yielding only a constant value. However, for the higher modes, the model is able to capture the behavior relatively better; again, including a significant phase-shift, but capturing the amplitude quite well, as observed in figures~\ref{fig:QUBO_r8_a3} -- \ref{fig:QUBO_r8_a4}. As the underlying state incorporates higher harmonics of the fundamental frequency in the higher-order modes, it is interesting that the QUBO model seems to have more success at predicting the harmonics of the state, rather than the fundamental frequency, which is a bit counter-intuitive given that most data-driven models (for instance, DMD) typically can capture the most energetic state of the system (which would correspond to the fundamental frequency here) and may struggle to accurately capture the higher harmonics for lower truncation levels. This is an interesting observation for differences in the quantum architecture with DMD as compared to the classical ROM methods. However, this could be sensitive to the number of snapshots including in identifying the POD subspace, and for longer times this may not be present.

Figure~\ref{fig:QUBO-QAOA_r8_results} shows the results for predictions generated from the quantum annealer QUBO-QAOA model for varying number of bits compared to predictions from an implicit DMD model with a truncation level of r = 8 for flow over a cylinder. Results were only obtained for p = 1 and 2 due to high computational costs. As before with the QUBO model predictions, the QUBO-QAOA model is able to generate better predictions for the higher-order modes (figures~\ref{fig:QUBO-QAOA_r8_a3} -- \ref{fig:QUBO-QAOA_r8_a4}) while struggling to capture the leading modes (figures~\ref{fig:QUBO-QAOA_r8_a1} -- \ref{fig:QUBO-QAOA_r8_a2}). In this case, however, the QUBO-QAOA model is able to pick up some portion of the second mode (see figure~\ref{fig:QUBO-QAOA_r8_a2}) that the QUBO model misses.

\begin{figure}[htb!]
\begin{subfigure}[!htb]{.5\textwidth}
 \centering
  \includegraphics[trim=0cm 0cm 0cm 0cm,clip,width=1\linewidth]{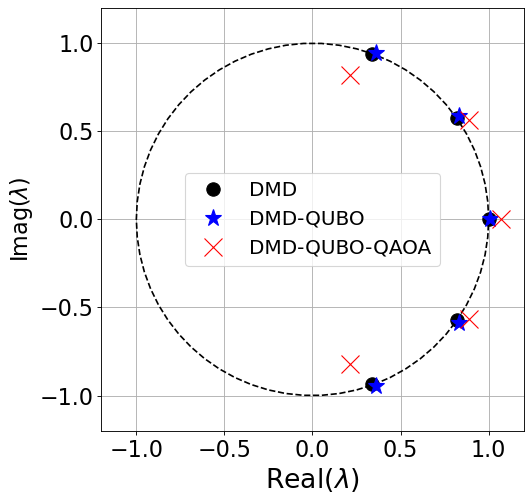}  \caption{}
  \label{fig:DMD_spectrum_discrete}
\end{subfigure}
\begin{subfigure}[!htb]{.5\textwidth}
 \centering
  \includegraphics[trim=0cm 0cm 0cm 0cm,clip,width=1\linewidth]{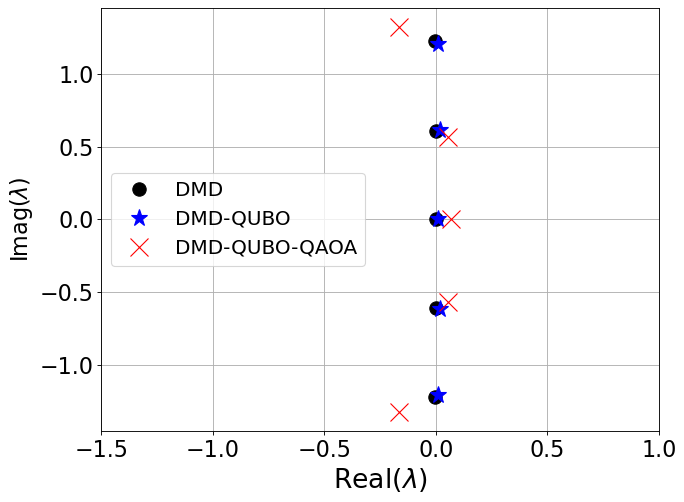}  \caption{}
  \label{fig:DMD_spectrum_continuous}
\end{subfigure}
\caption{Discrete (a) and continuous (b) spectrum computed from implicit DMD and DMD performed on the predicted state from the QUBO and QUBO-QAOA optimization procedures.}
\label{fig:DMD_spectrum_cyl_flow}
\end{figure}

Additionally, it was of interest to investigate the ability of the optimization methodologies to predict the spectral properties of the flowfield. Figure~\ref{fig:DMD_spectrum_discrete} shows the discrete eigenspectrum obtained from implicit DMD, as well as DMD models identified from the state predictions obtained from the optimization methodologies. 

\begin{figure}
 \centering
\begin{subfigure}[!htb]{.45\textwidth}
  \includegraphics[trim=0cm 0cm 0cm 0cm,clip,width=1\linewidth]{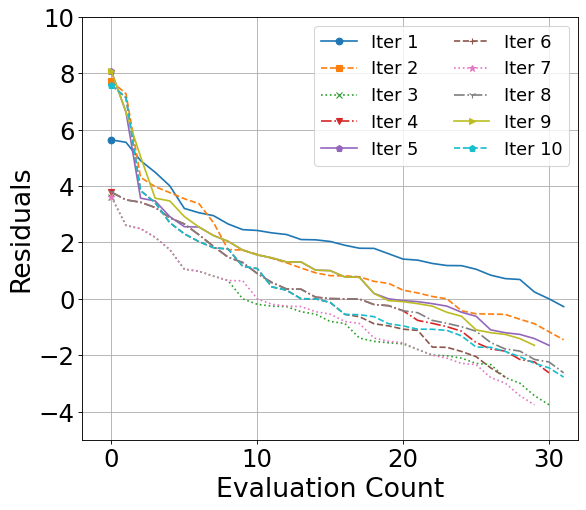}
  \caption{}
  \label{fig:QUBO-QAOA_converge_p1}
\end{subfigure}
\begin{subfigure}[!htb]{.45\textwidth}
 \centering
  \includegraphics[trim=0cm 0cm 0cm 0cm,clip,width=1\linewidth]{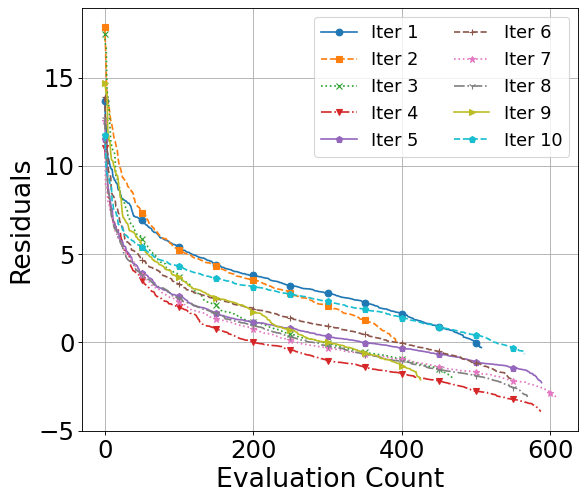}
  \caption{}
  \label{fig:QUBO-QAOA_converge_p2}
\end{subfigure}
\begin{subfigure}[!htb]{.45\textwidth}
 \centering
  \includegraphics[trim=0cm 0cm 0cm 0cm,clip,width=1\linewidth]{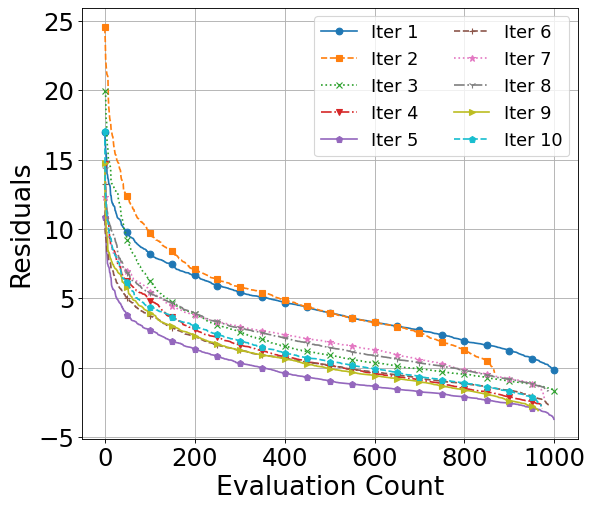}
  \caption{}
  \label{QUBO-QAOA_converge_p3}
\end{subfigure}
\caption{Number of iterations required for convergence for QUBO-QAOA for (a) p = 1, (b) p = 2, and (c) p = 3.}
\label{fig:QUBO-QAOA_convergence_crit}
\end{figure}

Figure~\ref{fig:DMD_spectrum_continuous} shows the continuous eigenspectrum obtained from implicit DMD, as well as DMD models identified from the state predictions obtained from the optimization methodologies. It can be observed that the optimization methodologies are able to accurately capture the fundamental frequency corresponding to vortex shedding at $\text{imag}(\lambda) = 0.57$, as well as the higher harmonics associated with vortex-shedding. We note that although the model predictions are unstable for QUBO-QAOA (a positive real component was found for each mode), the frequency prediction is of greater interest, as it can be verified that we are indeed capturing the fundamental physics inherent to these flowfields with the quantum optimization procedures introduced here.

The results shown in this section are for a quantum simulator, and it is possible to investigate the number of evaluations required to reach a certain convergence criteria for the variational methods of interest in this work. Figure~\ref{fig:QUBO-QAOA_convergence_crit} shows the results for the number of iterations required for convergence as a function of the number of bits utilized, and it can be observed that the number of iterations for convergence increases with the number of bits. 

\subsection{Flowfield Reconstruction for Cylinder Flow}\label{cylflow_results_recon}

\begin{figure}[htb!]
\begin{subfigure}{.5\textwidth}
 \centering
  \includegraphics[trim=0cm 0cm 0cm 0cm,clip,width=1\linewidth]{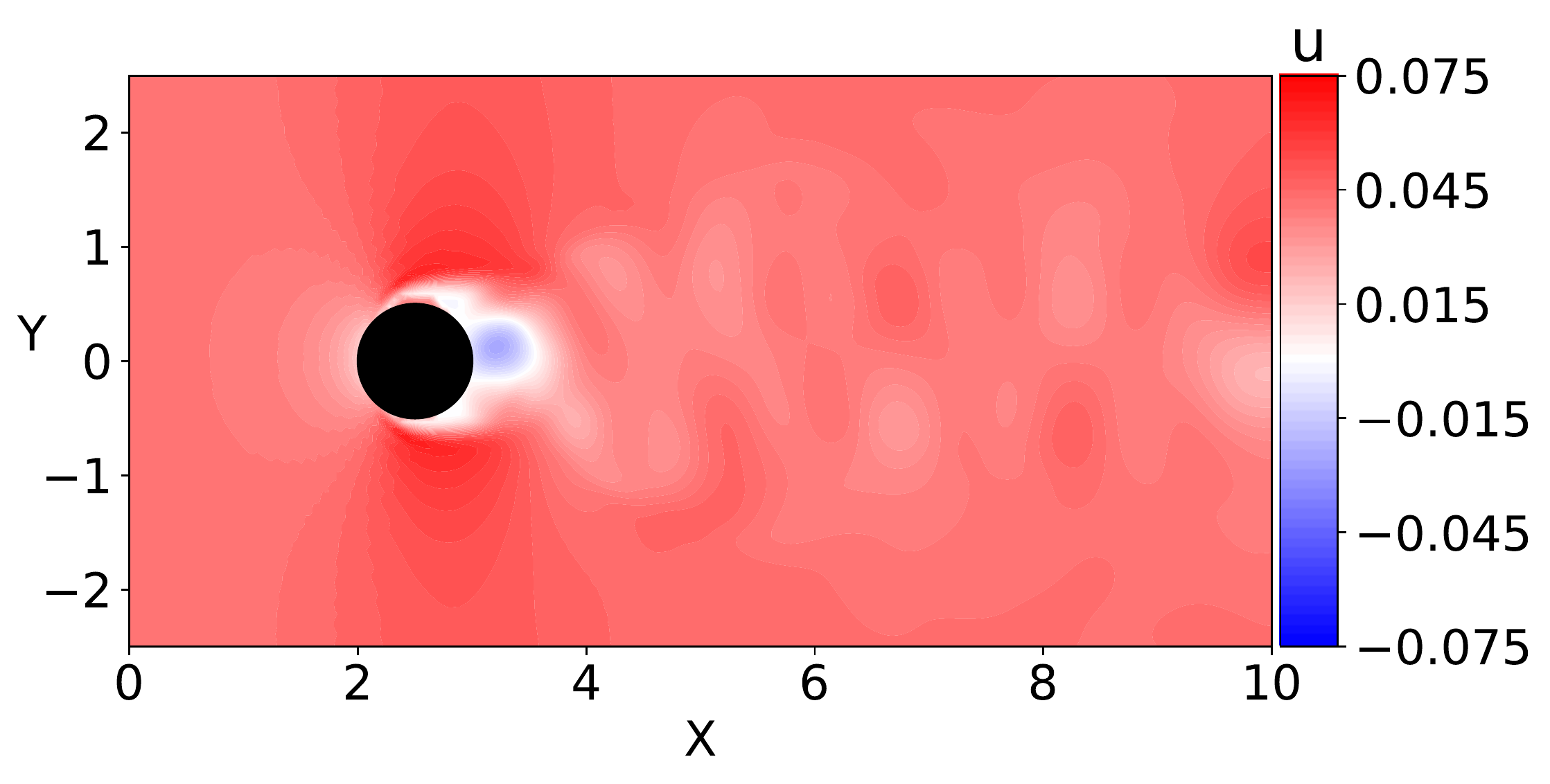}
  \caption{}
  \label{fig:DMD_u1}
\end{subfigure}
\begin{subfigure}{.5\textwidth}
 \centering
  \includegraphics[trim=0cm 0cm 0cm 0cm,clip,width=1\linewidth]{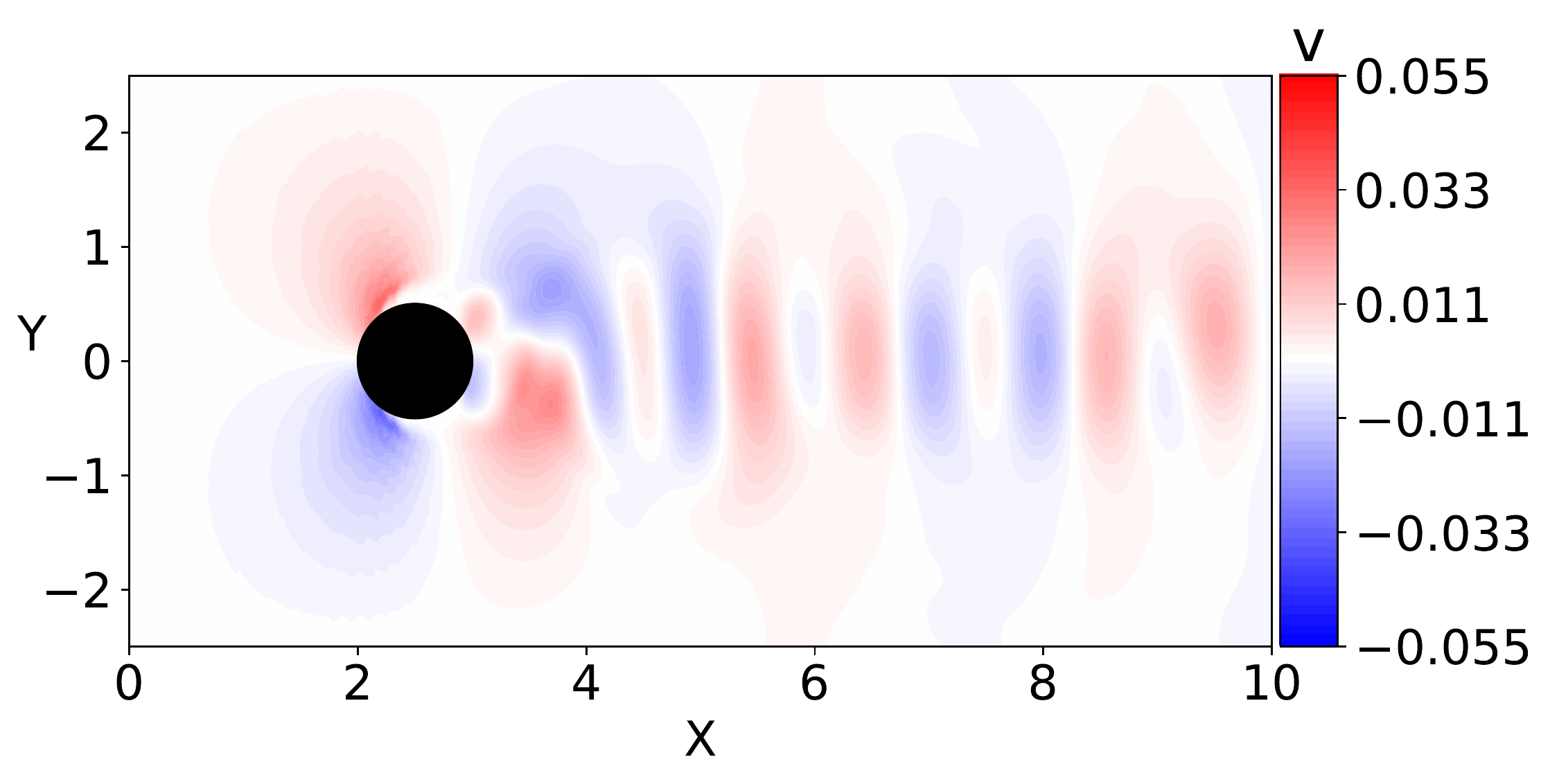}
  \caption{}
  \label{fig:DMD_v1}
\end{subfigure}
\begin{subfigure}{.5\textwidth}
 \centering
  \includegraphics[trim=0cm 0cm 0cm 0cm,clip,width=1\linewidth]{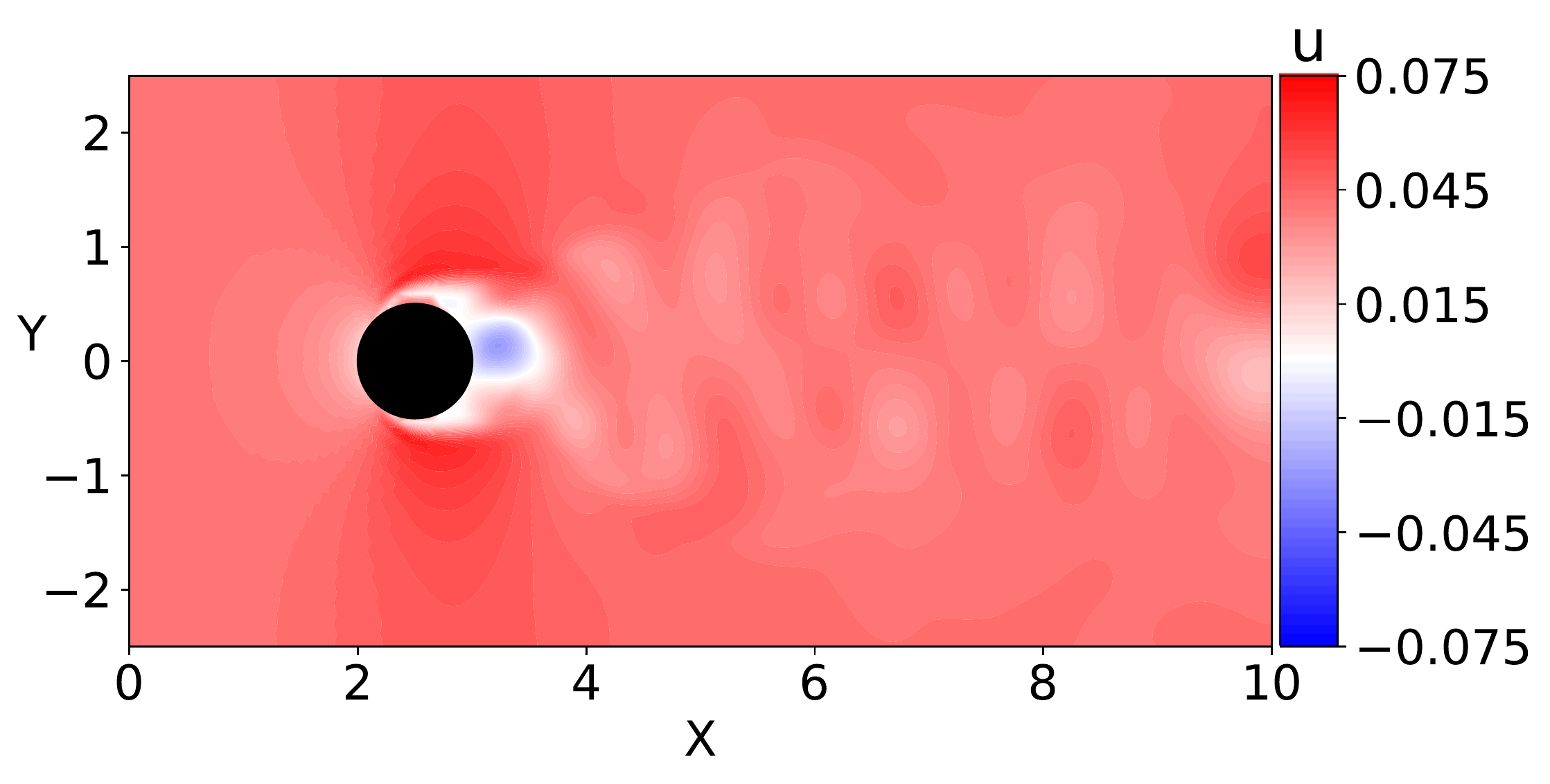}
  \caption{}
  \label{fig:QUBO_u1}
\end{subfigure}
\begin{subfigure}{.5\textwidth}
 \centering
  \includegraphics[trim=0cm 0cm 0cm 0cm,clip,width=1\linewidth]{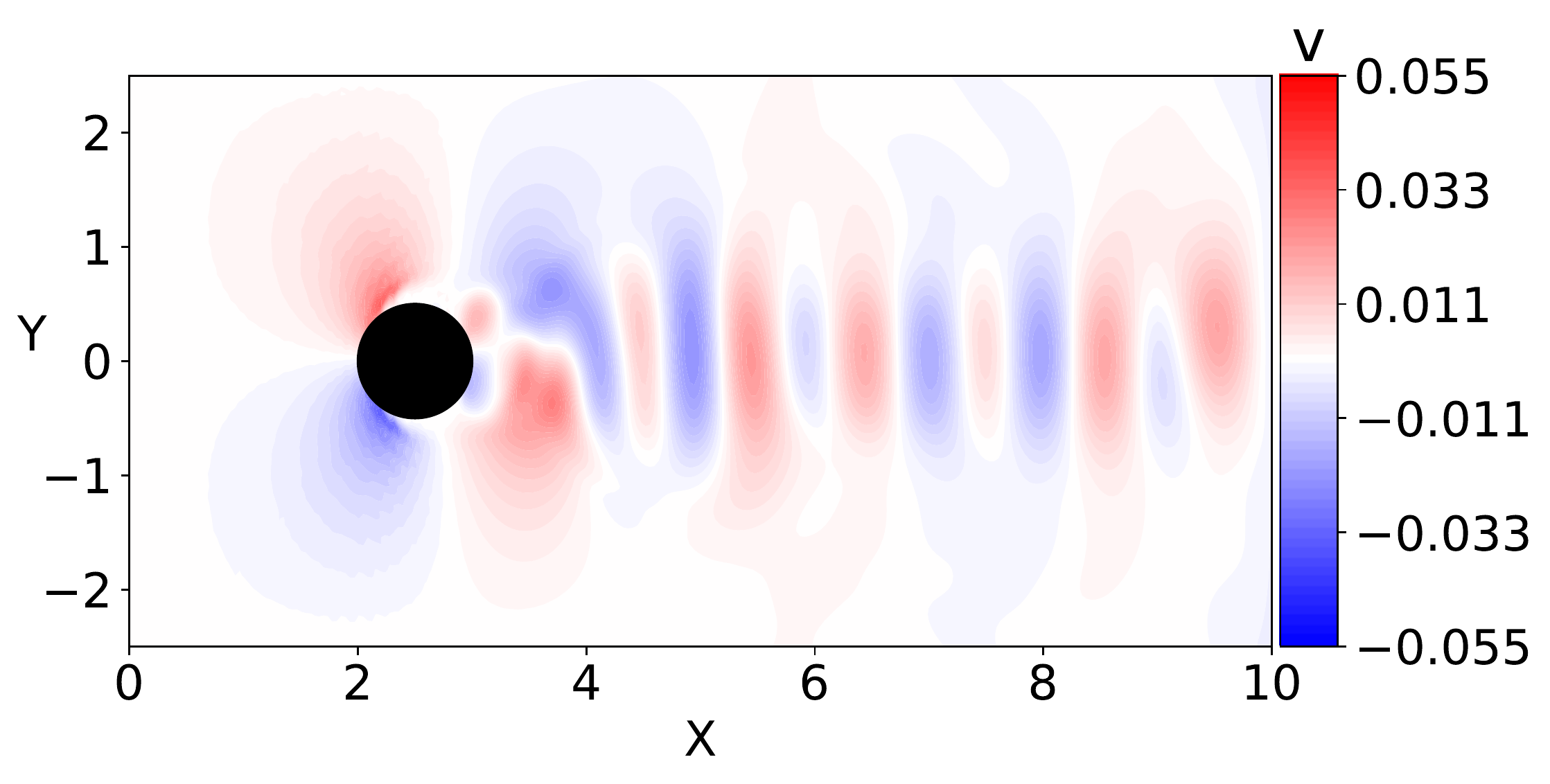}
  \caption{}
  \label{fig:QUBO_v1}
\end{subfigure}
\begin{subfigure}{.5\textwidth}
 \centering
  \includegraphics[trim=0cm 0cm 0cm 0cm,clip,width=1\linewidth]{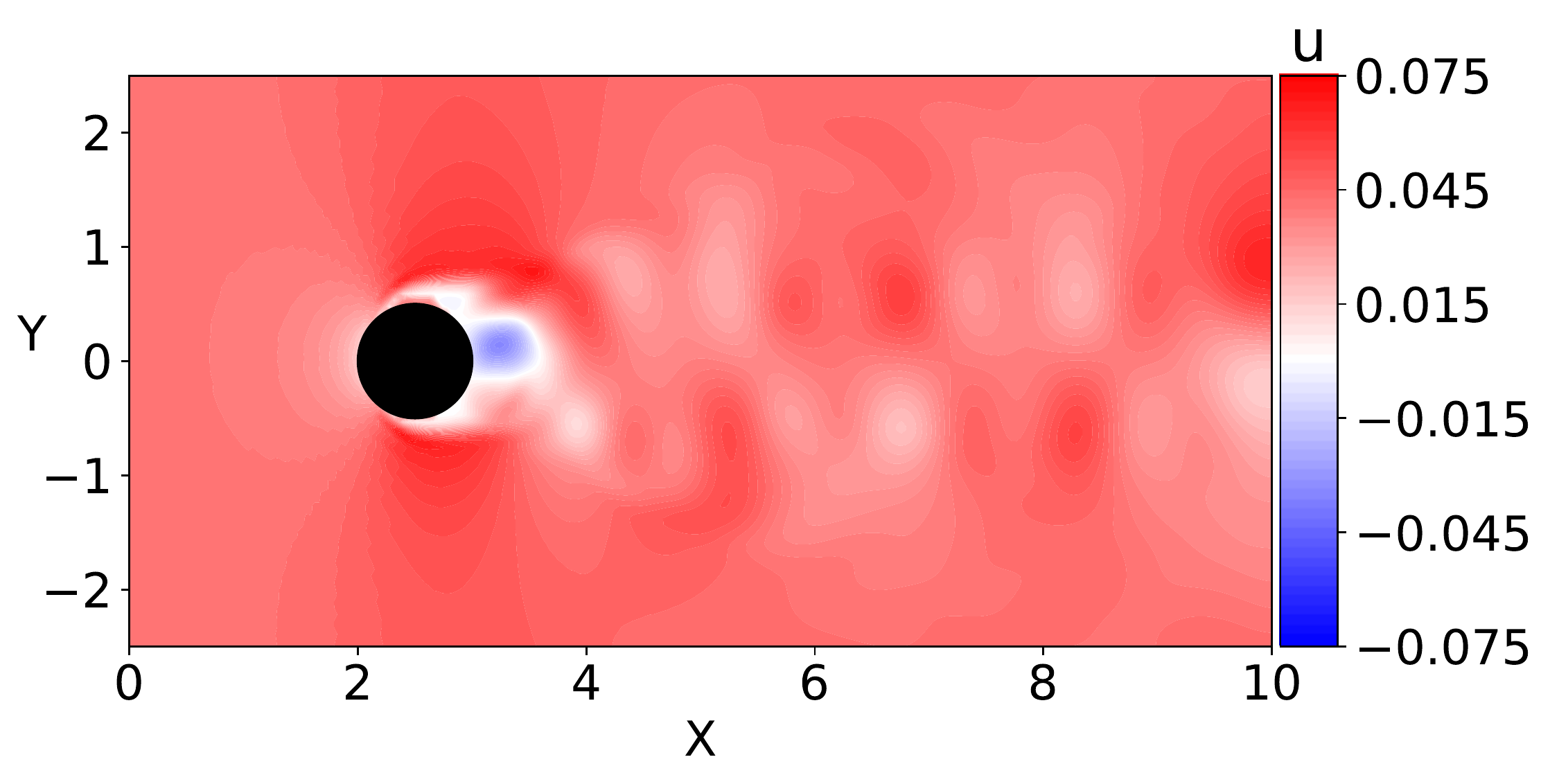}
  \caption{}
  \label{fig:QUBO_QAOA_u1}
\end{subfigure}
\begin{subfigure}{.5\textwidth}
 \centering
  \includegraphics[trim=0cm 0cm 0cm 0cm,clip,width=1\linewidth]{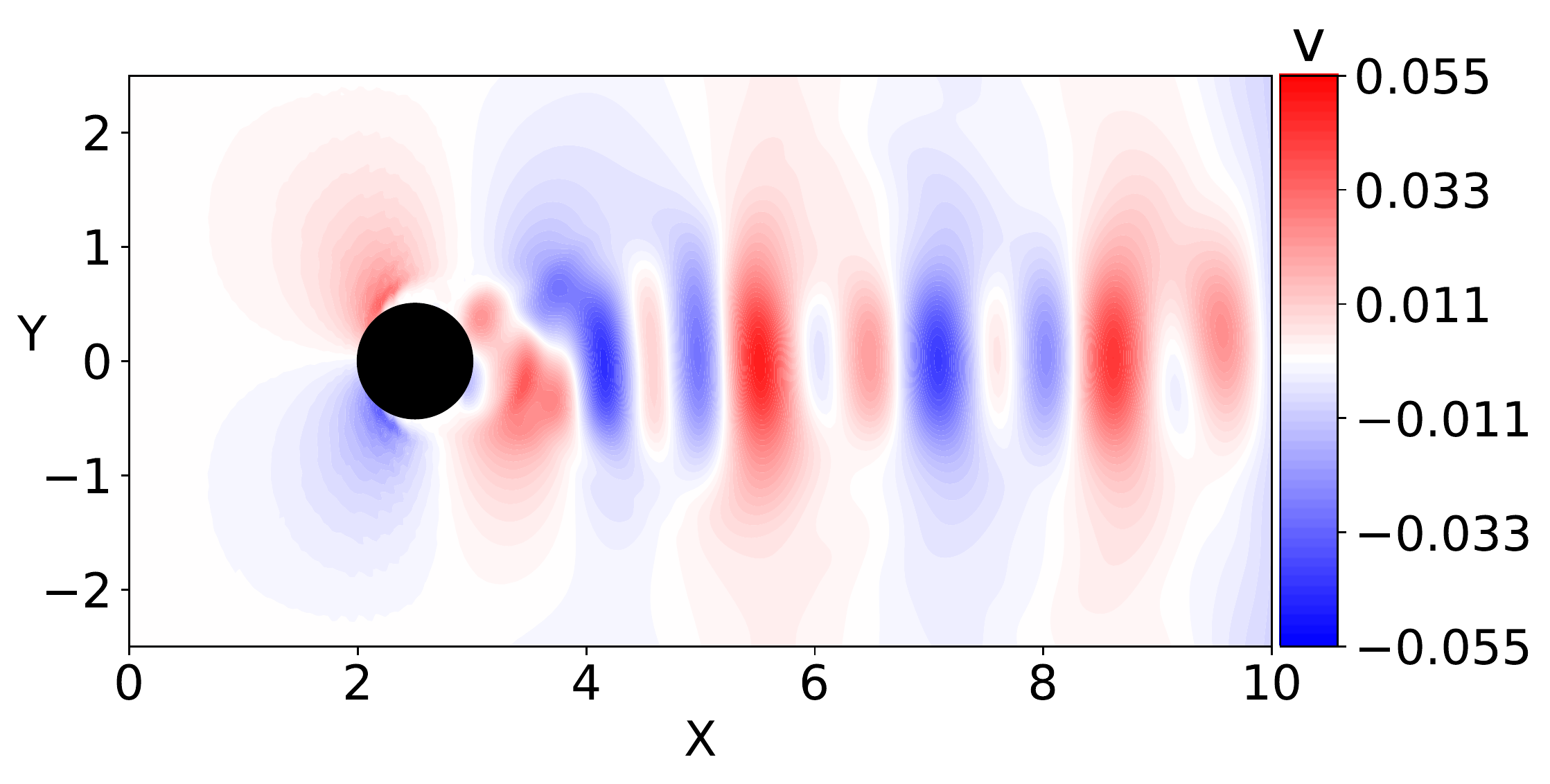}
  \caption{}
  \label{fig:QUBO_QAOA_v1}
\end{subfigure}
\caption{Reconstructed flowfield for both velocity components computed at t = 10 from (a) -- (b) DMD using r = 5 modes, (c) -- (d) QUBO using r = 5 modes and p = 5 bits, and (e) -- (f) QUBO-QAOA using r = 5 and p = 5 bits. The reconstructed flowfield computed from predictions generated by QUBO for a reduced-order model or r = 5 modes and p = 5 bits.}
\label{fig:DMD_QUBO_QAOA_reconstruct_r5_p5}
\end{figure}

The state predictions obtained for implicit DMD, as well as the optimization methodologies for QUBO and QUBO-QAOA, can be used to reconstruct the flowfield for a comparison on how well these ROMs do at predicting the evolution of the state. For each reconstruction, r = 5 POD modes (corresponding to $\approx90\%$ of the kinetic energy of the flowfield) was utilized. For the QUBO and QUBO-QAOA optimization procedures, the number of bits in the quantum-ROM utilized was p = 5. The reconstructed flowfield is shown in Figure~\ref{fig:DMD_QUBO_QAOA_reconstruct_r5_p5} at a timestep corresponding to t = 10. It can be observed that each model is able to capture the dominant pattern of structures of opposite signs in the wake behind the cylinder that is characteristic of vortex shedding. It can be observed in the u-component of the reconstructed velocity (fig~\ref{fig:DMD_u1}, \ref{fig:QUBO_u1}, and \ref{fig:QUBO_QAOA_u1}) that the vortex shedding is more disorganized than the true flowfield, with more dissipation in the immediate wake and more asymmetry in the flowfield patterns. It can also be observed that the model predictions are unstable, as the structures grow in magnitude in the wake (most predominantly seen in figure~\ref{fig:QUBO_QAOA_v1}). This was also observed in computing the eigenspectrum, as the QUBO-QAOA model found slightly unstable growth modes (see figure~\ref{fig:DMD_spectrum_continuous}). It should be emphasized that the quantum optimization procedures are able to capture the same fundamental physics of interest as the implicit DMD methodology for this flowfield, and that potentially more accurate predictions of the growth/decay rate for this system could be obtained by extending the number of snapshots in the original snapshot matrix. The predictions on the reconstructed flowfield verify that quantum optimization procedures can be used to create a ROM of flowfields of interest and obtain accurate predictions in which the physics of the original flowfield is preserved.

Finally, the reconstruction error was computing using a normalized inner product between the true and reconstructed error, computed via: 

\begin{equation}
    e = \left(1 - \dfrac{\langle E_{recon}, E_{true} \rangle}{\| E_{recon} \| \| E_{true} \|} \right) ,\label{recon_error}
\end{equation}
where $E_{recon}$ and $E_{true}$ represent the reconstructed and true energy of the flowfield, respectively. 

\begin{figure}[htb!]
 \centering
  \includegraphics[trim=0cm 0cm 0cm 0cm,clip,width=0.6\linewidth]{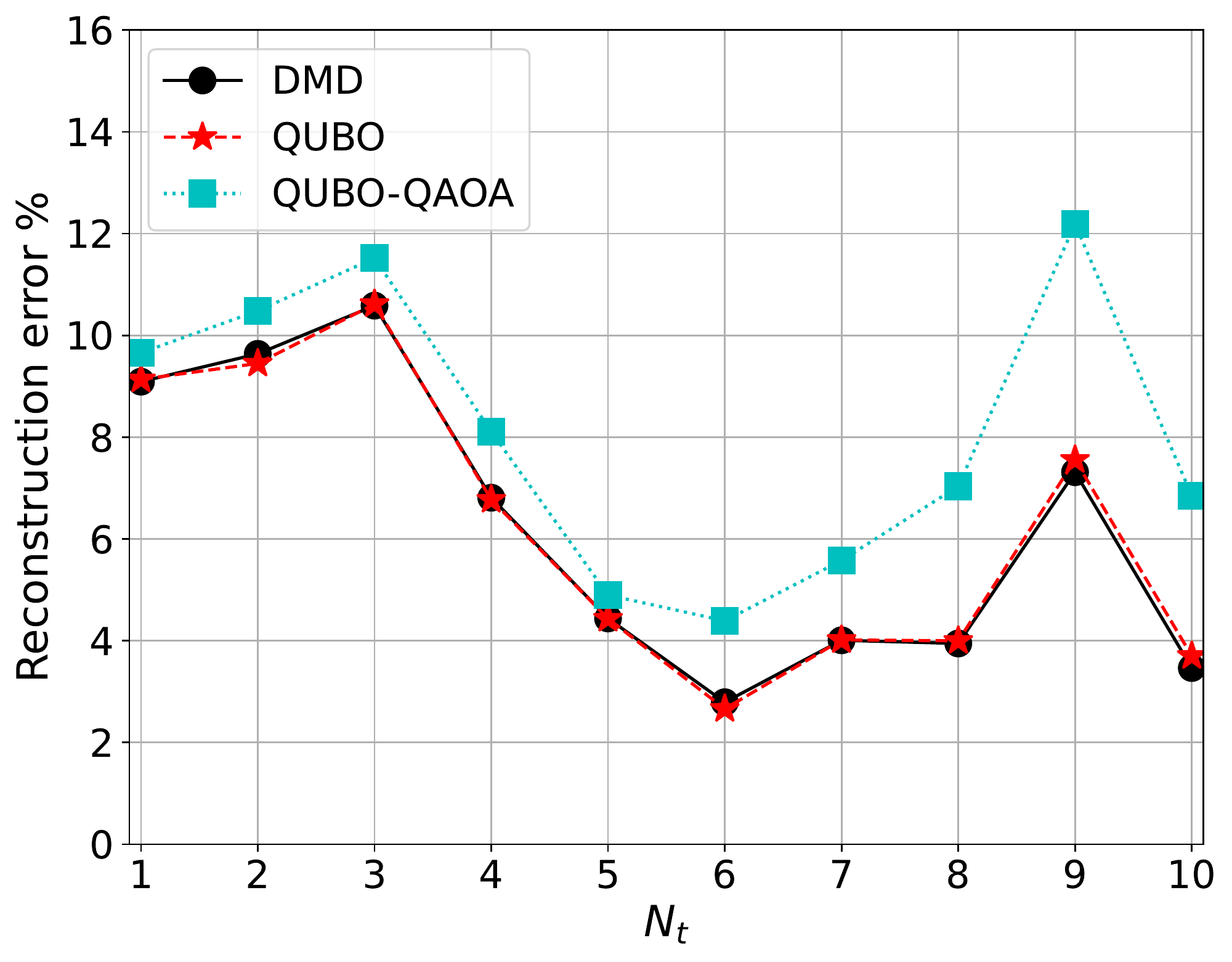}
\caption{Reconstruction error computed for implicit DMD, QUBO, and QUBO-QAOA optimization methodologies.}
\label{fig:error_reconstruct}
\end{figure}

Figure~\ref{fig:error_reconstruct} shows the results computed for a r = 5 model with p = 5 bits for the quantum optimization DMD algorithms. It can be observed that the QUBO optimization methodology is very close to the reconstruction error obtained with implicit DMD, while the QUBO-QAOA optimization procedure has a slightly higher reconstruction error than both, but that all three methods follow the same general trend.

Finally, the results for extending the number of timesteps at which a prediction from the quantum-ROM could be made are shown in section~\ref{results_cylflow_MS}.

\section{Quantum-ROM Predictions for Flow over NACA0009 Airfoil}\label{results_airfoil}

To further understand the performance of quantum ROMs, as well as to extend the applications to relevant engineering problems of interest, experiments were performed on data collected for flow over a NACA0009 airfoil at an angle of attack of 15$^\circ{}$ and a Reynolds number of 500, the details of which are provided in~\cite{asztalos2021modeling}. The dataset was generated to study active flow control, in which momentum was added to the flowfield in order to improve the aerodynamic performance in adverse operating conditions.

The formulation of DMD utilized in this work has been performed on mean-subtract and scaled data, as shown in section~\ref{results_cylflow}. Data is re-scaled such that $x \in \left[0,1 \right]$ for input into QUBO and QAOA. Performing DMD on centered data improves the performance of DMD, as discussed in-depth in Hirsh et al.~\cite{hirsh2020centering}. The methodology for centering data for inputs to the models follows Hirsh et al., and will be discussed briefly below.

\subsection{Overview on Centering Data for DMD}\label{centering_DMD}

Recalling the formulation of DMD in which a linear operator is identified from data, the DMD matrix can be approximated by $\mathbf{A} = \mathbf{Y} \mathbf{X}^\dagger$. If the mean is non-zero, then the DMD model can be improved through the inclusion of an additional affine term: 
\begin{equation*} 
    \mathbf{Y} = \mathbf{A}\mathbf{X} + b\mathbf{1}^T
\end{equation*}
where $b \in \mathbb{R}$ and $\mathbf{1}$ is a vector of length $T$ whose elements are all one. As described by Hirsh et al., the use of an affine term is equivalent to centering the data. The mean of the data can be computed as: 

\begin{equation*}
    \mu_1 = \frac{X1}{1^T1} \quad \& \quad \mu_2 = \frac{Y1}{1^T1}
\end{equation*}
with mean-subtraction of the data computed as: 

\begin{gather}
    \bar{\mathbf{X}} = \mathbf{X} - \mu_1 1^T \label{X1bar} \\
    \bar{\mathbf{Y}} = \mathbf{Y} - \mu_2 1^T. \label{X2bar}
\end{gather}
The unbiased regression problem can now be solved as: 
\begin{equation}
    \bar{\mathbf{Y}} = \bar{\mathbf{A}} \bar{\mathbf{X}}, \label{DMD_unbiased}
\end{equation}
where predictions can be computed from $\bar{\mathbf{A}}$ on the unbiased data. To transform back and solve the biased problem:

\begin{gather}
    \tilde{\mathbf{A}} = \bar{\mathbf{A}} \label{Abar} \\
    \tilde{b} = \mu_2 - \bar{\mathbf{A}} \mu_1 ,\label{btilde}
\end{gather}
where predictions on the original biased data can be computed from:

\begin{equation}
    \mathbf{Y} = \tilde{\mathbf{A}}X + \tilde{b}1^T.\label{DMD_biased}
\end{equation}

The formulation of inputs to QUBO and QUBO-QAOA require the data to be scaled between $x \in \left[0, 1 \right]$. Therefore, we are interested in a DMD model that can perform accurate predictions for scaled data with a non-zero mean. To accomplish this, the unbiased DMD matrix $\bar{\mathbf{A}}$ can be utilized using equation~\ref{DMD_unbiased}, with the mean of the data added to predictions on the state made such that $\mathbf{Y} = \mathbf{Y} + \mu_2$.

\subsection{Spectrum Analysis}\label{airfoil_spectrum_results}

To better understand the importance of performing DMD on centered data for inputs to the quantum-ROM, consider applying DMD on the data ($A_{\text{unscaled}}$), DMD on the scaled data ($A_{\text{scaled}}$) with a non-zero mean and without inclusion of an additional affine term, and DMD on the unbiased scaled data ($A_{\text{scaled}}$, or $\bar{A}$ in previous notation) performed with centering of the data. 

\begin{figure}[htb!]
\begin{subfigure}{.4\textwidth}
 \centering
  \includegraphics[trim=0cm 0cm 0cm 0cm,clip,width=1\linewidth]{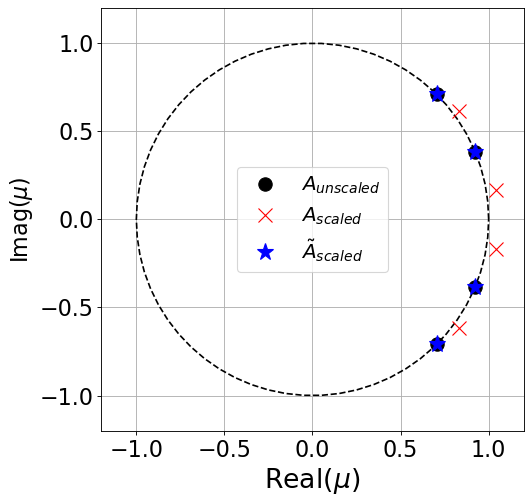}
  \caption{}
  \label{fig:DMD_operator_compare_discrete}
\end{subfigure}
\begin{subfigure}{.5\textwidth}
 \centering
  \includegraphics[trim=0cm 0cm 0cm 0cm,clip,width=1\linewidth]{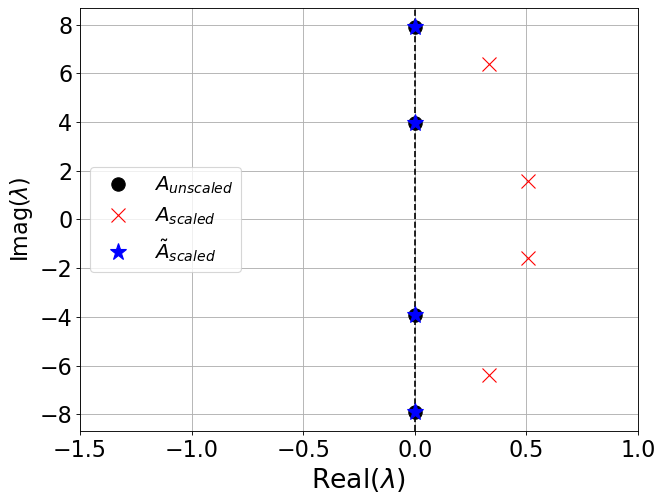}
  \caption{}
  \label{fig:DMD_operator_compare_continuous}
\end{subfigure}
\caption{The (a) discrete and (b) continuous spectrum computed from DMD operators computed on unscaled, scaled, and scaled data with an additional affine term identified. Note that the label $\mu$ applied in (a) corresponds to the discrete DMD eigenvalues.}
\label{fig:operator_spectrum_compare}
\end{figure}

Figure~\ref{fig:DMD_operator_compare_discrete} shows the discrete DMD eigenvalues computed from these test cases, with figure~\ref{fig:DMD_operator_compare_continuous} showing the continuous eigenvalues. DMD has been computed on a r = 4 model for airfoil data in this case. Note that when applying DMD directly to the scaled data without the addition of an additional affine term, the real and imaginary components of the DMD eigenvalues differ in comparison to the ``true" DMD results for the original data. Flow over an airfoil at Re = 500 corresponds to an oscillator-type system, in which the eigenvalues are marginally stable (i.e., $real({\lambda}) = 0$). Performing DMD on the scaled data with a non-zero mean without centering results in an unstable system with positive growth rates. The frequency of oscillation is also affected, as the imaginary component of the eigenvalues are damped. However, performing DMD on the unbiased scaled data with centering of the data results in an accurate spectrum computed from the DMD $\bar{\mathbf{A}}$ operator. Therefore, in order to accurately capture the dynamics of the true flowfield, the DMD operator that must be utilized for performing predictions on the state is $\bar{\mathbf{A}}$, with predictions computed on the biased scaled data achieved through addition of the non-zero mean. For the results presented in this section, the formulation for computing the DMD operator is achieved through equation~(\ref{DMD_unbiased}), with the mean added to predictions of the state as $\mathbf{Y} = \mathbf{Y} + \mu_2$.

\subsection{Dimensionality Reduction}\label{airfoil_pod_results}

\begin{figure}[htb!]
\begin{subfigure}{.5\textwidth}
 \centering
  \includegraphics[trim=0cm 0cm 0cm 0cm,clip,width=1\linewidth]{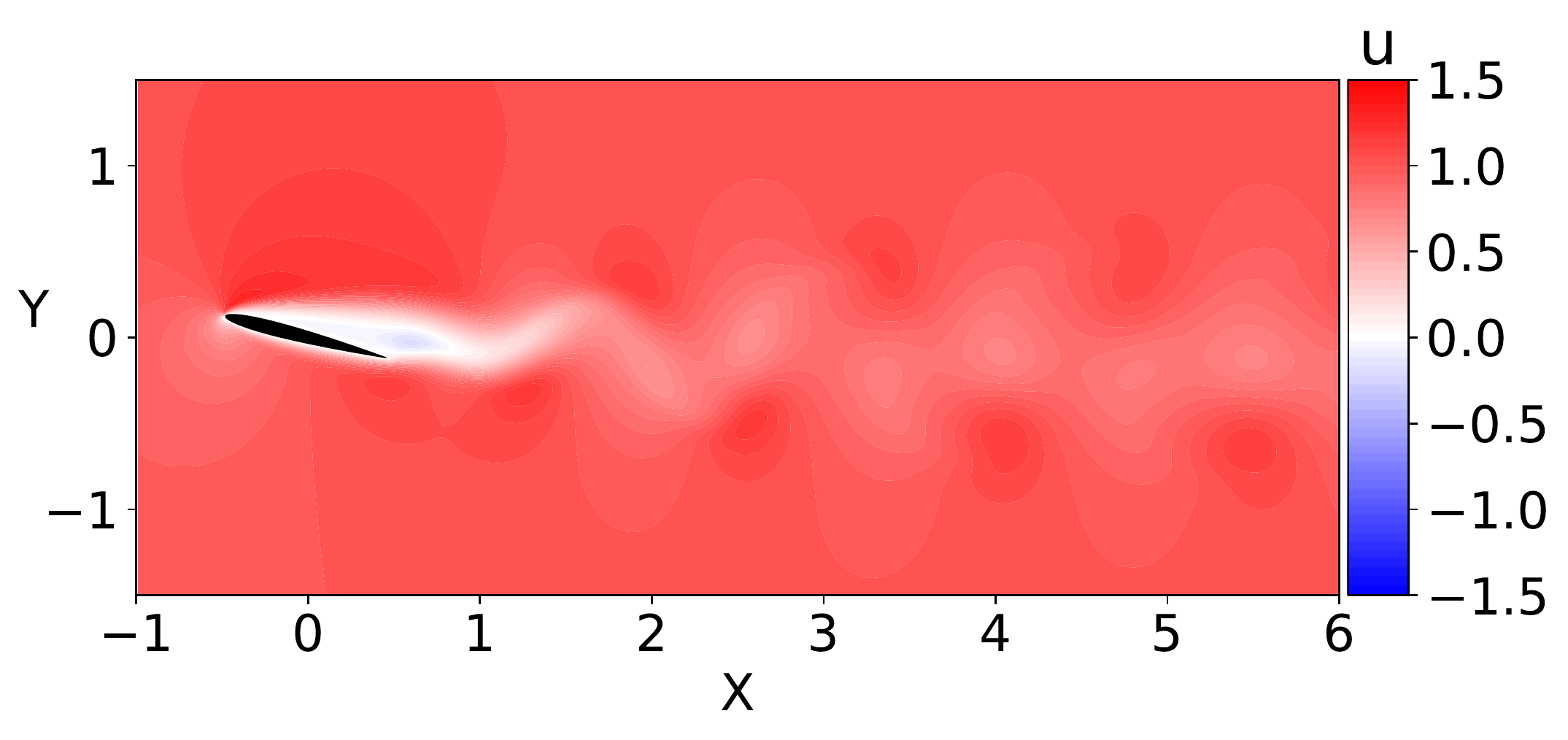}
  \caption{}
  \label{fig:true_u1_naca0009}
\end{subfigure}
\begin{subfigure}{.5\textwidth}
 \centering
  \includegraphics[trim=0cm 0cm 0cm 0cm,clip,width=1\linewidth]{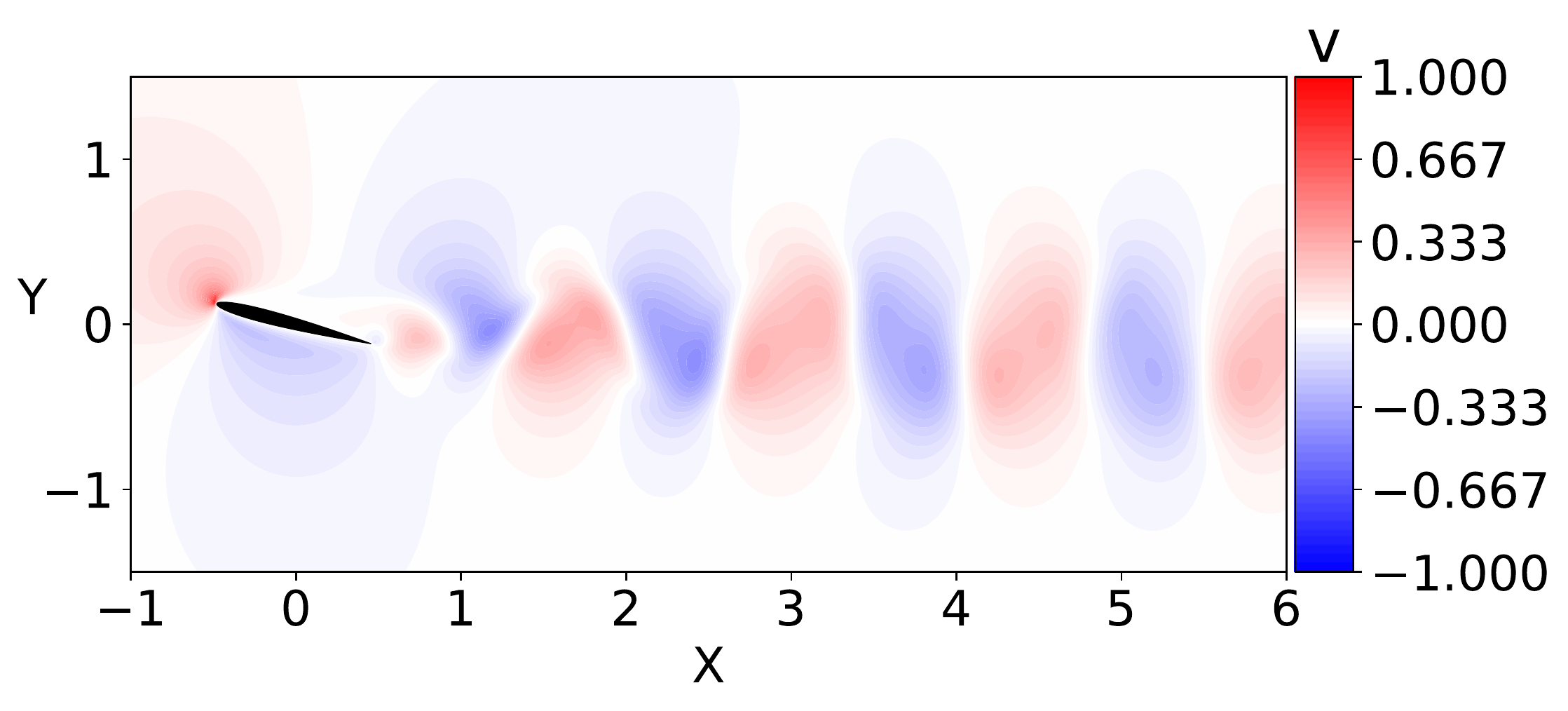}
  \caption{}
  \label{fig:true_v1_naca0009}
\end{subfigure}
\begin{subfigure}{.5\textwidth}
 \centering
  \includegraphics[trim=0cm 0cm 0cm 0cm,clip,width=1\linewidth]{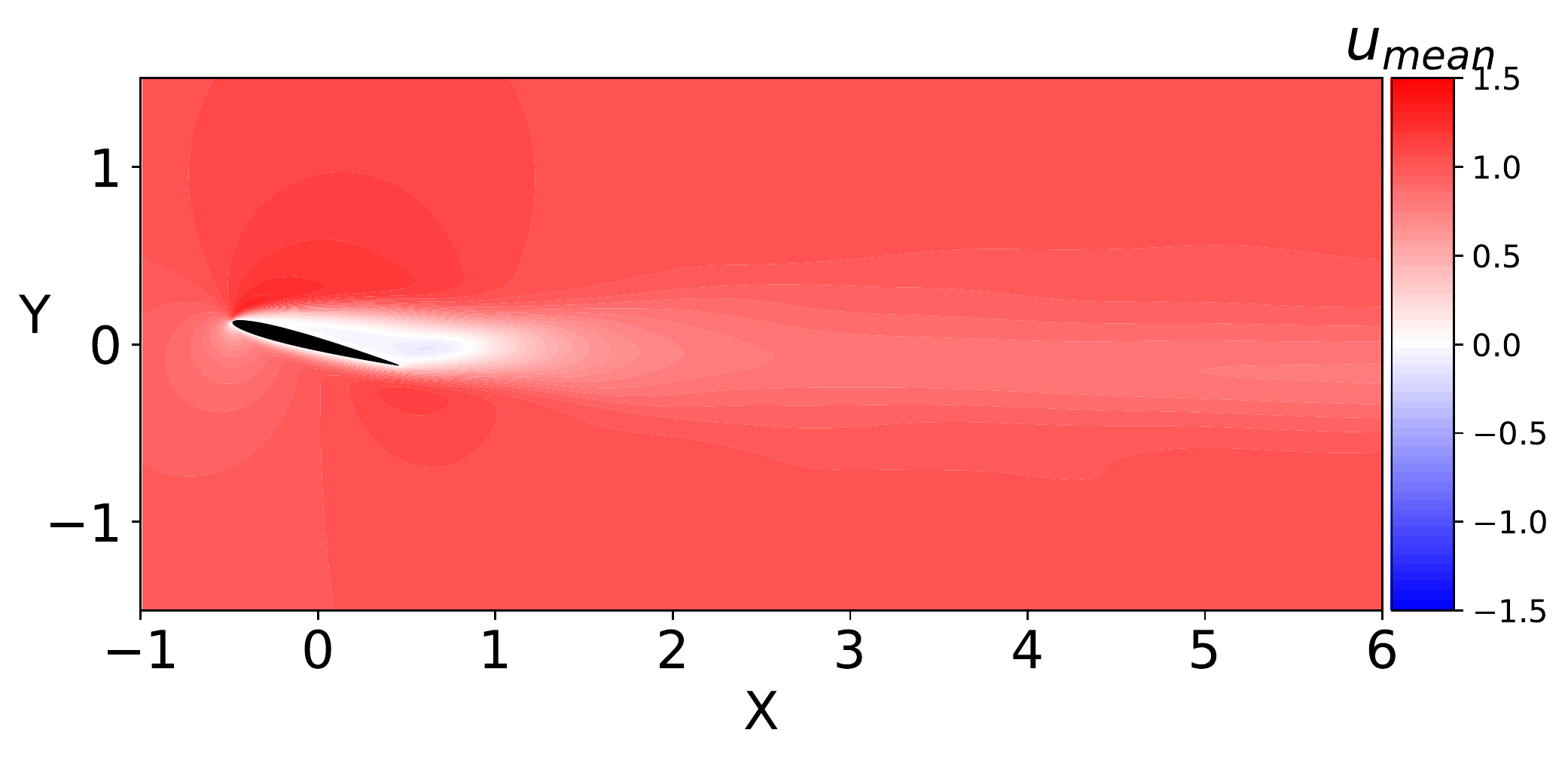}
  \caption{}
  \label{fig:true_umean_naca0009}
\end{subfigure}
\begin{subfigure}{.5\textwidth}
 \centering
  \includegraphics[trim=0cm 0cm 0cm 0cm,clip,width=1\linewidth]{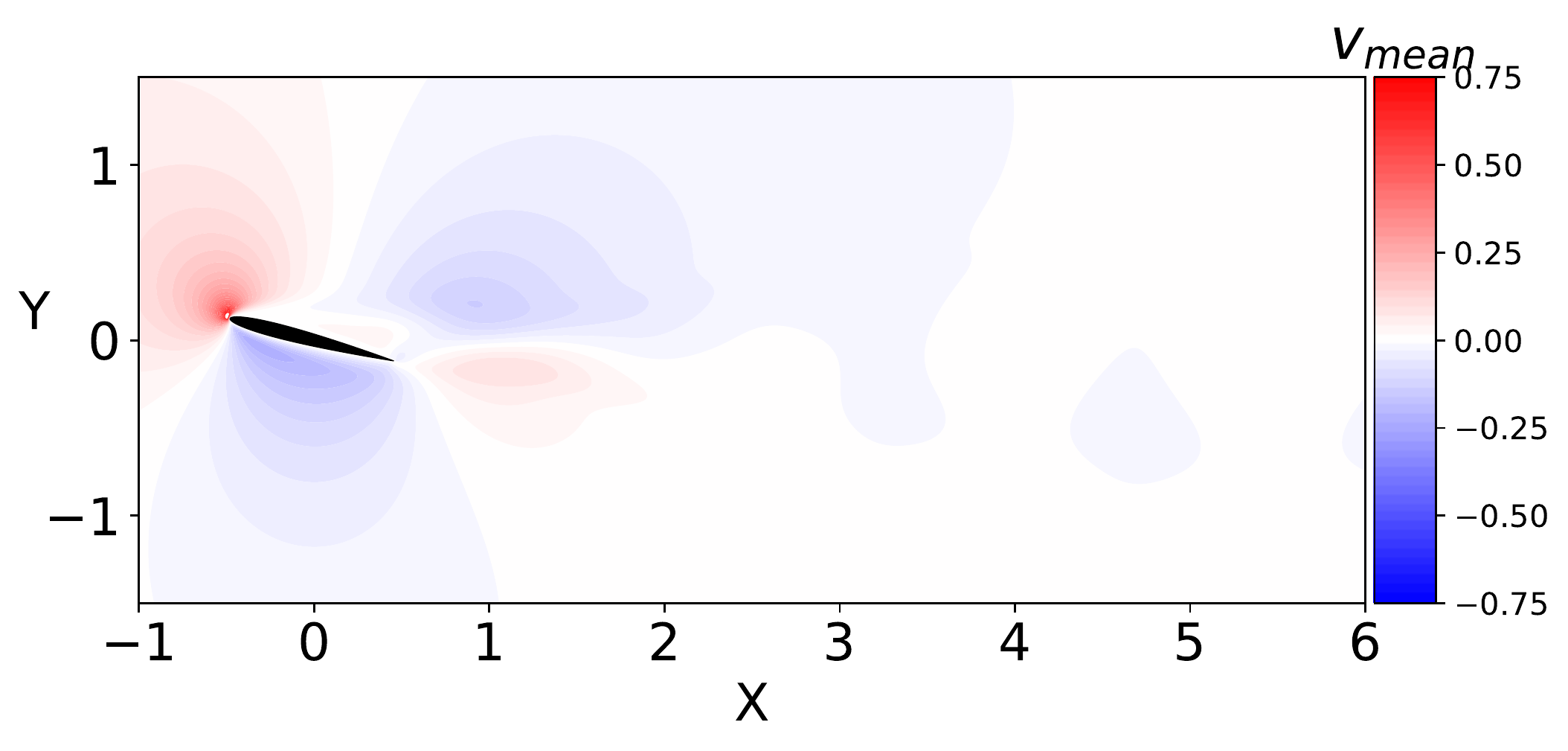}
  \caption{}
  \label{fig:true_vmean_naca0009}
\end{subfigure}
\begin{subfigure}{.5\textwidth}
 \centering
  \includegraphics[trim=0cm 0cm 0cm 0cm,clip,width=1\linewidth]{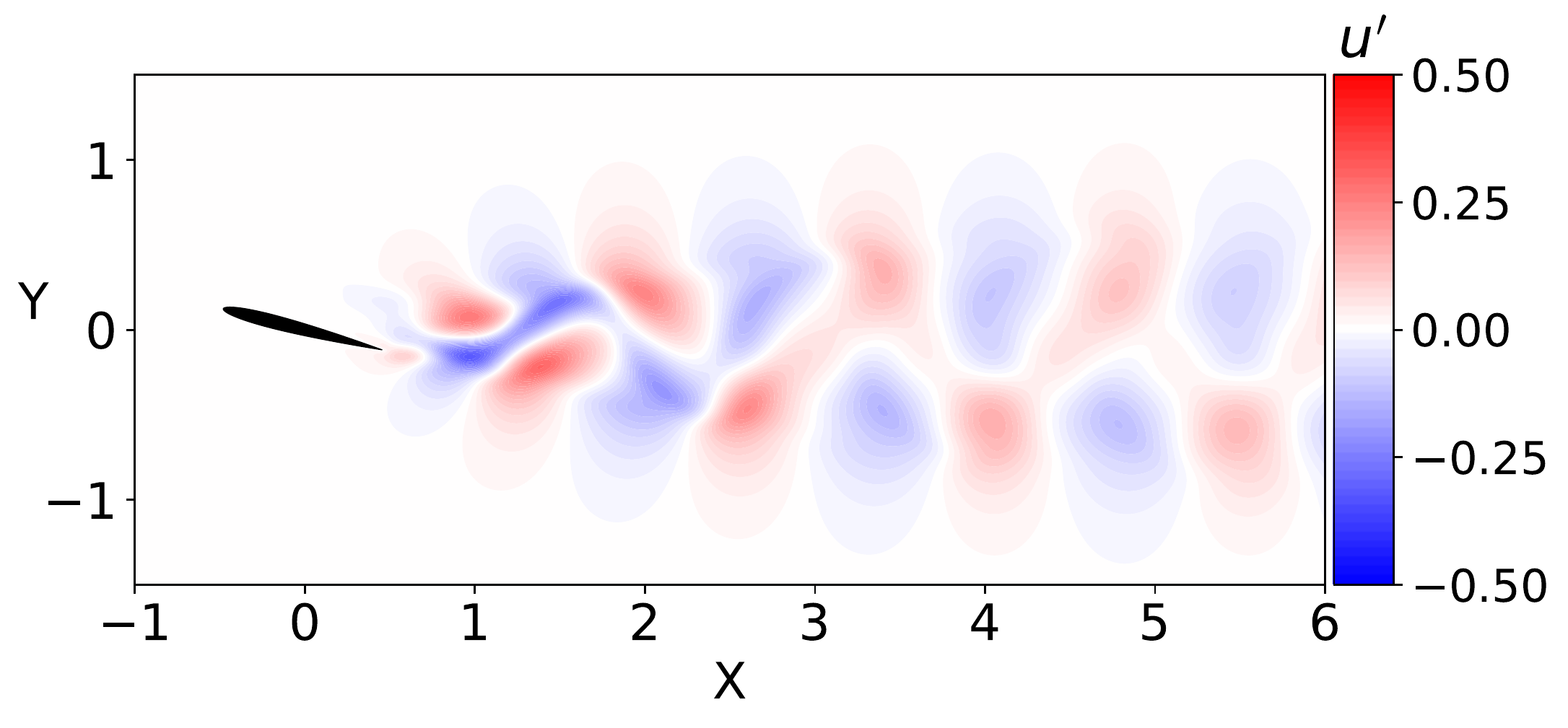}
  \caption{}
  \label{fig:true_uMS_naca0009}
\end{subfigure}
\begin{subfigure}{.5\textwidth}
 \centering
  \includegraphics[trim=0cm 0cm 0cm 0cm,clip,width=1\linewidth]{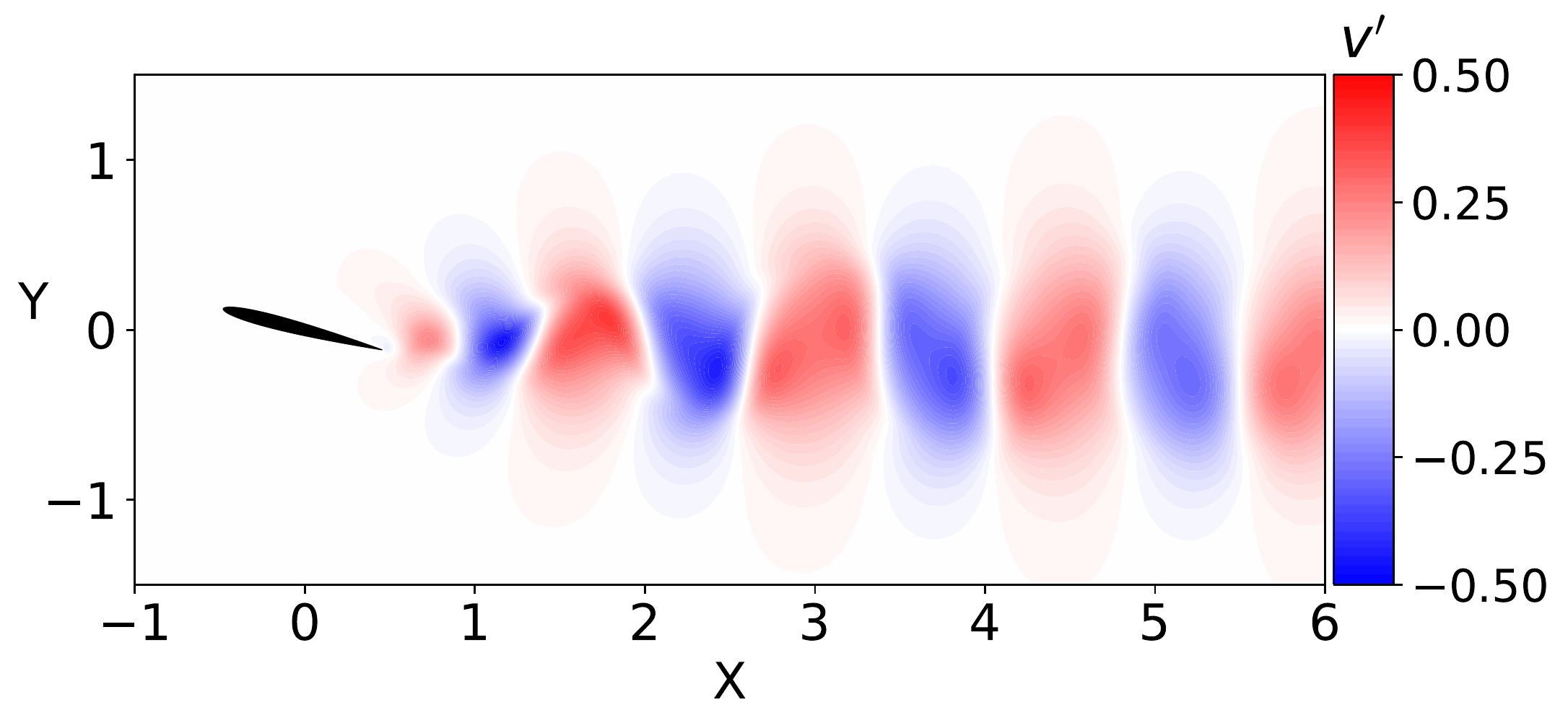}
  \caption{}
  \label{fig:true_vMS_naca0009}
\end{subfigure}
\caption{The true flowfield for flow over a NACA0009 airfoil at $\alpha = 15^\circ{}$ and $Re = 500$ plotted at $t^{+}$ = 10 showing (a) the u-component and (b) the v-component of velocity, with the mean components shown in (c) and (d), and the mean-subtracted components shown in (e) and (f).}
\label{fig:true_flowfield_NACA0009}
\end{figure}

The true flowfield is plotted in figure~\ref{fig:true_flowfield_NACA0009}, where the established vortex street in the airfoil wake can be clearly seen, especially in the u-component of the velocity (figure~\ref{fig:true_u1_naca0009}). The mean velocity components can be seen in figures~\ref{fig:true_umean_naca0009} and \ref{fig:true_vmean_naca0009}, with the fluctuating components (mean-subtracted flowfield) shown in figures~\ref{fig:true_uMS_naca0009} and \ref{fig:true_vMS_naca0009}.

\begin{figure}[htb!]
\begin{subfigure}{.5\textwidth}
 \centering
  \includegraphics[trim=0cm 0cm 0cm 0cm,clip,width=1\linewidth]{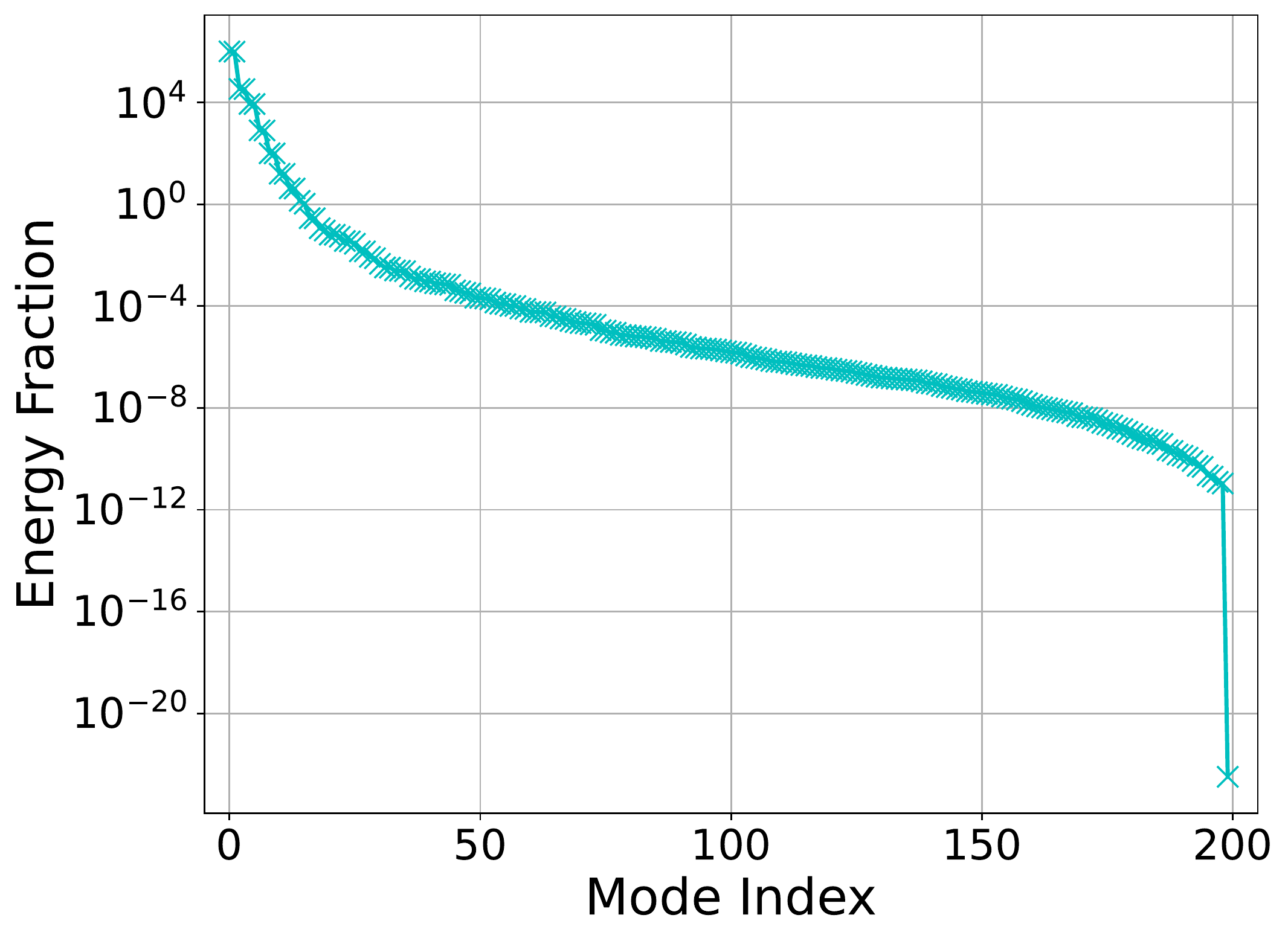}
  \caption{}
  \label{fig:true_sigma_energy}
\end{subfigure}
\begin{subfigure}{.5\textwidth}
 \centering
  \includegraphics[trim=0cm 0cm 0cm 0cm,clip,width=1\linewidth]{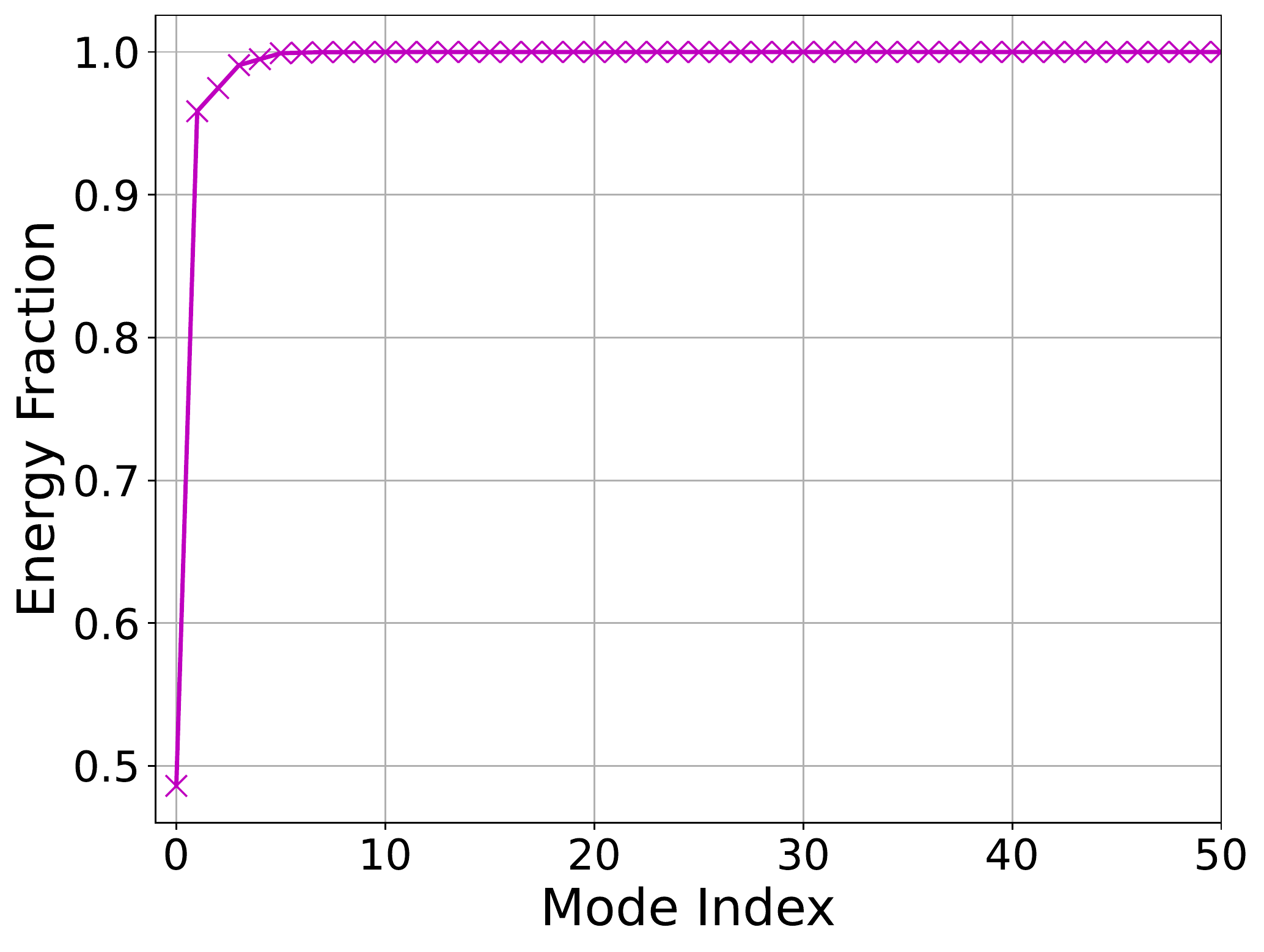}
  \caption{}
  \label{fig:true_sigma_cum_energy}
\end{subfigure}
\caption{Results showing the (a) energy fraction and (b) cumulative energy fraction computed from the singular values computed from POD on data collected for flow over a NACA0009 airfoil at $\alpha = 15^\circ{}$ and $Re = 500$.}
\label{fig:sigma_NACA0009}
\end{figure}

\begin{figure}[htb!]
 \centering
  \includegraphics[trim=0cm 0cm 0cm 0cm,clip,width=0.6\linewidth]{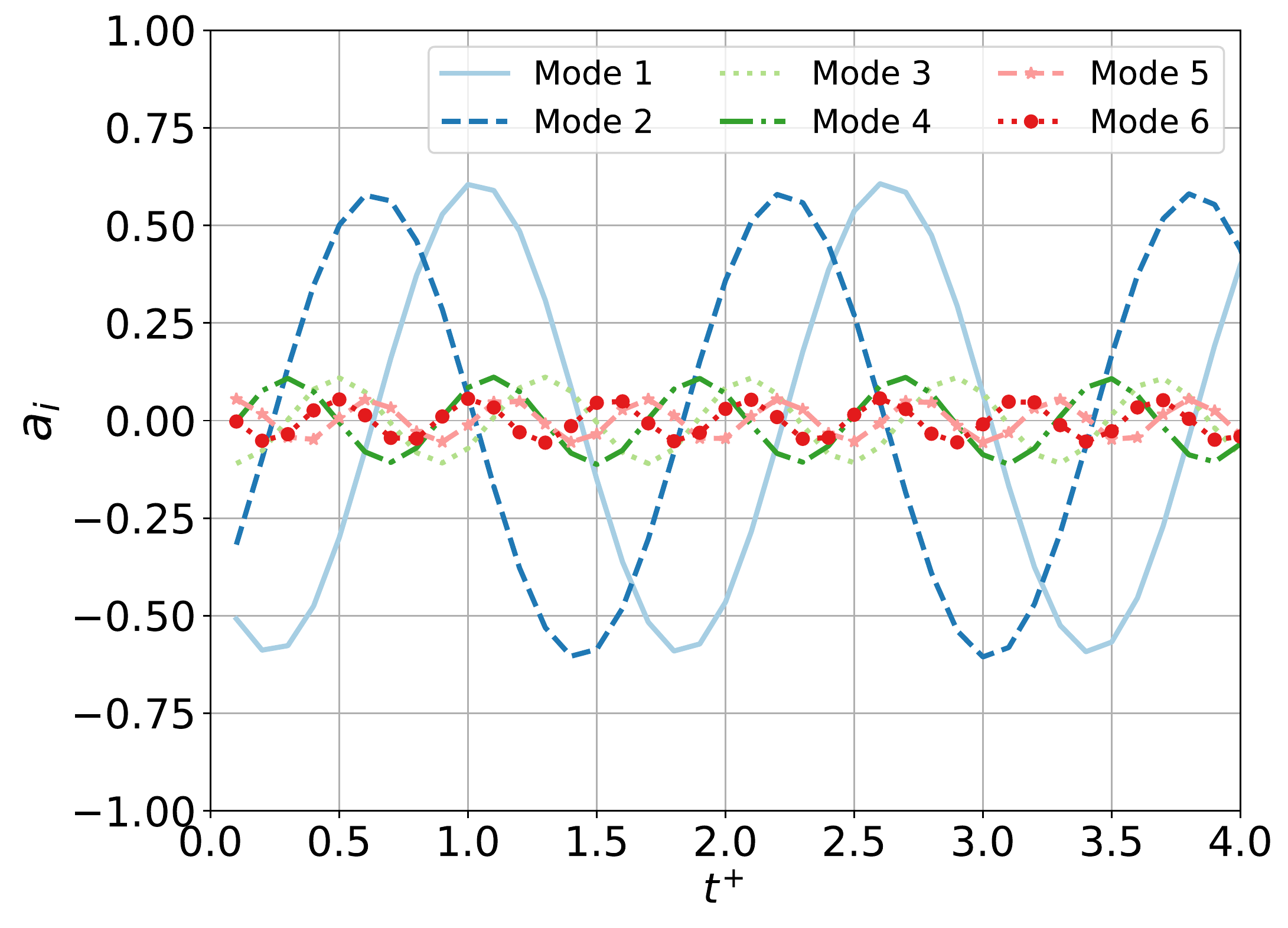}
\caption{Leading POD mode coefficients computed from flow over a NACA0009 airfoil.}
\label{fig:NACA0009_modecoeffs}
\end{figure}

POD was performed on the data collected during the steady limit cycle corresponding to vortex shedding from the trailing edge of the airfoil, for a total snapshot length of $N_t$ = 200 with a sampling frequency of 10 Hz. Figure~\ref{fig:sigma_NACA0009} shows a plot of the energy fraction and cumulative energy fraction computed from the POD singular values. Note that as observed from figure~\ref{fig:true_sigma_cum_energy}, the first six modes capture 99\% of the energy of the flowfield.

The first six POD mode coefficients are shown in Figure~\ref{fig:NACA0009_modecoeffs}. It can be seen that the higher POD mode coefficients correspond to the higher harmonics of the fundamental frequency, which corresponds to the vortex shedding observed in the wake behind the airfoil. 

\subsection{QC-ROM Prediction Results}\label{airfoil_pred_results}

\begin{figure}[!htb]
 \centering
\begin{subfigure}[!htb]{.45\textwidth}
 \centering
  \includegraphics[trim=0cm 0cm 0cm 0cm,clip,width=1\linewidth]{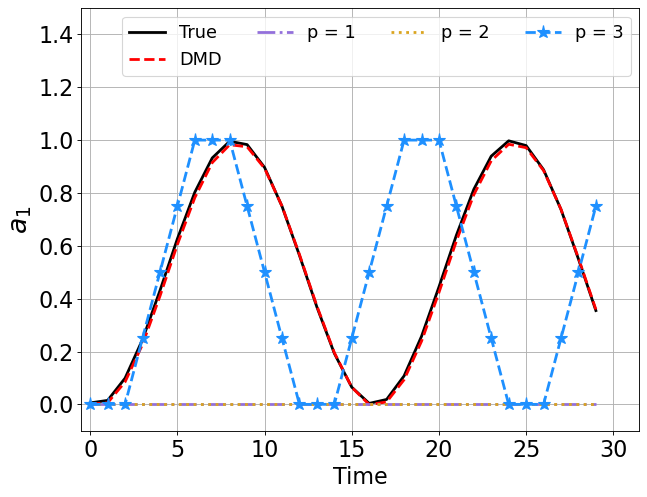}
  \caption{}
  \label{fig:QUBO_naca0009_r6_a1}
\end{subfigure}
\begin{subfigure}[!htb]{.45\textwidth}
 \centering
  \includegraphics[trim=0cm 0cm 0cm 0cm,clip,width=1\linewidth]{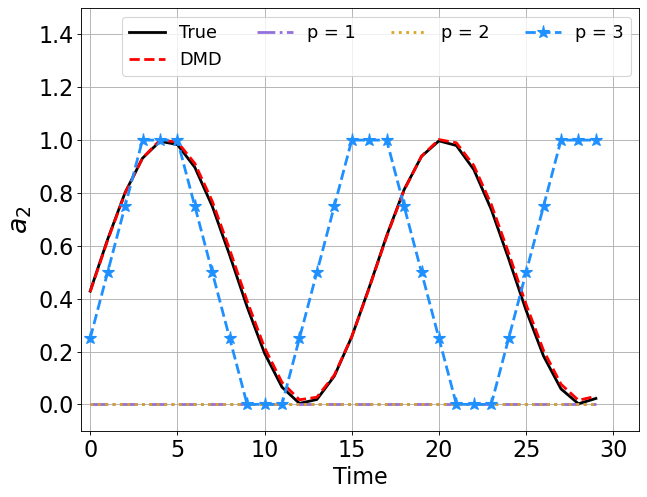}
  \caption{}
  \label{fig:QUBO_naca0009_r6_a2}
\end{subfigure}
\begin{subfigure}[!htb]{.45\textwidth}
 \centering
  \includegraphics[trim=0cm 0cm 0cm 0cm,clip,width=1\linewidth]{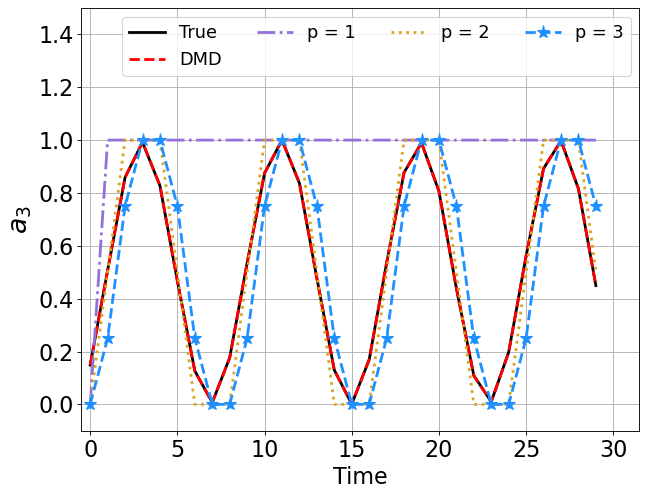}
  \caption{}
  \label{fig:QUBO_naca0009_r6_a3}
\end{subfigure}
\begin{subfigure}[!htb]{.45\textwidth}
 \centering
  \includegraphics[trim=0cm 0cm 0cm 0cm,clip,width=1\linewidth]{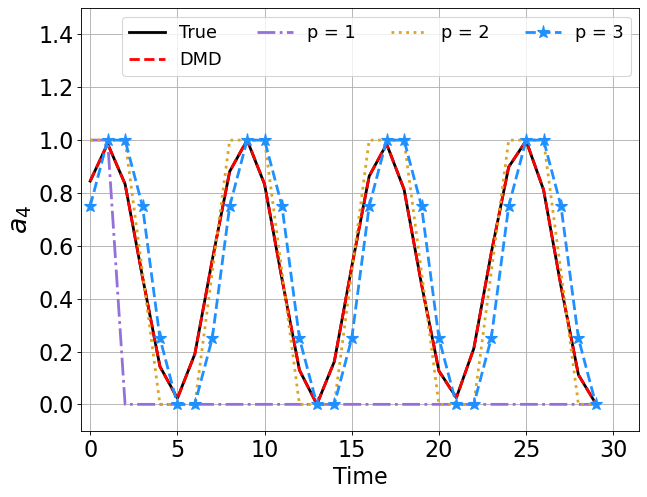}
  \caption{}
  \label{fig:QUBO_naca0009_r6_a4}
\end{subfigure}
\begin{subfigure}[!htb]{.45\textwidth}
 \centering
  \includegraphics[trim=0cm 0cm 0cm 0cm,clip,width=1\linewidth]{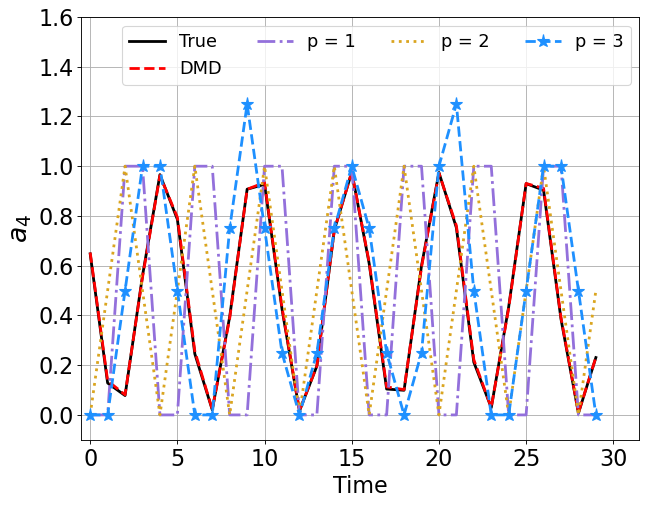}
  \caption{}
  \label{fig:QUBO_naca0009_r6_a5}
\end{subfigure}
\begin{subfigure}[!htb]{.45\textwidth}
 \centering
  \includegraphics[trim=0cm 0cm 0cm 0cm,clip,width=1\linewidth]{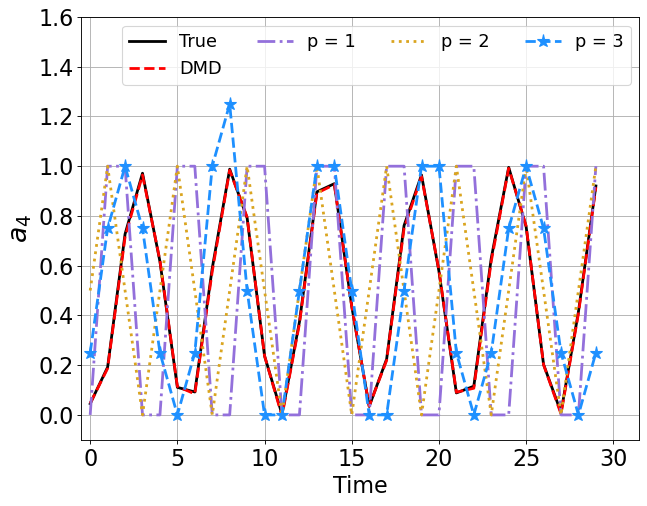}
  \caption{}
  \label{fig:QUBO_naca0009_r6_a6}
\end{subfigure}
\caption{QUBO predictions compared to a DMD model with r = 6 for mode coefficients (a) $a_1$ - (f) $a_6$ for varying number of bits for flow over a NACA0009 airfoil.}
\label{fig:QUBO_r6_results_naca0009}
\end{figure}

It is of interest to evaluate the evolution of the state using the quadratic unconstrained binary optimization methodology (QUBO) and annealing-based DMD methodology (QUBO-QAOA) to compare with implicit DMD for the more complex system considered in this section. For the implicit DMD results shown in this section, a r = 6 model was utilized, which corresponds to retaining 99\% of the kinetic energy of the flowfield. It was challenging to increase the number of bits $p > 3$ for a model using six modes; if the truncation level was further reduced, results could be obtained for higher number of bits such as p = 4 and 5. However, implicit DMD itself obtains an inaccurate prediction of the dynamics in the long-time for a model with less than 6 modes, and therefore results are shown here for the r = 6 model. 

Figure~\ref{fig:QUBO_r6_results_naca0009} shows the results for predictions generated from a QUBO simulator for varying precision levels compared to predictions from a implicit DMD model with a truncation level of r = 6 for flow over a NACA0009 airfoil at an angle of attack of $15^\circ{}$ and Re = 500.

It can be observed in figures~\ref{fig:QUBO_naca0009_r6_a1} and \ref{fig:QUBO_naca0009_r6_a2} that p = 3 bits are necessary to obtain a prediction on the leading pair of modes, while it can be seen in figures ~\ref{fig:QUBO_naca0009_r6_a5} and \ref{fig:QUBO_naca0009_r6_a6} that when using only 1 bit, the model is able to obtain a reasonably accurate prediction of the amplitude and frequency, although it is only the p = 3 model that predicts an accurate phase in the higher-order modes. 

\begin{figure}[!htb]
\centering
\begin{subfigure}[!htb]{.45\textwidth}
 \centering
  \includegraphics[trim=0cm 0cm 0cm 0cm,clip,width=1\linewidth]{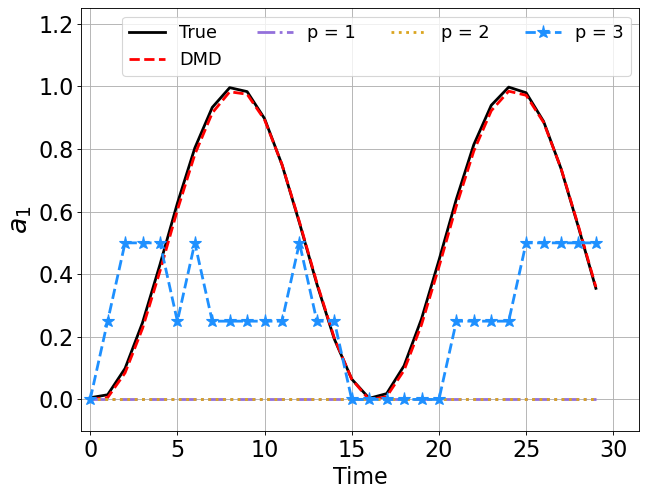}
  \caption{}
  \label{fig:QUBO-QAOA_naca0009_r6_a1}
\end{subfigure}
\begin{subfigure}[!htb]{.45\textwidth}
 \centering
  \includegraphics[trim=0cm 0cm 0cm 0cm,clip,width=1\linewidth]{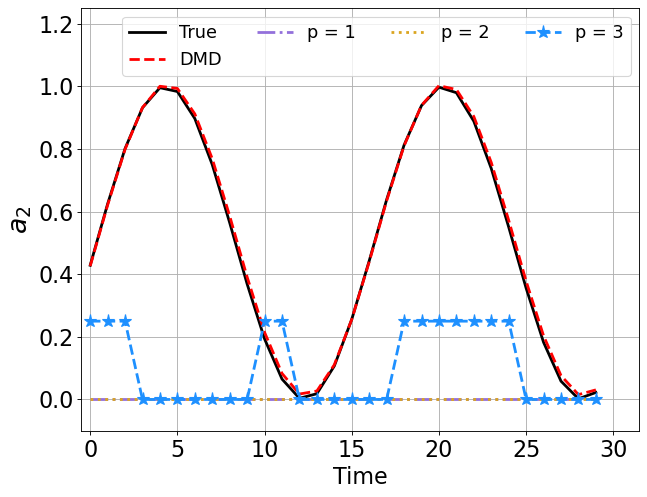}
  \caption{}
  \label{fig:QUBO-QAOA_naca0009_r6_a2_naca}
\end{subfigure}
\begin{subfigure}[!htb]{.45\textwidth}
 \centering
  \includegraphics[trim=0cm 0cm 0cm 0cm,clip,width=1\linewidth]{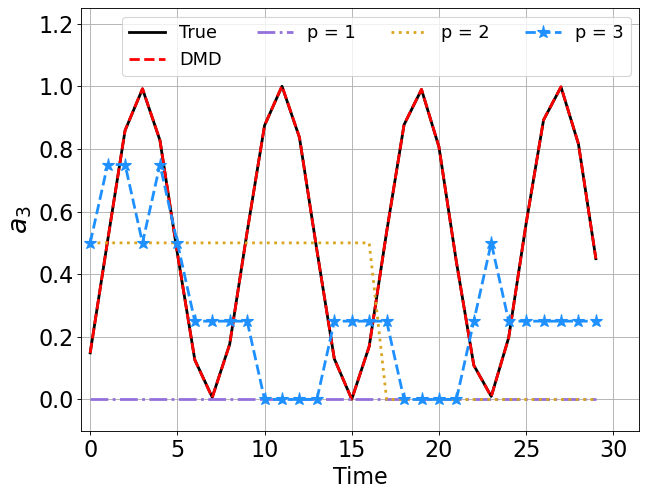}
  \caption{}
  \label{fig:QUBO-QAOA_naca0009_r6_a3_naca}
\end{subfigure}
\begin{subfigure}[!htb]{.45\textwidth}
 \centering
  \includegraphics[trim=0cm 0cm 0cm 0cm,clip,width=1\linewidth]{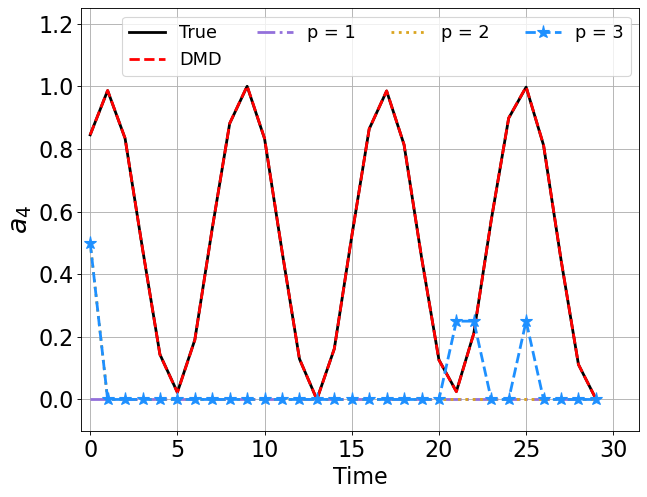}
  \caption{}
  \label{fig:QUBO-QAOA_naca0009_r6_a4_naca}
\end{subfigure}
\begin{subfigure}[!htb]{.45\textwidth}
 \centering
  \includegraphics[trim=0cm 0cm 0cm 0cm,clip,width=1\linewidth]{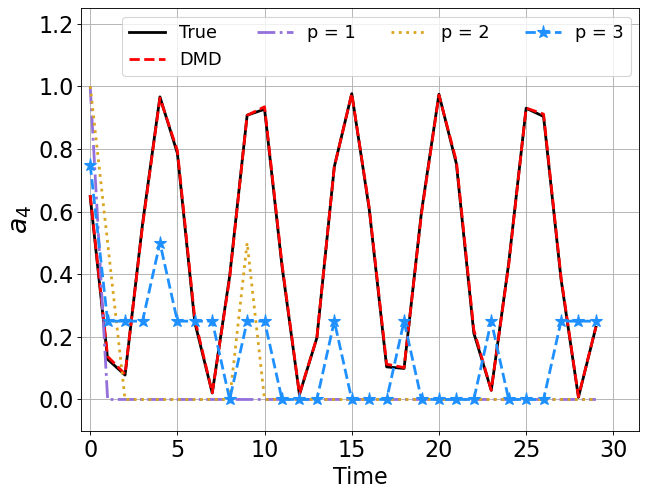}
  \caption{}
  \label{fig:QUBO-QAOA_naca0009_r6_a5_naca}
\end{subfigure}
\begin{subfigure}[!htb]{.45\textwidth}
 \centering
  \includegraphics[trim=0cm 0cm 0cm 0cm,clip,width=1\linewidth]{QUBO-QAOA_NACA0009_compare_r6_a5_pfigs}
  \caption{}
  \label{fig:QUBO-QAOA_naca0009_r6_a6_naca}
\end{subfigure}
\caption{QUBO-QAOA predictions compared to a DMD model with r = 6 for mode coefficients (a) $a_1$ - (f) $a_6$ for varying the number of bits for flow over a NACA0009 airfoil.}
\label{fig:QUBO-QAOA_r6_results_naca0009}
\end{figure}

Figure~\ref{fig:QUBO-QAOA_r6_results_naca0009} shows the results for predictions generated from the quantum annealer QUBO-QAOA model for different numbers of bits compared to predictions from a classical DMD model with a truncation level of r = 6 for flow over a NACA0009 airfoil at an angle of attack of $15^\circ{}$ and Re = 500.

The QUBO-QAOA model struggles to obtain an accurate prediction of the state trajectory in terms of both amplitude and frequency for p = 3, with p = 1 and p = 2 models unable to attain a meaningful prediction. As seen in figures~\ref{fig:QUBO-QAOA_naca0009_r6_a5_naca} and \ref{fig:QUBO-QAOA_naca0009_r6_a6_naca}, the model is able to capture an attempt at the frequency for the higher-order modes, but the amplitude is significantly off and the periodicity observed in this system is not captured. Higher numbers of bits are hypothesized to obtain a better prediction for the QUBO-QAOA model, but exceeded the availability of current computational resources.

\subsection{Flowfield Reconstruction for Airfoil Flow}\label{airfoil_results_recon}

The flowfield was reconstructed from implicit DMD and compared to the results from QUBO and QUBO-QAOA to evaluated how well these ROMs perform at predicting the evolution of the dynamics of the system.

\begin{figure}[!htb]
\centering
\begin{subfigure}[!htb]{.45\textwidth}
 \centering
  \includegraphics[trim=0cm 0cm 0cm 0cm,clip,width=1\linewidth]{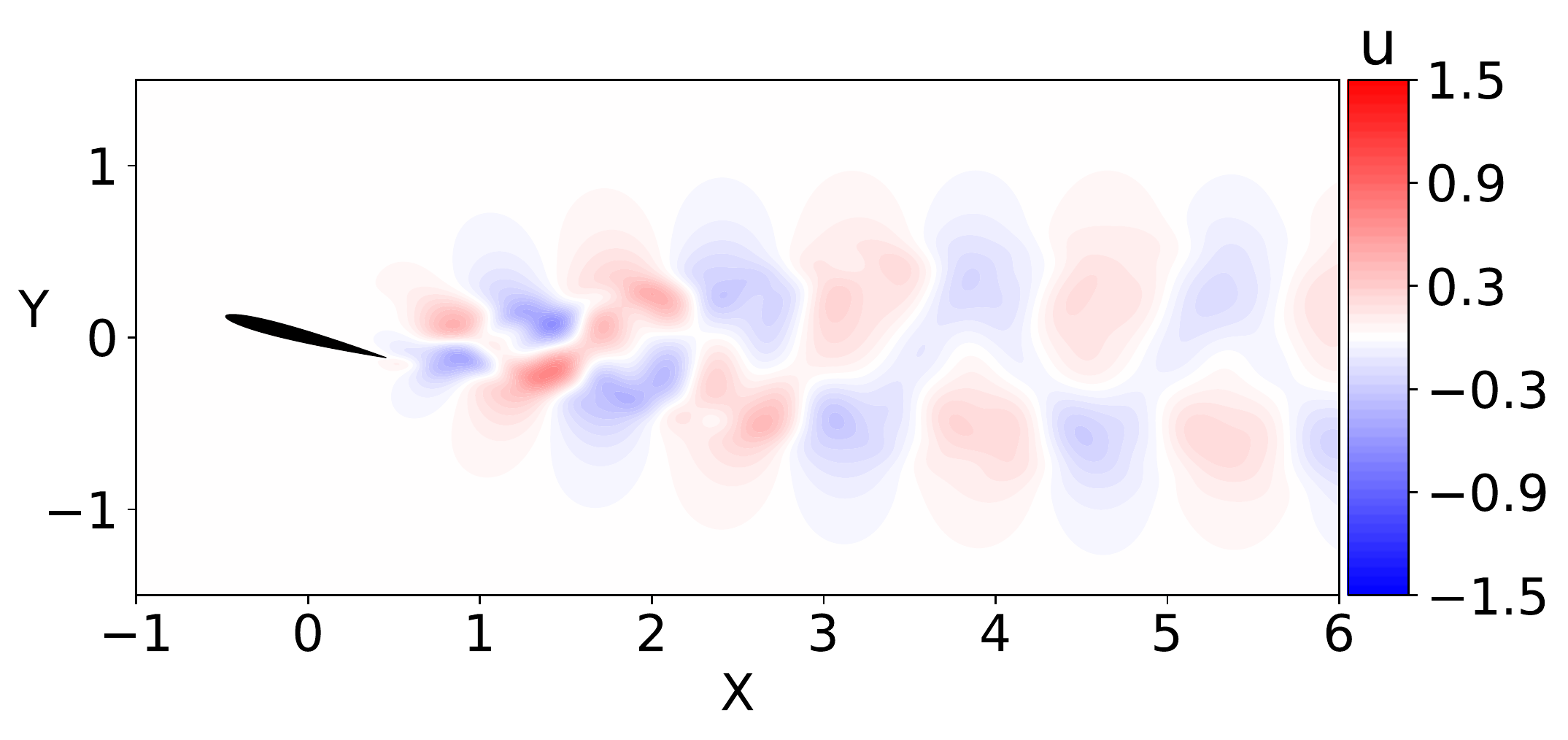}
  \caption{}
  \label{fig:DMD_naca_u1}
\end{subfigure}
\begin{subfigure}[!htb]{.45\textwidth}
 \centering
  \includegraphics[trim=0cm 0cm 0cm 0cm,clip,width=1\linewidth]{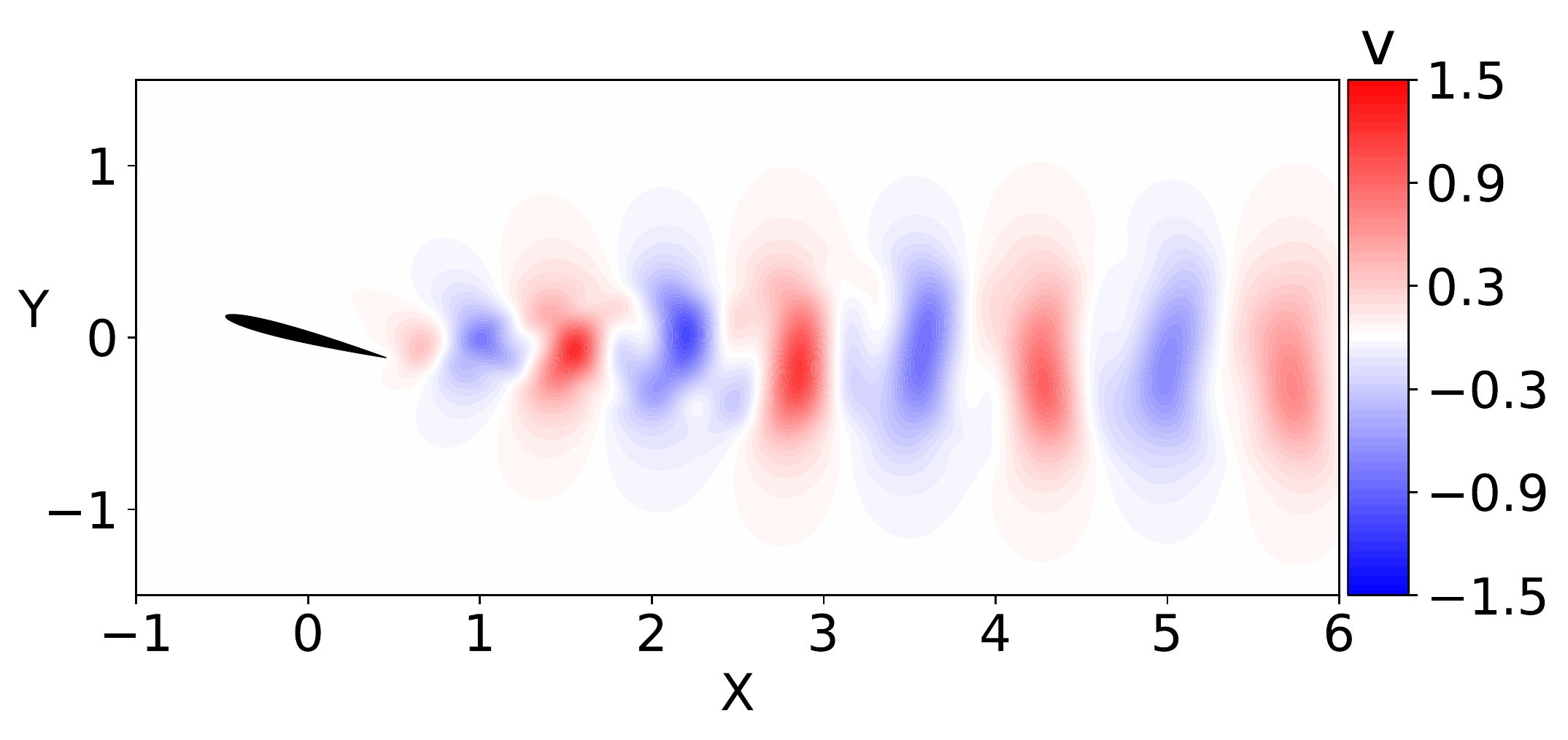}
  \caption{}
  \label{fig:DMD_naca_v1}
\end{subfigure}
\begin{subfigure}[!htb]{.45\textwidth}
 \centering
  \includegraphics[trim=0cm 0cm 0cm 0cm,clip,width=1\linewidth]{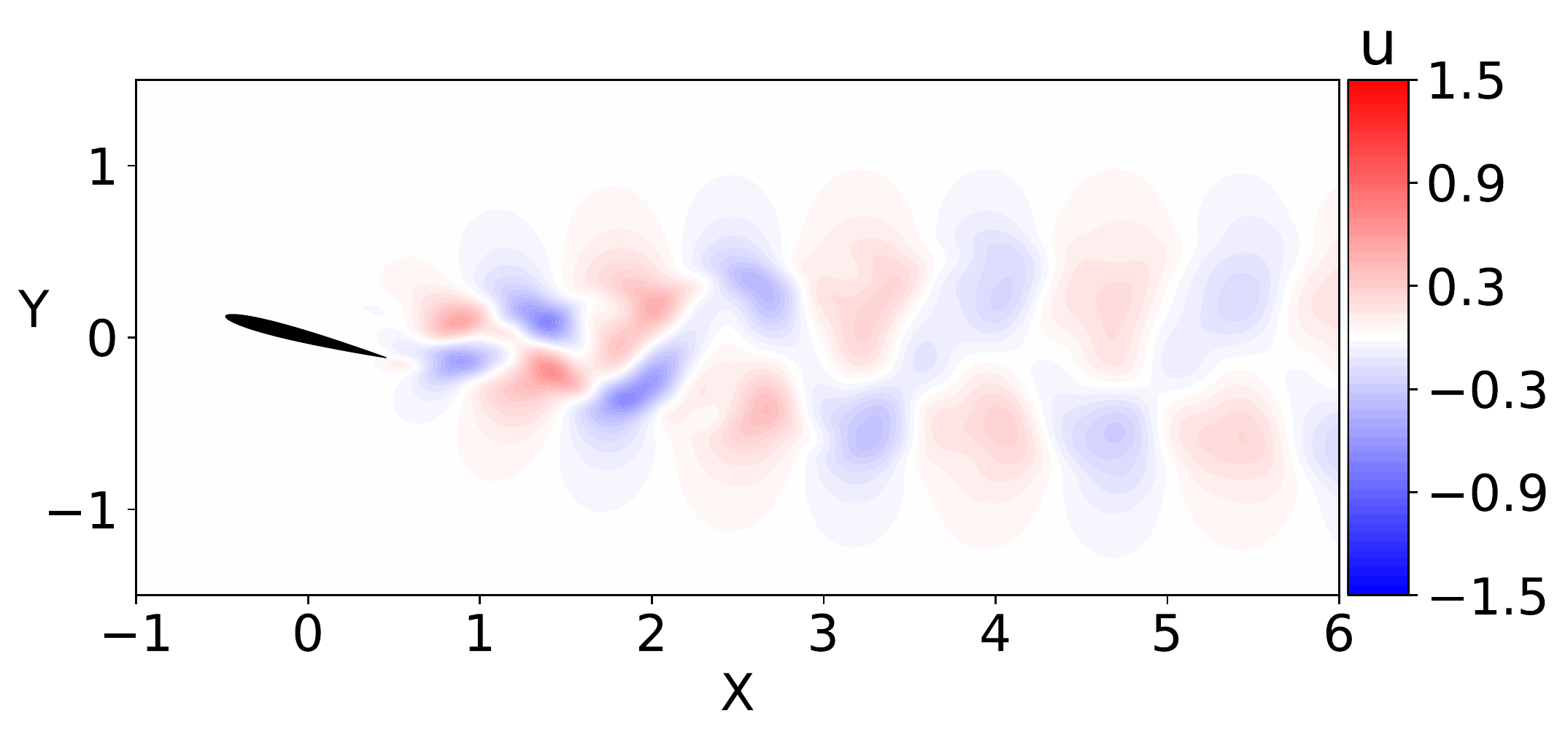}
  \caption{}
  \label{fig:QUBO_naca_u1}
\end{subfigure}
\begin{subfigure}[!htb]{.45\textwidth}
 \centering
  \includegraphics[trim=0cm 0cm 0cm 0cm,clip,width=1\linewidth]{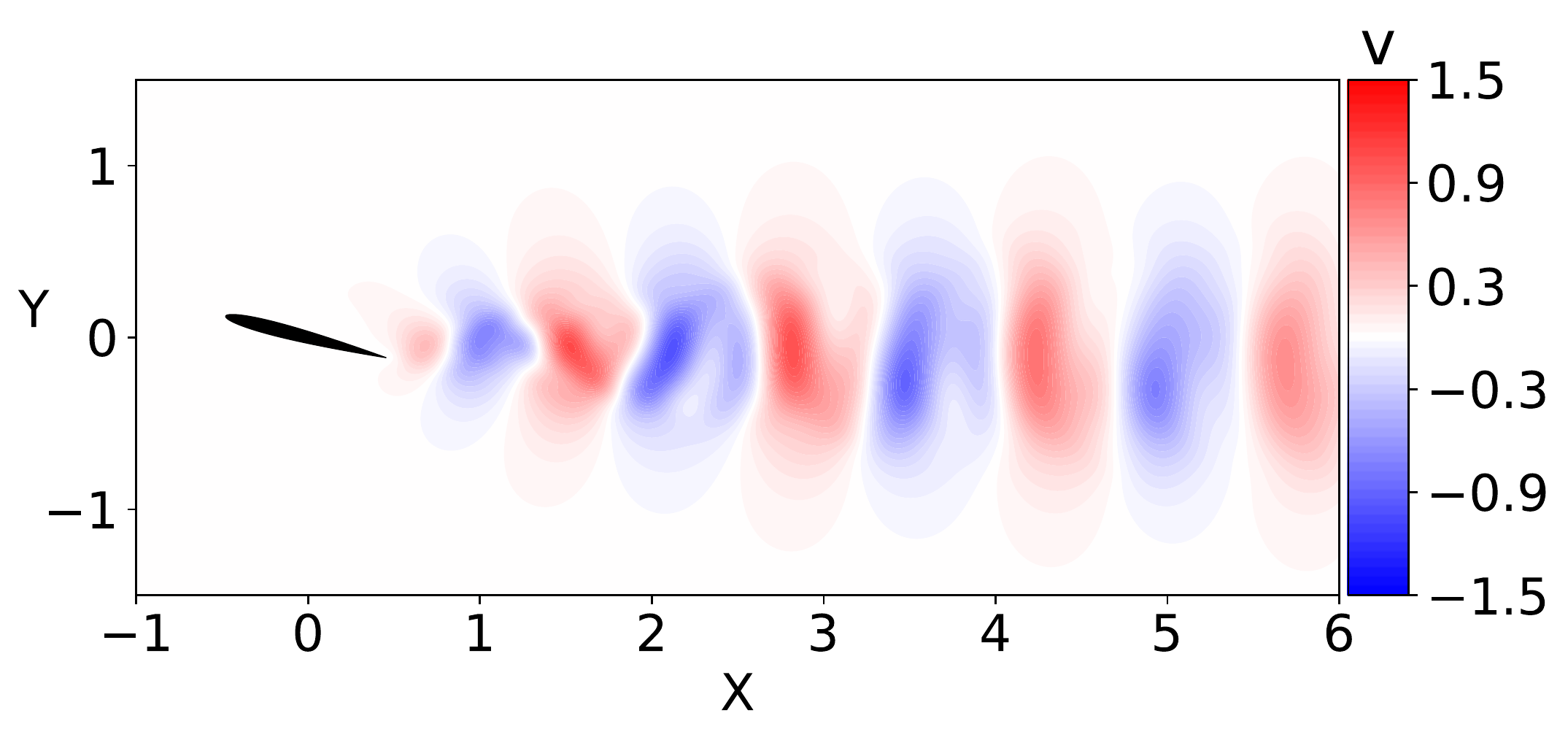}
  \caption{}
  \label{fig:QUBO_naca_v1}
\end{subfigure}
\begin{subfigure}[!htb]{.45\textwidth}
 \centering
  \includegraphics[trim=0cm 0cm 0cm 0cm,clip,width=1\linewidth]{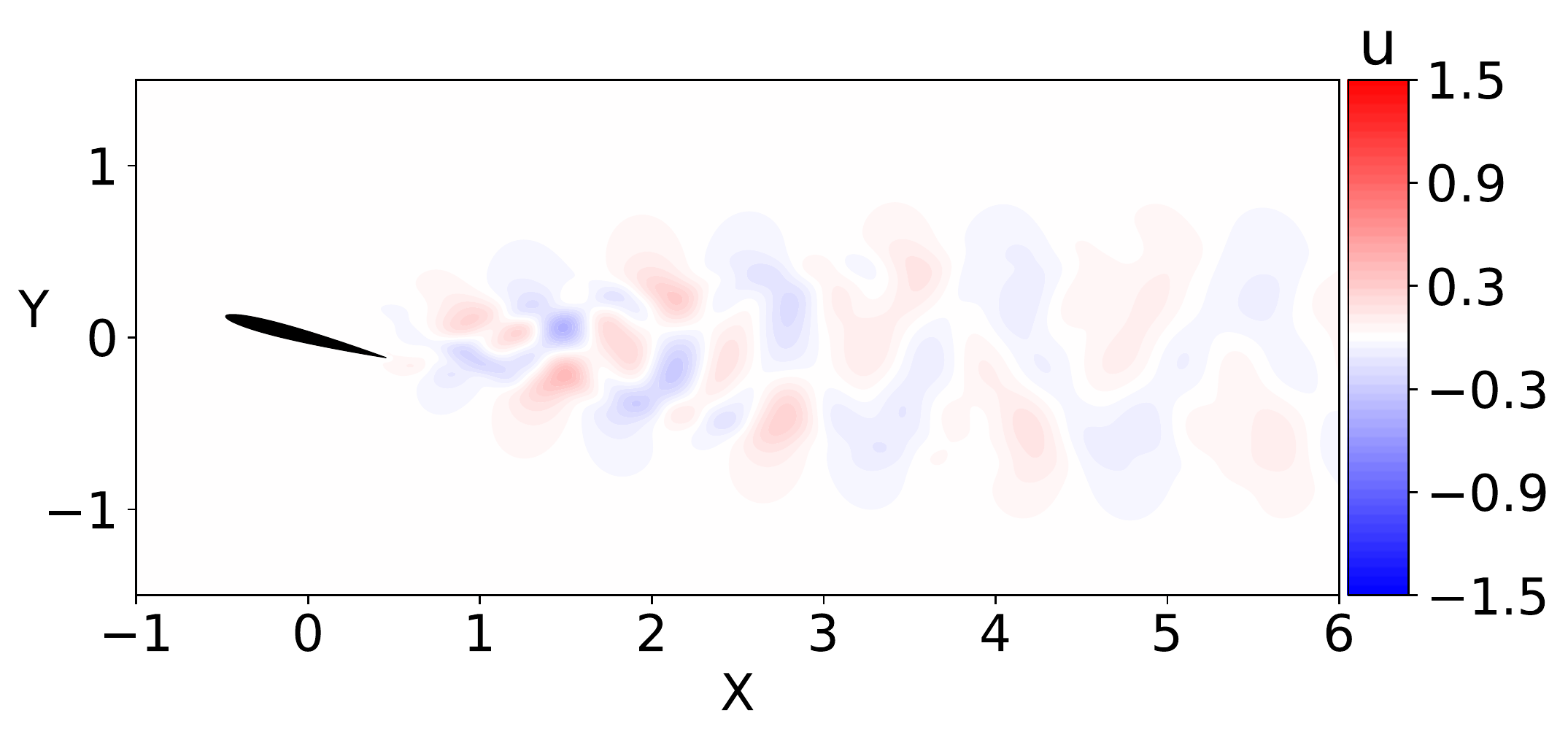}
  \caption{}
  \label{fig:QUBO_QAOA_naca_u1}
\end{subfigure}
\begin{subfigure}[!htb]{.45\textwidth}
 \centering
  \includegraphics[trim=0cm 0cm 0cm 0cm,clip,width=1\linewidth]{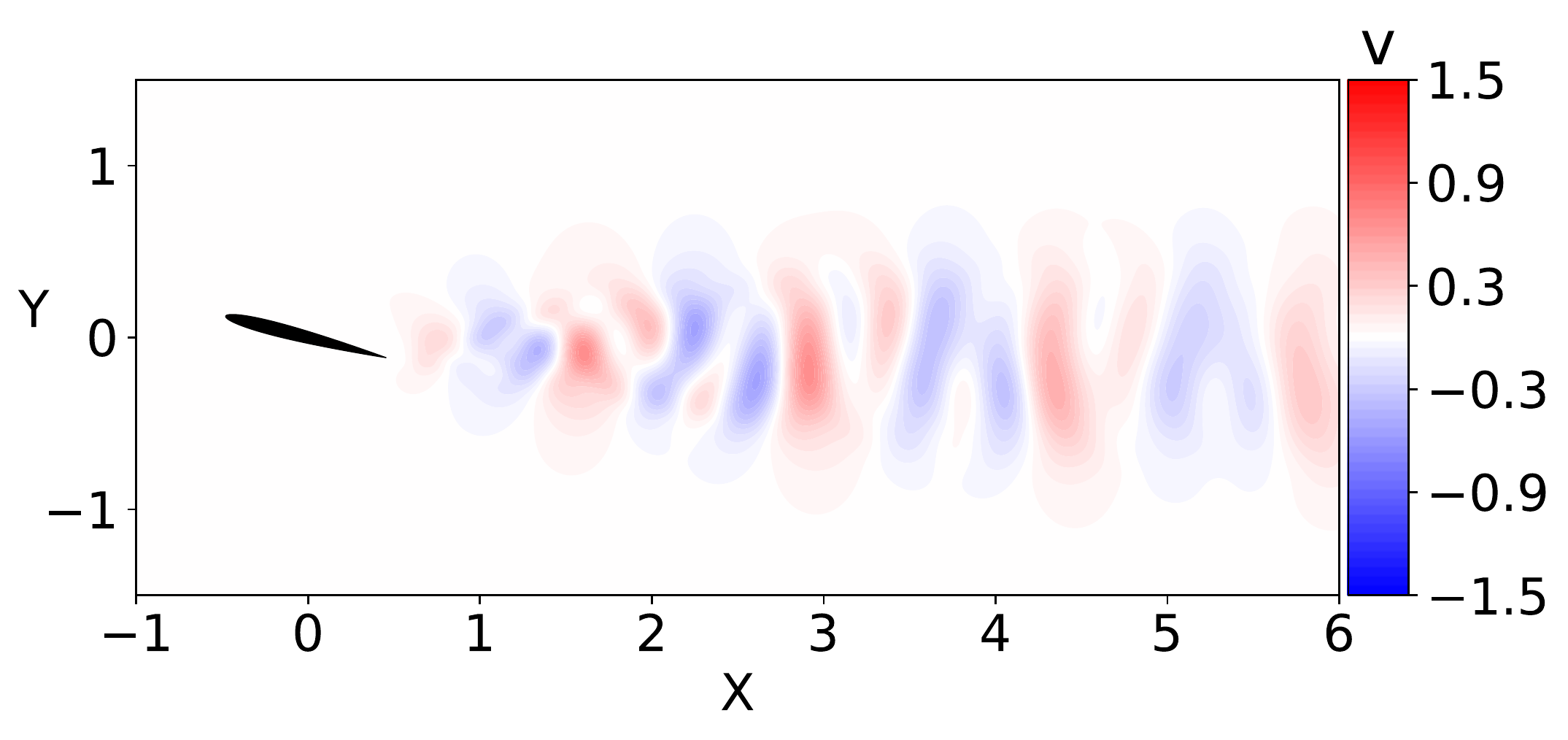}
  \caption{}
  \label{fig:QUBO_QAOA_naca_v1}
\end{subfigure}
\caption{Reconstructed flowfield for both velocity components computed at t = 10 from (a) -- (b) DMD using r = 6 modes, (c) -- (d) QUBO using r = 6 modes and a precision of p = 3, and (e) -- (f) QUBO-QAOA using r = 6 and a precision of p = 3. The reconstructed flowfield computed from predictions generated by QUBO for a reduced-order model or r = 6 modes and p = 3 bits.}
\label{fig:DMD_QUBO_QAOA_naca_reconstruct_r6_p3}
\end{figure}

For each reconstruction, r = 6 POD modes (corresponding to $\approx99\%$ of the kinetic energy of the flowfield) was utilized. The mean-subtracted flowfield was computed, to discern the models ability to capture the vortex-shedding patterns in the wake. For the QUBO and QUBO-QAOA optimization procedures, p = 3 bits was utilized. The reconstructed flowfield is shown in Figure~\ref{fig:DMD_QUBO_QAOA_naca_reconstruct_r6_p3} at a timestep corresponding to t = 7. 

Each model is able to capture the majority of the dominant patterns in the wake of the airfoil, namely the pattern of opposite-sign structures corresponding to vortex shedding. It can be seen in the u-component of the reconstructed velocity that the wake is more disorganized for QUBO-QAOA (figure~\ref{fig:QUBO_QAOA_naca_u1}) than for the DMD model (figure~\ref{fig:DMD_naca_u1}), with the QUBO results (figure~\ref{fig:QUBO_naca_u1}) matching the DMD results more closely. In general, the DMD (figure~\ref{fig:DMD_naca_v1}) and QUBO (figure~\ref{fig:QUBO_naca_v1}) models attain a close prediction in comparison to each other, with the QUBO-QAOA model (figure~\ref{fig:QUBO_QAOA_naca_v1}) more damped, owing to the model significantly under-predicting the amplitude of the leading mode coefficients.

\begin{figure}[htb!]
 \centering
  \includegraphics[trim=0cm 0cm 0cm 0cm,clip,width=0.75\linewidth]{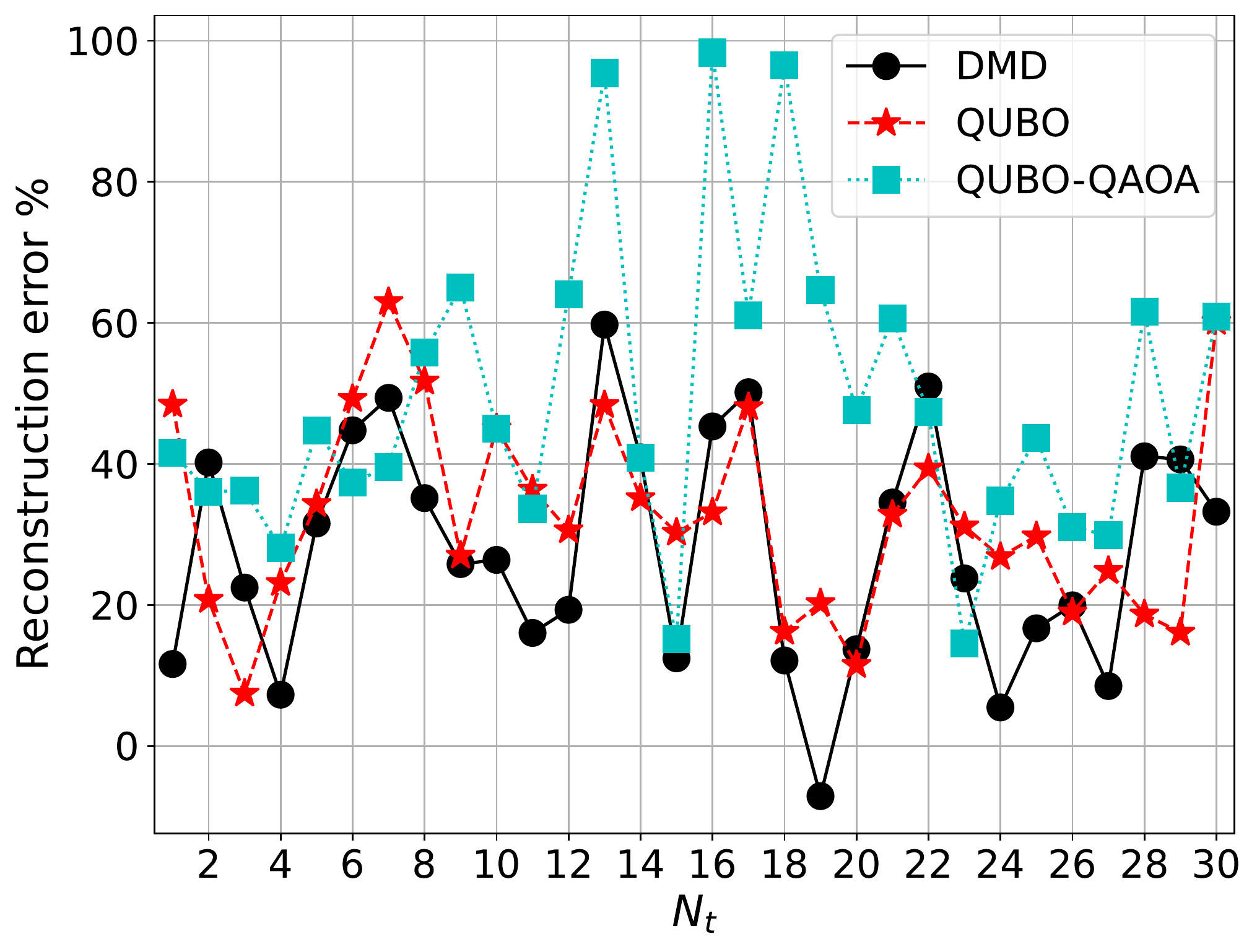}
\caption{Reconstruction error computed for implicit DMD, QUBO, and QUBO-QAOA optimization methodologies.}
\label{fig:error_reconstruct_naca}
\end{figure}

Figure~\ref{fig:error_reconstruct_naca} shows the results computed for a r = 6 model with p = 3 bits for the quantum optimization DMD algorithms. It can be observed that the QUBO optimization methodology is closer to the reconstruction error obtained with implicit DMD, while the QUBO-QAOA optimization procedure has a higher reconstruction error than both.

\section{Complexity Analysis}\label{sect_complexity}

In the computational experiments detailed in the previous sections, the sensitivity to varying truncation levels and number of bits utilized in a fixed-point representation was investigated. The size of the considered linear system is a function of these numerical parameters. A key research question we aim to answer in this work is how the quantum computer resources needed for the considered reduced-order models depend on the two parameters.

\subsection{DMD - system size as a function of $r$ and $p$}\label{complexity_DMD}

For a `single time-step' update, DMD involves solving the system,

\begin{align}
    J = \min_{\mathbf{x}^{n+1}} || (\mathbf{A}^T \mathbf{A})^{-1} \mathbf{A}^T \mathbf{x}^{n+1} - \mathbf{x}^n||
\end{align}
with $\mathbf{A}$ and $(\mathbf{A}^T \mathbf{A})$ being $r\times r$ matrices, where $r$ is the truncation level corresponding to the number of modes retained in the model (here, we perform POD on the space of POD mode coefficients; recovering the full state size is achieved through projection onto the POD modes) and $p$ is the number of bits used in a fixed-point representation. For ``multiple time-step" updates, the system size grows linearly with the number of time steps considered, $N_{step}$.

\subsection{2D cylinder flow - values for $r$ and $p$ required}\label{complexity_qubo}

Experiments using a QUBO simulator performed at $r=5$ for flow over a cylinder and varying the number of bits required for a fixed-point representation showed that increasing the number of bits (while maintaining a constant truncation level) had a similar effect as increasing the truncation level for implicit DMD does. It was found that for the highest number of bits considered, i.e. $p=5$, the QUBO model is able to accurately capture all of the modes, including the leading modes. In this case, predictions from higher values of bits (p $>$ 5) were not obtained due to their computational cost. In contrast, the corresponding experiments using QAOA simulator showed that QAOA-based model is unable to completely capture the state for higher bit levels ($p=5$), and at lower bit values ($p = 1$ and $2$) the results were quite inaccurate. For low $p$, results for QUBO and QAOA had similar poor levels of agreement with ``regular" DMD.

For the truncation level of $r=8$, due to computational limitations, only experiments with low number of bits ($p=1$ and $p=2$) could be performed.

\clearpage

\section{Conclusions}
\label{sect_conclusions}
The application of DMD on near-term Quantum Computers was investigated. For this type of machine, the small qubit count and severe limitations on acceptable quantum-circuit depth motivated the formulation of time-stepping in terms of optimization problems. Two approaches were investigated in detail. The first involved quantum annealing based on the reformulation of the linear system in terms of a QUBO. The second approach similarly employs a QUBO, followed by a QAOA-based solution method. Dynamic mode decomposition is reformulated as an optimization problem to propagate the state of the linearized dynamical system on a quantum computer. A quantum-ROM is obtained using the two aforementioned approaches and utilized to generate predictions of the state trajectories. Two examples are shown for bluff body flows: flow over a 2D cylinder and flow over a 2D NACA0009 airfoil. The reconstructed flowfield is compared with the true flowfield, and qualitative and quantitative comparisons find that the quantum-ROM is able to recover the key features of these flowfields. In particular, we find that the accuracy of the quantum-ROMs depends on the truncation level of the DMD model as well as the number of bits utilized for a fixed point representation with QUBO. 


Future work will include the application of physics-informed DMD~\cite{baddoo2023physics} on a (more) fault-tolerant Quantum Computer, using a direct quantum-circuit implementation of the DMD matrix-vector multiplication. This facilitates a comparison of the prospects of quantum-ROM on NISQ-era hardware (as studied here) as well as future quantum hardware with fewer restrictions on quantum circuit depth.


Furthermore, interesting applications of the methodology introduced here for quantum-ROM could be applied to control problems. Particularly, extensions for accounting for inputs would be well-suited for considering actuated systems. 

It is of great interest to apply the methodology discussed in this problem to a real quantum computer, and is one of the immediate goals following this work.

On-going work is involved with parallelizing the code to be able to increase the number of bits utilized for a fixed point representation with QUBO, which will also be beneficial for extending this application to more complex problems in which the truncation level of the DMD model might also increase to capture all of the relevant physics necessary for an accurate representation of the inherent dynamics of the problem. 


\section{Acknowledgements}

This work was supported by the U.S. Department of Energy (DOE), Office of Science, Office of Advanced Scientific Computing Research (ASCR), under Contract No. DEAC02–06CH11357, at Argonne National Laboratory. We acknowledge support from the Argonne National Laboratory LCRC (Laboratory Computing Resource Center) for the numerical experiments in this article.

\clearpage
\newpage



\appendix

\section{Performing Multiple Steps in QUBO}\label{results_cylflow_MS}

\begin{figure}[!htb]
\centering
\begin{subfigure}[!htb]{.4\textwidth}
 \centering
  \includegraphics[trim=0cm 0cm 0cm 0cm,clip,width=1\linewidth]{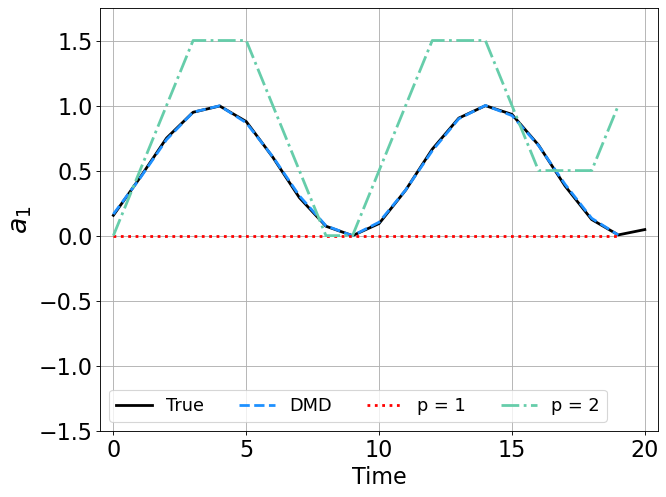}
  \caption{}
  \label{fig:QUBO_MS_r5_a1}
\end{subfigure}
\begin{subfigure}[!htb]{.4\textwidth}
 \centering
  \includegraphics[trim=0cm 0cm 0cm 0cm,clip,width=1\linewidth]{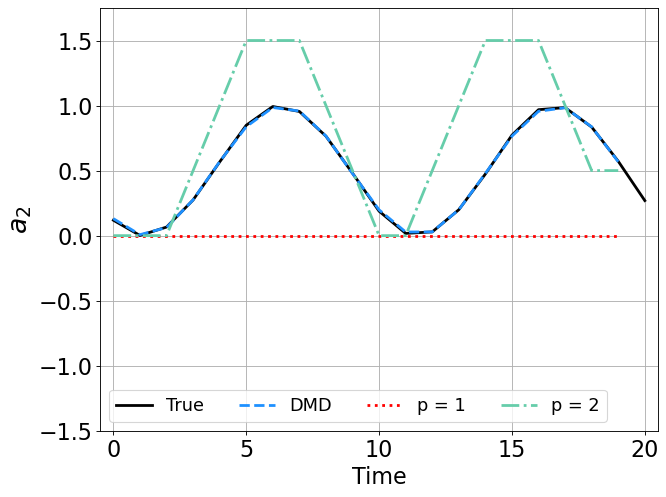}
  \caption{}
  \label{fig:QUBO_MS_r5_a2}
\end{subfigure}
\begin{subfigure}[!htb]{.4\textwidth}
 \centering
  \includegraphics[trim=0cm 0cm 0cm 0cm,clip,width=1\linewidth]{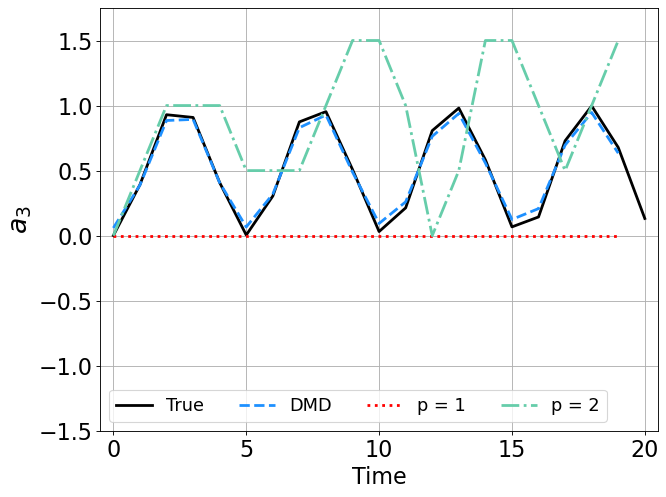}
  \caption{}
  \label{fig:QUBO_MS_r5_a3}
\end{subfigure}
\begin{subfigure}[!htb]{.4\textwidth}
 \centering
  \includegraphics[trim=0cm 0cm 0cm 0cm,clip,width=1\linewidth]{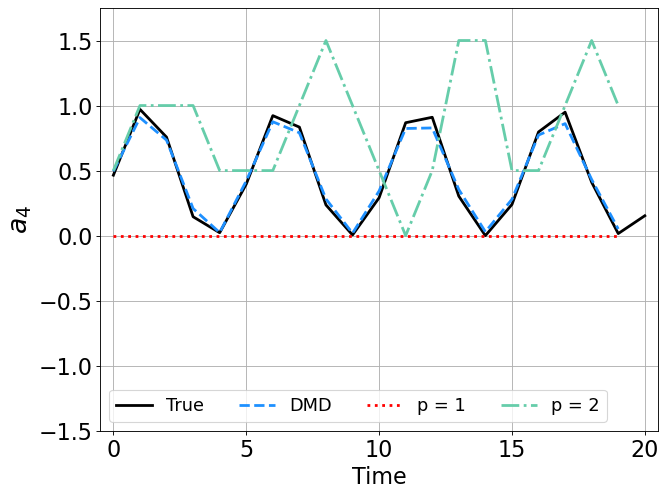}
  \caption{}
  \label{fig:QUBO_MS_r5_a4}
\end{subfigure}
\caption{QUBO predictions compared to a DMD model with r = 5 for mode coefficients (a) $a_1$ - (d) $a_4$ for varying the number of bits utilized with $N_{steps} = 2$.}
\label{fig:QUBO_MS_r5_results}
\end{figure}

Figure~\ref{fig:QUBO_MS_r5_results} shows the results for predictions generated from a QUBO simulator for varying the number of bits utilized for a fixed-point representation compared to predictions from an implicit DMD model with a truncation level of r = 5 for flow over a cylinder, with predictions made for $N_{steps} = 2$.

\begin{figure}[!htb]
\centering
\begin{subfigure}[!htb]{.4\textwidth}
 \centering
  \includegraphics[trim=0cm 0cm 0cm 0cm,clip,width=1\linewidth]{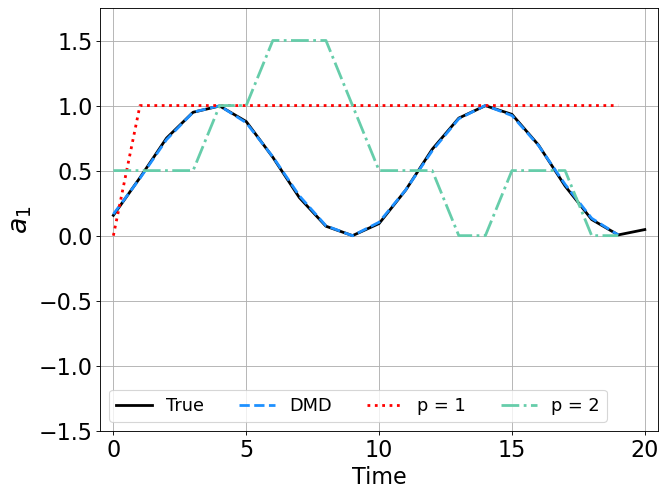}
  \caption{}
  \label{fig:QUBO-QAOA_MS_r5_a1}
\end{subfigure}
\begin{subfigure}[!htb]{.4\textwidth}
 \centering
  \includegraphics[trim=0cm 0cm 0cm 0cm,clip,width=1\linewidth]{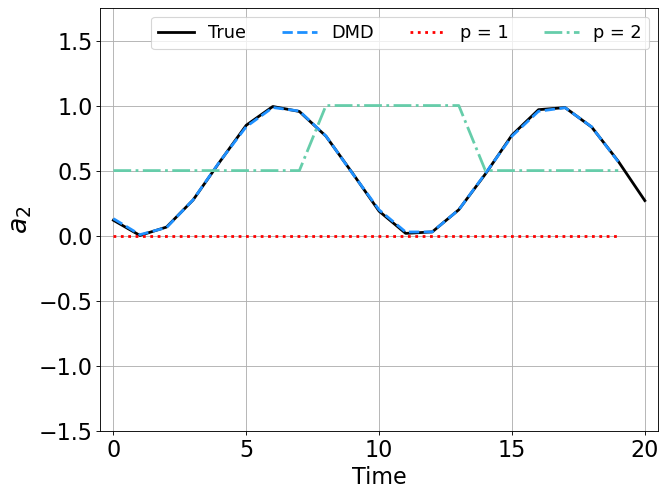}
  \caption{}
  \label{fig:QUBO-QAOA_MS_r5_a2}
\end{subfigure}
\begin{subfigure}[!htb]{.4\textwidth}
 \centering
  \includegraphics[trim=0cm 0cm 0cm 0cm,clip,width=1\linewidth]{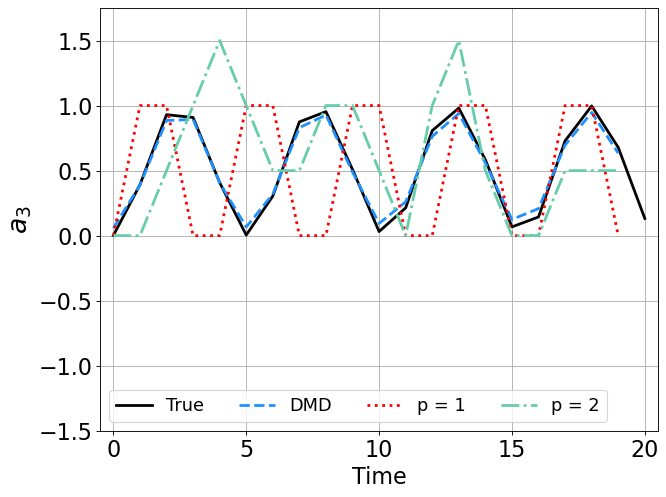}
  \caption{}
  \label{fig:QUBO-QAOA_MS_r5_a3}
\end{subfigure}
\begin{subfigure}[!htb]{.4\textwidth}
 \centering
  \includegraphics[trim=0cm 0cm 0cm 0cm,clip,width=1\linewidth]{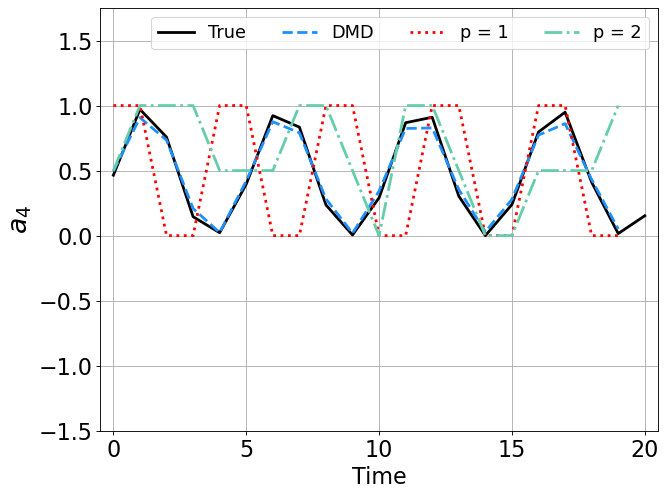}
  \caption{}
  \label{fig:QUBO-QAOA_MS_r5_a4}
\end{subfigure}
\caption{QUBO-QAOA predictions compared to a DMD model with r = 5 for mode coefficients (a) $a_1$ - (d) $a_4$ for varying the number of bits utilized with $N_{steps} = 2$.}
\label{fig:QUBO-QAOA_MS_r5_results}
\end{figure}

Figure~\ref{fig:QUBO-QAOA_MS_r5_results} shows the results for predictions generated from the quantum annealer QUBO-QAOA model for varying the number of bits utilized compared to predictions from an implicit DMD model with a truncation level of r = 5 for flow over a cylinder, with $N_{steps} = 2$.

\begin{figure}[htb!]
\centering
\begin{subfigure}{.4\textwidth}
 \centering
  \includegraphics[trim=0cm 0cm 0cm 0cm,clip,width=1\linewidth]{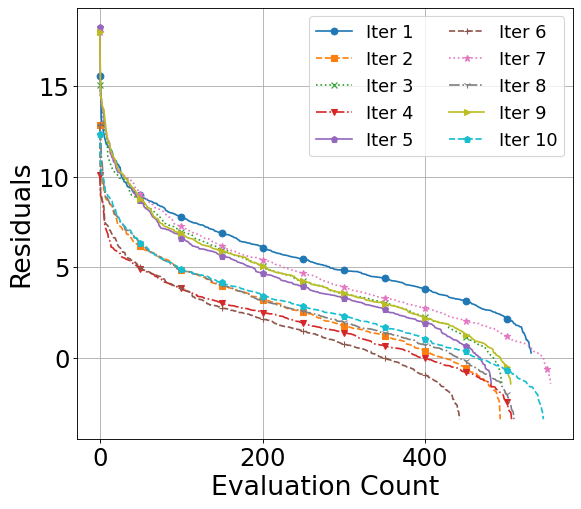}
  \caption{}
  \label{fig:QUBO-QAOA_MS_converge_p1}
\end{subfigure}
\begin{subfigure}{.4\textwidth}
 \centering
  \includegraphics[trim=0cm 0cm 0cm 0cm,clip,width=1\linewidth]{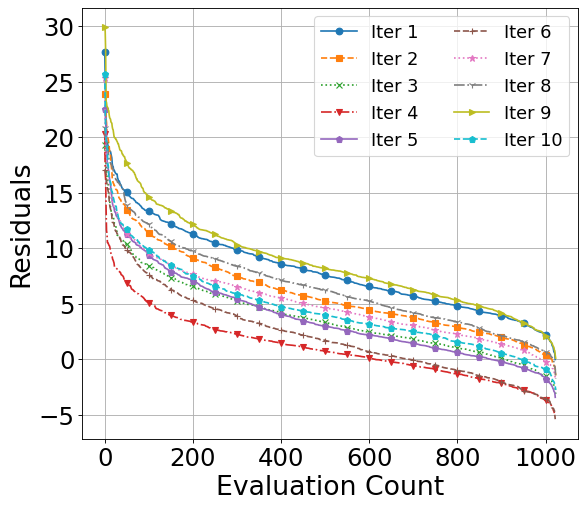}
  \caption{}
  \label{fig:QUBO-QAOA_MS_converge_p2}
\end{subfigure}
\caption{Number of iterations required for convergence for QUBO-QAOA computed for $N_{steps} = 2$ for (a) p = 1 and (b) p = 2.}
\label{fig:QUBO-QAOA_convergence_crit_MS}
\end{figure}

Figure~\ref{fig:QUBO-QAOA_convergence_crit_MS} shows the results for the number of iterations required for convergence as a function of the number of bits utilized for a model with $N_{steps} = 2$, and the same trend as before can be observed, that the number of iterations for convergence increases with the number of bits used. It can be seen that for $N_{steps} \geq 1$ and the same precision level, a greater number of iterations for convergence is required.

\subsection{Flowfield Reconstruction with Multiple Timestep Predictions}

\begin{figure}[!htb]
\centering
\begin{subfigure}[!htb]{.45\textwidth}
 \centering
  \includegraphics[trim=0cm 0cm 0cm 0cm,clip,width=1\linewidth]{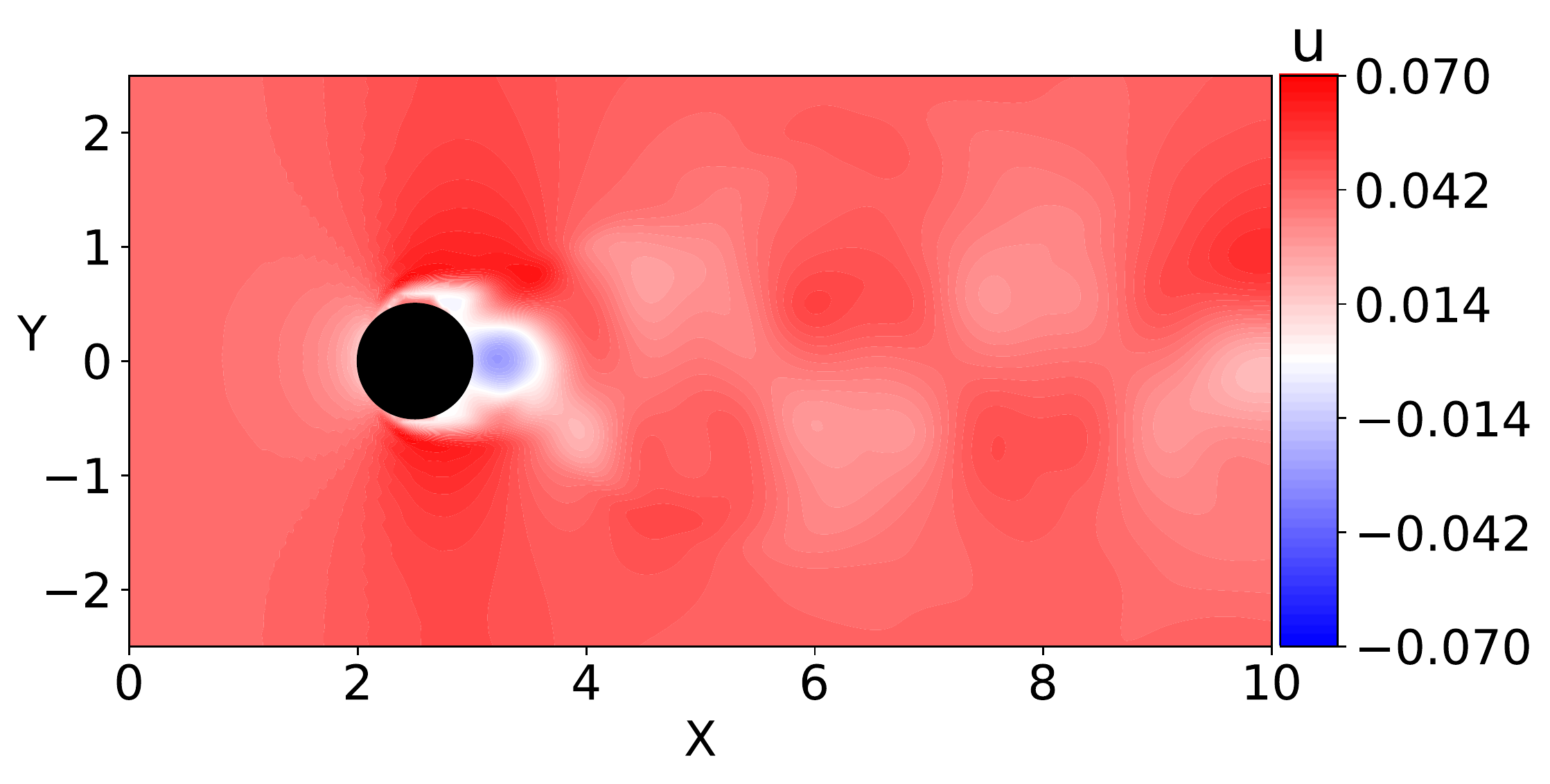}
  \caption{}
  \label{fig:DMD_u1_MS}
\end{subfigure}
\begin{subfigure}[!htb]{.45\textwidth}
 \centering
  \includegraphics[trim=0cm 0cm 0cm 0cm,clip,width=1\linewidth]{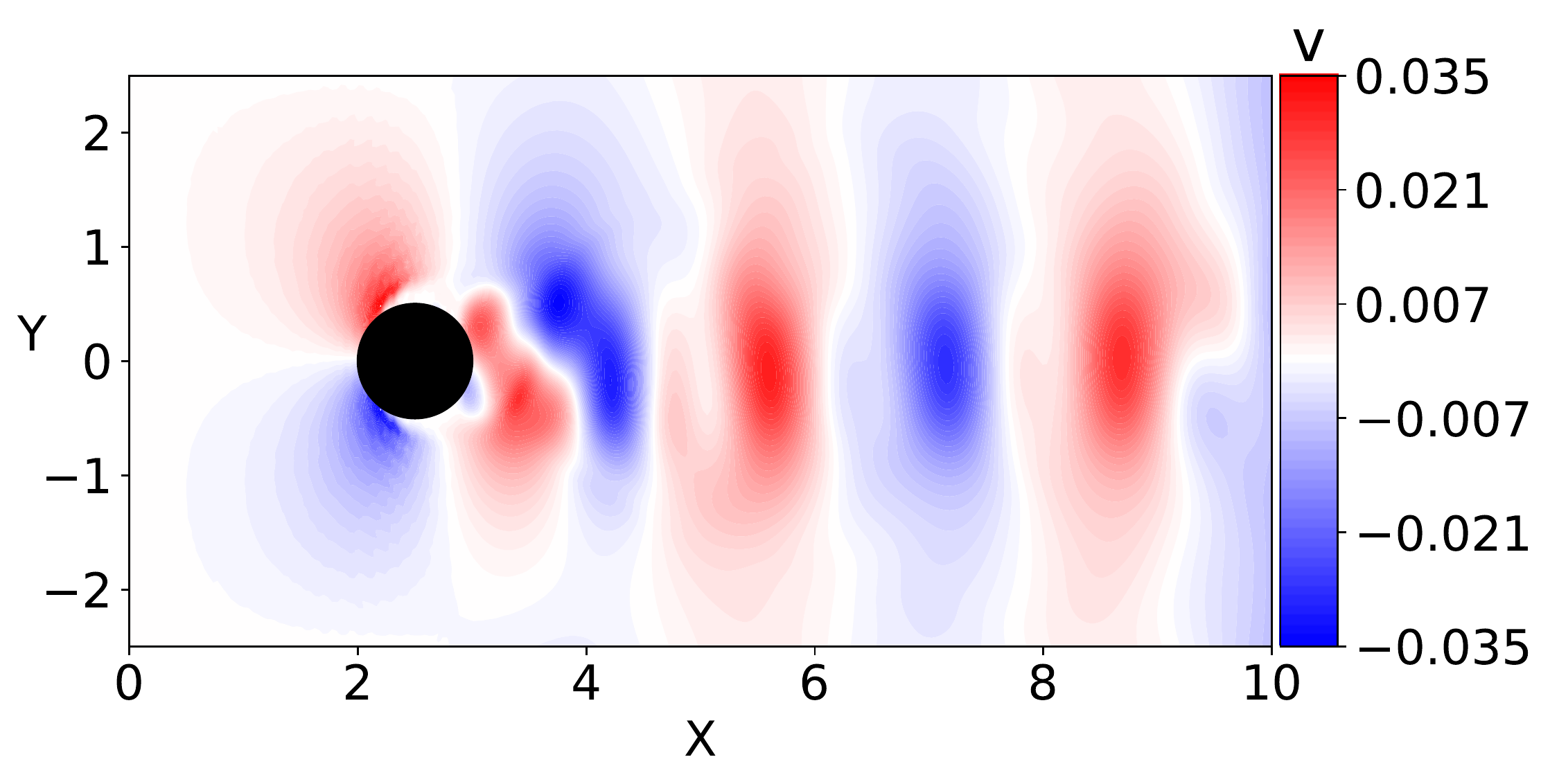}
  \caption{}
  \label{fig:DMD_v1_MS}
\end{subfigure}
\begin{subfigure}[!htb]{.45\textwidth}
 \centering
  \includegraphics[trim=0cm 0cm 0cm 0cm,clip,width=1\linewidth]{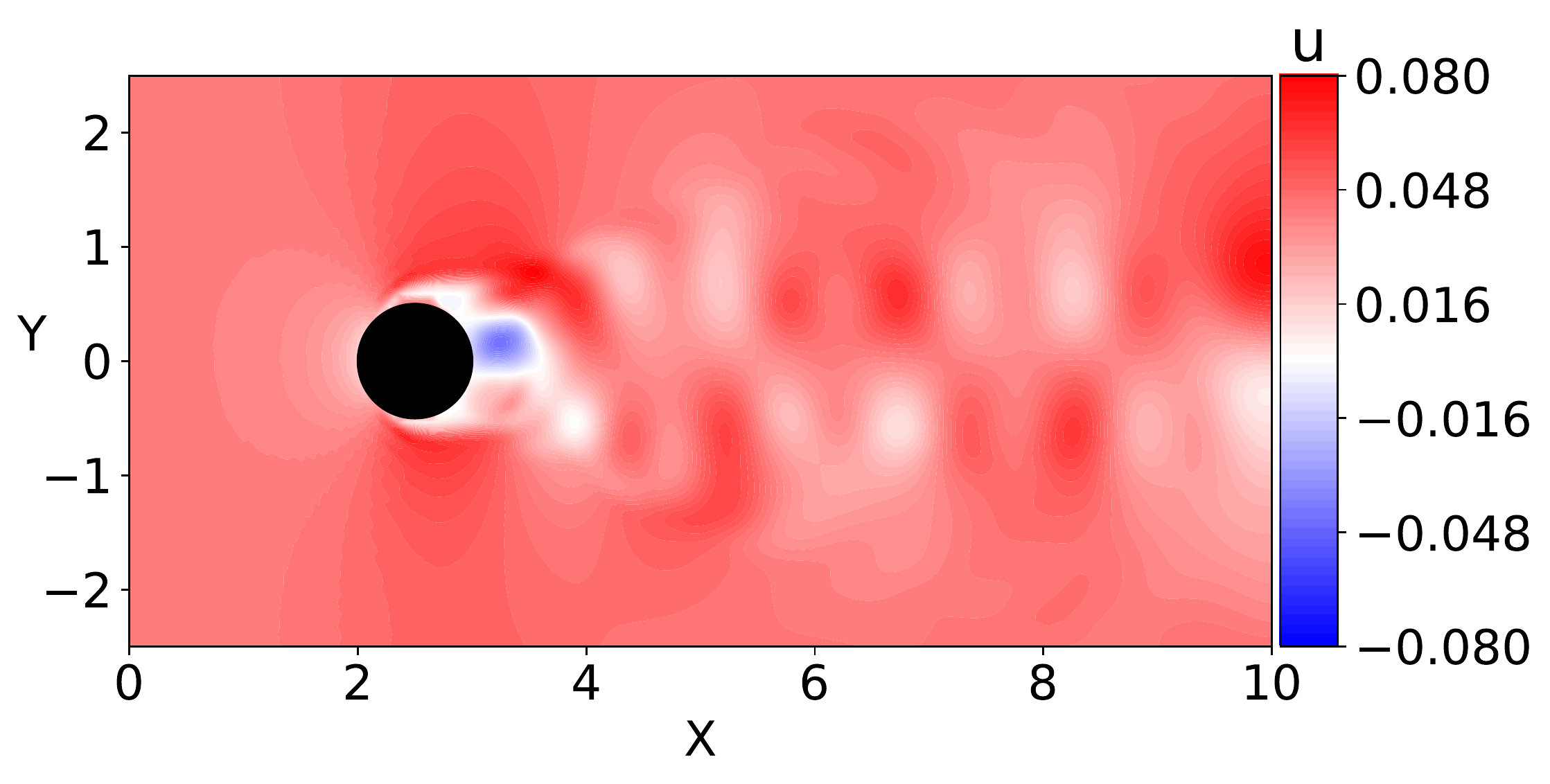}
  \caption{}
  \label{fig:QUBO_u1_MS}
\end{subfigure}
\begin{subfigure}[!htb]{.45\textwidth}
 \centering
  \includegraphics[trim=0cm 0cm 0cm 0cm,clip,width=1\linewidth]{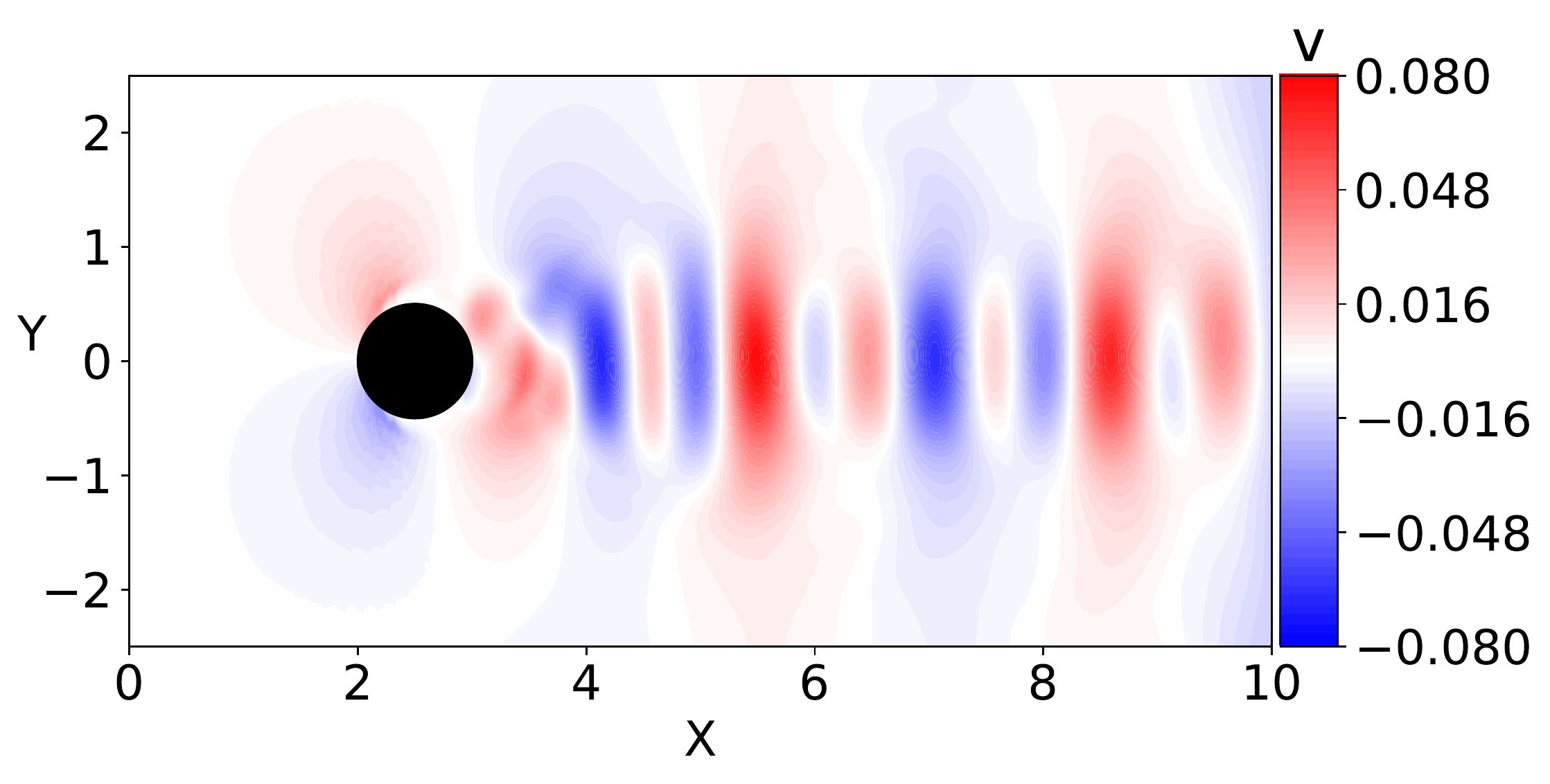}
  \caption{}
  \label{fig:QUBO_v1_MS}
\end{subfigure}
\begin{subfigure}[!htb]{.45\textwidth}
 \centering
  \includegraphics[trim=0cm 0cm 0cm 0cm,clip,width=1\linewidth]{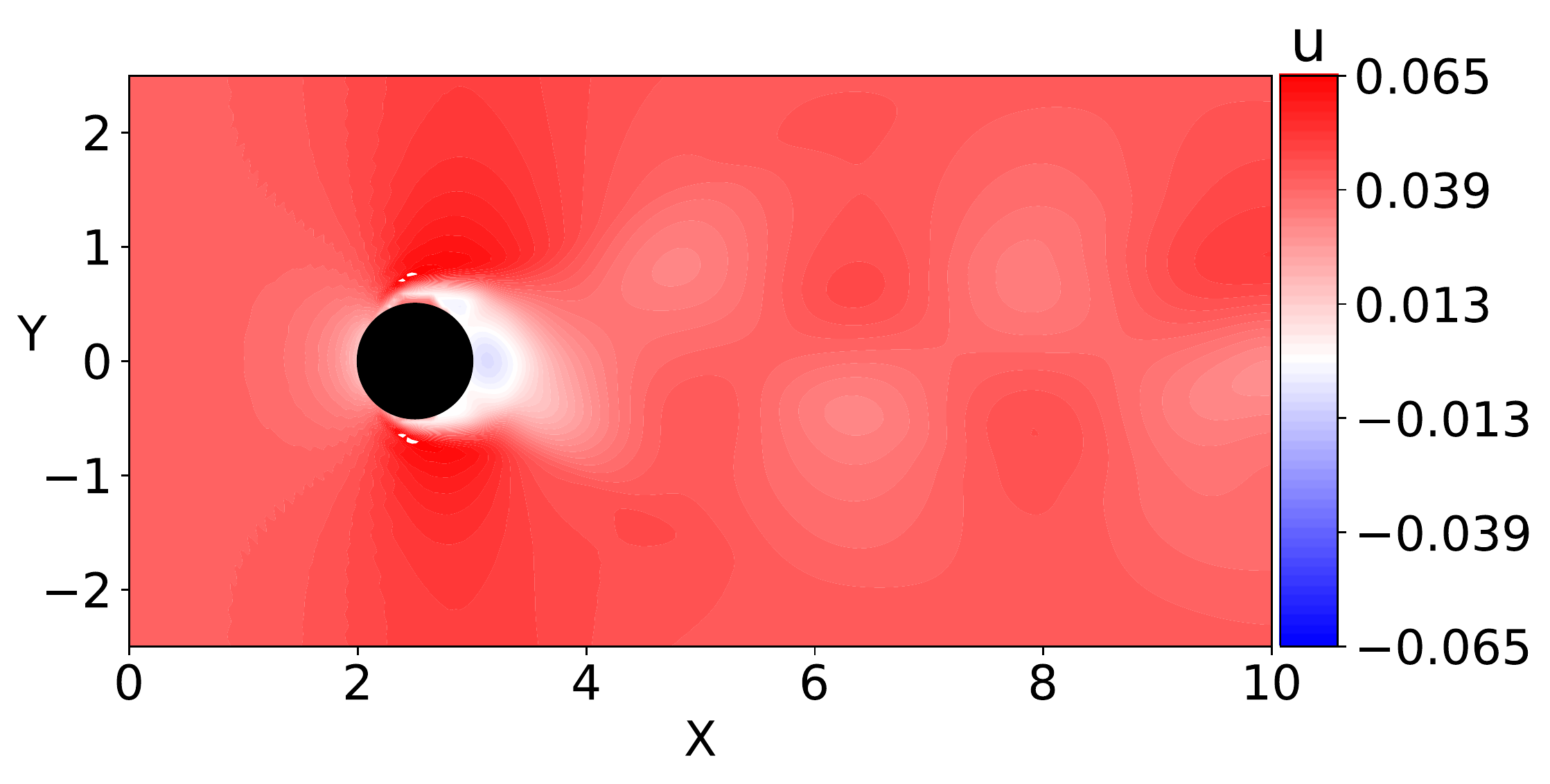}
  \caption{}
  \label{fig:QUBO_QAOA_u1_MS}
\end{subfigure}
\begin{subfigure}[!htb]{.45\textwidth}
 \centering
  \includegraphics[trim=0cm 0cm 0cm 0cm,clip,width=1\linewidth]{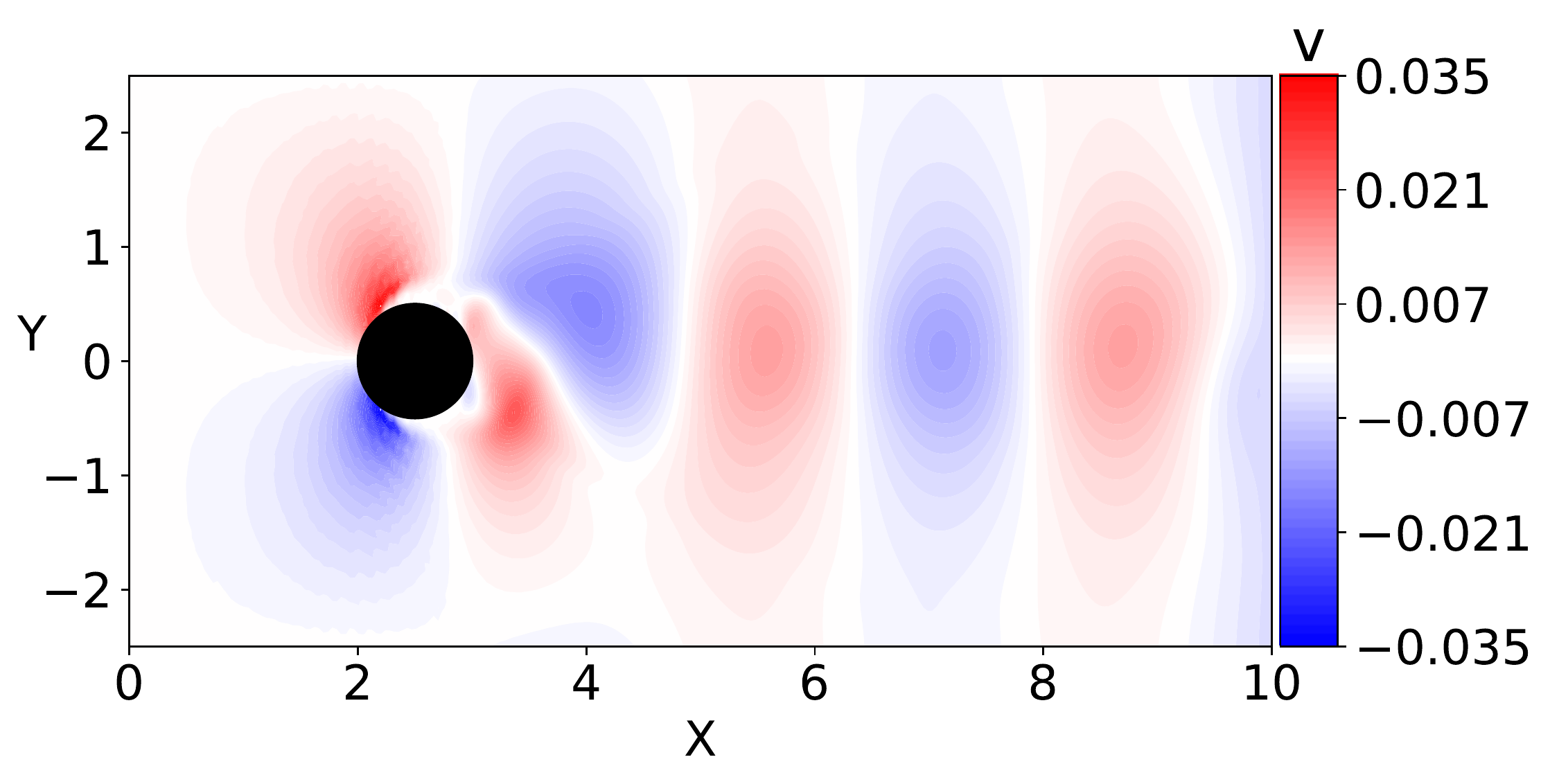}
  \caption{}
  \label{fig:QUBO_QAOA_v1_MS}
\end{subfigure}
\caption{Reconstructed flowfield for both velocity components computed at t = 10 from (a) -- (b) DMD using r = 5 modes, (c) -- (d) QUBO using r = 5 modes and p = 2 bits, and (e) -- (f) QUBO-QAOA using r = 5 and p = 2 bits. The reconstructed flowfield computed from predictions generated by QUBO for a reduced-order model or r = 5 modes and p = 2 bits for $N_{steps} = 2$.}
\label{fig:DMD_QUBO_QAOA_MS_reconstruct_r5_p2}
\end{figure}

\begin{figure}[htb!]
 \centering
  \includegraphics[trim=0cm 0cm 0cm 0cm,clip,width=0.6\linewidth]{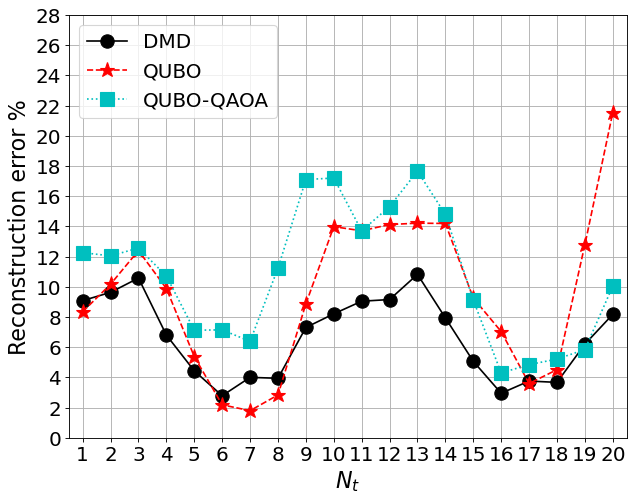}
\caption{Reconstruction error computed for implicit DMD, QUBO, and QUBO-QAOA optimization methodologies for $N_{steps} = 2$.}
\label{fig:error_reconstruct_MS}
\end{figure}

The reconstructed flowfield is shown in figure~\ref{fig:DMD_QUBO_QAOA_MS_reconstruct_r5_p2} at a timestep corresponding to t = 16, corresponding to a minimum in the reconstruction error as shown in figure~\ref{fig:error_reconstruct_MS}.

The reconstruction error can be seen in figure~\ref{fig:error_reconstruct_MS} computed for a r = 5 and p = 2 model.

\clearpage



\bibliographystyle{elsarticle-num}
\bibliography{qc-dmd}





\end{document}